\documentclass[runningheads,openany]{svmult}

\usepackage{palatino}
\usepackage{euler}
\usepackage{helvet}

\usepackage{graphicx}  
\usepackage{physprbb}  

\usepackage{proc}  
\usepackage{epsfig}

\usepackage{amssymb}

\makeindex             

\usepackage{amsmath}   
\usepackage{slashed}
\let\tocauthor\authorrunning

\begin{document}

\frontmatter

\thispagestyle{empty}
\parindent=0pt

{\Large\sc Blejske delavnice iz fizike \hfill Letnik~3, \v{s}t. 4}

\smallskip

{\large\sc Bled Workshops in Physics \hfill Vol.~3, No.~4}

\smallskip

\hrule

\hrule

\hrule

\vspace{0.5mm}

\hrule

\medskip
{\sc ISSN 1580--4992}

\vfill

\bigskip\bigskip
\begin{center}

{\bfseries 
{\Huge  Proceedings to the workshops\\
What comes beyond the Standard model 2000, 2001, 2002}

\vspace{5mm}
\centerline{\Large  Volume 2} 
\vspace{5mm}
\centerline{\Huge Proceedings --- PART II}  }

\vfill

{\bfseries\large
Edited by

\vspace{5mm}
Norma Manko\v c Bor\v stnik\rlap{$^{1,2}$}

\smallskip

Holger Bech Nielsen\rlap{$^{3}$}

\smallskip

Colin D. Froggatt\rlap{$^{4}$}

\smallskip

Dragan Lukman\rlap{$^2$}

\bigskip

{\em\normalsize $^1$University of Ljubljana, $^2$PINT, %
$^3$ Niels Bohr Institute, $^4$ Glasgow University}

\vspace{12pt}

\vspace{3mm}

\vrule height 1pt depth 0pt width 54 mm}

\vspace*{3cm}

{\large {\sc  DMFA -- zalo\v{z}ni\v{s}tvo} \\[6pt]
{\sc Ljubljana, december 2002}}
\end{center}
\newpage
\thispagestyle{empty}
\parindent=0pt
\begin{flushright}
{\parskip 6pt
{\bfseries\large
                  The 5th Workshop \textit{What Comes Beyond  
                  the Standard Model}}

\bigskip\bigskip

{\bf was organized by}

{\parindent8pt
\textit{Department of Physics, Faculty of Mathematics and Physics,
University of Ljubljana}

\textit{Primorska Institute of Natural Sciences and Technology, Koper}}

\bigskip

{\bf and sponsored by}

{\parindent8pt
\textit{Ministry of Education, Science and Sport of Slovenia}

\textit{Department of Physics, Faculty of Mathematics and Physics,
University of Ljubljana}

\textit{Primorska Institute of Natural Sciences and Technology, Koper}

\textit{Society of Mathematicians, Physicists and Astronomers
of Slovenia}}}
\bigskip
\medskip

{\bf Organizing Committee}

\medskip

{\parindent9pt
\textit{Norma Manko\v c Bor\v stnik}

\textit{Colin D. Froggatt}

\textit{Holger Bech Nielsen}}

\end{flushright}

\setcounter{tocdepth}{0}
\tableofcontents

\mainmatter
\parindent=20pt
\setcounter{page}{106}
\title*{%
Renormalization of Coupling Constants in the Minimal SUSY Models\thanks{%
Editors' note: This contribution was intended for the Holger Bech Nielsen's 
Festschrift (Vol. 1 of this Proceedings), but was received late, so 
we include it here.}}
\author{%
R. B. Nevzorov${}^{\dag,\ddag}$, K. A. Ter-Martirosyan${}^{\dag}$ %
and M. A. Trusov${}^{\dag}$}
\institute{%
${}^{\dag}$ITEP, Moscow, Russia,\\
${}^{\ddag}$DESY Theory, Hamburg, Germany}

\authorrunning{R. B. Nevzorov, K. A. Ter-Martirosyan and M. A. Trusov}
\titlerunning{Renormalization of Coupling Constants in the Minimal SUSY Models}
\maketitle

\begin{abstract}
The considerable part of the parameter space in the MSSM
corresponding to the infrared quasi fixed point scenario is
excluded by LEP\,II bounds on the lightest Higgs boson mass. In
the NMSSM the mass of the lightest Higgs boson reaches its maximum
value in the strong Yukawa coupling limit when Yukawa couplings
are essentially larger than gauge ones at the Grand Unification
scale. In this case the renormalization group flow of Yukawa
couplings and soft SUSY breaking terms is investigated. The
quasi--fixed and invariant lines and surfaces are briefly
discussed. The coordinates of the quasi--fixed points, where all
solutions are concentrated, are given.
\end{abstract}

\section{Introduction}

The search for the Higgs boson remains one of the top priorities
for existing accelerators as well as for those still at the design
stage. This is because this boson plays a key role in the Standard
Model which describes all currently available experimental data
with a high degree of accuracy. As a result of the spontaneous
symmetry breaking $SU(2)\otimes U(1)$ the Higgs scalar acquires a
nonzero vacuum expectation value without destroying the Lorentz
invariance, and generates the masses of all fermions and vector
bosons. An analysis of the experimental data using the Standard
Model has shown that there is a $95\%$ probability that its mass
will not exceed $210\text{~GeV}$ \cite{A1}. At the same time,
assuming that there are no new fields and interactions and also no
Landau pole in the solution of the renormalization group equations
for the self-action constant of Higgs fields up to the scale
$M_\text{Pl}\approx 2.4\cdot 10^{18}\text{~GeV}$, we can show that
$m_h<180\text{~GeV}$ \cite{A2},\cite{A3}. In this case, physical
vacuum is only stable provided that the mass of the Higgs boson is
greater than $135\text{~GeV}$ \cite{A2}-\cite{A6}. However, it
should be noted that this simplified model does not lead to
unification of the gauge constants \cite{A7} and a solution of the
hierarchy problem \cite{A8}. As a result, the construction of a
realistic theory which combines all the fields and interactions is
extremely difficult in this case.

Unification of the gauge constants occurs naturally on the  scale
$M_X\approx 3\cdot 10^{16}\text{~GeV}$ within the supersymmetric
generalisation of the Standard Model, i.e., the Minimal
Supersymmetric Standard Model (MSSM) \cite{A7}. In order that all
the fundamental fermions acquire mass in the MSSM, not one but two
Higgs doublets $H_1$ and $H_2$ must be introduced in the theory,
each acquiring the nonzero vacuum expectation value $v_1$ and
$v_2$ where $v^2=v_1^2+v_2^2=(246\text{~GeV})^2$. The spectrum of
the Higgs sector of the MSSM contains four massive states: two
CP--even, one CP--odd, and one charged. An important
distinguishing feature of the supersymmetric model is the existing
of a light Higgs boson in the CP--even sector. The upper bound on
its mass is determined to a considerable extent by the value
$\tan\beta=v_2/v_1$. In the tree-level approximation the mass of
the lightest Higgs boson in the MSSM does not exceed the mass of
the Z-boson ($M_Z\approx 91.2\text{~GeV}$): $m_h\le M_Z|\cos
2\beta|$ \cite{A9}. Allowance for the contribution of loop
corrections to the effective interaction potential of the Higgs
fields from a $t$--quark and its superpartners significantly
raises the upper bound on its mass:
\begin{equation}
m_h\le\sqrt{M_Z^2\cos^2 2\beta+\Delta}\, . \label{A1}
\end{equation}
Here $\Delta$ are the loop corrections \cite{A10},\cite{A11}. The
values of these corrections are proportional to $m_t^4$, where
$m_t$ is the running mass of $t$--quark which depends
logarithmically on the supersymmetry breaking scale $M_S$ and is
almost independent of the choice of $\tan\beta$. In
\cite{A3},\cite{A5},\cite{A6} bounds on the mass of the Higgs
boson were compared in the Minimal Standard and Supersymmetric
models. The upper bound on the mass of the light CP--even Higgs
boson in the MSSM increases with increasing $\tan\beta$ and for
$\tan\beta\gg 1$ in realistic supersymmetric models with $M_S\le
1000\text{~GeV}$ reaches $125-128\text{~GeV}$.

However, a considerable fraction of the solutions of the system of
MSSM renormalization group equations is focused near the infrared
quasi-fixed point at $\tan\beta\sim 1$. In the region of parameter
space of interest to us ($\tan\beta\ll 50$) the Yukawa constants
of a $b$--quark ($h_b$) and a $\tau$--lepton ($h_\tau$) are
negligible so that an exact analytic solution can be obtained for
the one--loop renormalization group equations \cite{A12}. For the
Yukawa constants of a $t$--quark $h_t(t)$ and the gauge constants
$g_i(t)$ its solution has the following form:
\begin{equation}
\begin{gathered}
Y_t(t)=\frac{\dfrac{E(t)}{6F(t)}}{1+\dfrac{1}{6Y_t(0)F(t)}},\quad
\tilde{\alpha}_i(t)=\frac{\tilde{\alpha}_i(0)}{1+b_i\tilde{\alpha}_i(0)t},
\\
E(t)=\left[\frac{\tilde{\alpha}_3(t)}{\tilde{\alpha}_3(0)}\right]^{16/9}
\left[\frac{\tilde{\alpha}_2(t)}{\tilde{\alpha}_2(0)}\right]^{-3}
\left[\frac{\tilde{\alpha}_1(t)}{\tilde{\alpha}_1(0)}\right]^{-13/99}
,\quad F(t)=\int\limits_0^t E(t')dt',
\end{gathered}
\label{A2}
\end{equation}
where the index $i$ has values between 1 and 3,
\begin{gather*}
b_1=33/5,\quad b_2=1,\quad b_3=-3\\
\tilde\alpha_i(t)=\left(\frac{g_i(t)}{4\pi}\right)^2,\quad
Y_i(t)=\left(\frac{h_t(t)}{4\pi}\right)^2.
\end{gather*}
The variable $t$ is determined by a standard method
$t=\ln(M_X^2/q^2)$. The boundary conditions for the
renormalization group equations are usually set at the grand
unification scale $M_X$ ($t=0$) where the values of all three
Yukawa constants are the same:
$\tilde\alpha_1(0)=\tilde\alpha_2(0)=\tilde\alpha_3(0)=\tilde\alpha_0.$
On the electroweak scale where $h_t^2(0)\gg 1$ the second term in
the denominator of the expression describing the evolution of
$Y_t(t)$ is much smaller than unity and all the solutions are
concentrated in a narrow interval near the quasi-fixed point
$Y_\text{QFP}(t)=E(t)/6F(t)$ \cite{A13}. In other words in the
low-energy range the dependence of $Y_t(t)$ on the initial
conditions on the scale $M_X$ disappears. In addition to the
Yukawa constant of the $t$--quark, the corresponding trilinear
interaction constant of the scalar fields $A_t$ and the
combination of the scalar masses
$\mathfrak{M}^2_t=m_Q^2+m_U^2+m_2^2$ also cease to depend on
$A_t(0)$ and $\mathfrak{M}_t^2(0)$ as $Y_t(0)$ increases. Then on
the electroweak scale near the infrared quasi--fixed point
$A_t(t)$ and $\mathfrak{M}_t^2(t)$ are only expressed in terms of
the gaugino mass on the Grand Unification scale. Formally this
type of solution can be obtained if $Y_t(0)$ is made to go to
infinity. Deviations from this solution are determined by ratio
$1/6F(t)Y_t(0)$ which is of the order of $1/10h_t^2(0)$ on the
electroweak scale.

The properties of the solutions of the system of MSSM
renormalization group equations and also the particle spectrum
near the infrared quasi-fixed point for $\tan\beta\sim 1$ have
been studied by many authors \cite{A14},\cite{A15}. Recent
investigations \cite{A15}-\cite{A17} have shown that for solutions
$Y_t(t)$ corresponding to the quasi-fixed point regime the value
of $\tan\beta$ is between $1.3$ and $1.8$. These comparatively low
values of $\tan\beta$ yield significantly more stringent bounds on
the mass of the lightest Higgs boson. The weak dependence of the
soft supersymmetry breaking parameters $A_t(t)$ and
$\mathfrak{M}_t^2(t)$ on the boundary conditions near the
quasi-fixed point means that the upper bound on its mass can be
calculated fairly accurately. A theoretical analysis made in
\cite{A15},\cite{A16} showed that $m_h$ does not exceed $94\pm
5\text{~GeV}$. This bound is $25-30\text{~GeV}$ below the absolute
upper bound in the Minimal Supersymmetric Model. Since the lower
bound on the mass of the Higgs boson from LEP\,II data is
$113\text{~GeV}$ \cite{A1}, which for the spectrum of heavy
supersymmetric particles is the same as the corresponding bound on
the mass of the Higgs boson in the Standard Model, a considerable
fraction of the solutions which come out to a quasi--fixed point
in the MSSM, are almost eliminated by existing experimental data.
This provides the stimulus for theoretical analyses of the Higgs
sector in more complex supersymmetric models.

The simplest expansion of the MSSM which can conserve the
unification of the gauge constants and raise the upper bound on
the mass of the lightest Higgs boson is the Next--to--Minimal
Supersymmetric Standard Model (NMSSM) \cite{A18}-\cite{A20}. By
definition the superpotential of the NMSSM is invariant with
respect to the discrete transformations $y'_\alpha=e^{2\pi
i/3}y_\alpha$ of the $Z_3$ group \cite{A19} which means that we
can avoid the problem of the $\mu$-term in supergravity models.
The thing is that the fundamental parameter $\mu$ should be of the
order of $M_\text{Pl}$ since this scale is the only dimensional
parameter characterising the hidden (gravity) sector of the
theory. In this case, however, the Higgs bosons $H_1$ and $H_2$
acquire an enormous mass $m^2_{H_1,H_2}\sim \mu^2\sim
M^2_\text{Pl}$ and no breaking of $SU(2)\otimes U(1)$ symmetry
occurs. In the NMSSM the term $\mu(\hat{H}_1\hat{H}_2)$ in the
superpotential is not invariant with respect to discrete
transformations of the $Z_3$ group and for this reason should be
eliminated from the analysis ($\mu=0$). As a result of the
multiplicative nature of the renormalization of this parameter,
the term $\mu(q)$ remains zero on any scale $q\le M_X\div
M_\text{Pl}$. However, the absence of mixing of the Higgs doublets
on electroweak scale has the result that $H_1$ acquires no vacuum
expectation value as a result of the spontaneous symmetry breaking
and $d$--type quarks and charged leptons remain massless. In order
to ensure that all quarks and charged leptons acquire nonzero
masses, an additional singlet superfield $\hat{Y}$ with respect to
gauge $SU(2)\otimes U(1)$ transformations is introduced in the
NMSSM. The superpotential of the Higgs sector of the Nonminimal
Supersymmetric Model \cite{A18}-\cite{A20} has the following form:
\begin{equation}
W_h=\lambda
\hat{Y}(\hat{H}_1\hat{H}_2)+\frac{\varkappa}{3}\hat{Y}^3.
\label{A4}
\end{equation}
As a result of the spontaneous breaking of $SU(2)\otimes U(1)$
symmetry, the field $Y$ acquires a vacuum expectation value
($\langle Y\rangle=y/\sqrt{2}$) and the effective $\mu$-term
($\mu=\lambda y/\sqrt{2}$) is generated.

In addition to the Yukawa constants $\lambda$ and $\varkappa$, and
also the Standard Model constants, the Nonminimal Supersymmetric
Model contains a large number of unknown parameters. These are the
so-called soft supersymmetry breaking parameters which are
required to obtain an acceptable spectrum of superpartners of
observable particles form the phenomenological point of view. The
hypothesis on the universal nature of these constants on the Grand
Unification scale allows us to reduce their number in the NMSSM to
three: the mass of all the scalar particles $m_0$, the gaugino
mass $M_{1/2}$, and the trilinear interaction constant of the
scalar fields $A$. In order to avoid strong CP--violation and also
spontaneous breaking of gauge symmetry at high energies
($M_\text{Pl}\gg E\gg m_t$) as a result of which the scalar
superpartners of leptons and quarks would require nonzero vacuum
expectation values, the complex phases of the soft supersymmetry
breaking parameters are assumed to be zero and only positive
values of $m_0^2$ are considered. Naturally universal
supersymmetry breaking parameters appear in the minimal
supergravity model \cite{A31} and also in various string models
\cite{A29},\cite{A32}. In the low-energy region the hypothesis of
universal fundamental parameters allows to avoid the appearance of
neutral currents with flavour changes and can simplify the
analysis of the particle spectrum as far as possible. The
fundamental parameters thus determined on the Grand Unification
scale should be considered as boundary conditions for the system
of renormalization group equations which describes the evolution
of these constants as far as the electroweak scale or the
supersymmetry breaking scale. The complete system of the
renormalization group equations of the Nonminimal Supersymmetric
Model can be found in \cite{A33}, \cite{A34}. These experimental
data impose various constraints on the NMSSM parameter space which
were analysed in \cite{A35},\cite{A36}.

The introduction of the neutral field $Y$  in the NMSSM potential
leads to the appearance of a corresponding $F$--term in the
interaction potential of the Higgs fields. As a consequence, the
upper bound on the mass of the lightest Higgs boson is increased:
\begin{equation}
m_h\le\sqrt{\frac{\lambda^2}{2}v^2\sin^2 2\beta+M_Z^2\cos^2
2\beta+\Delta^{(1)}_{11}+\Delta^{(2)}_{11}} . \label{A5}
\end{equation}
The relationship (\ref{A5}) was obtained in the tree-level
approximation ($\Delta_{11}=0$) in \cite{A20}. However, loop
corrections to the effective interaction potential of the Higgs
fields from the $t$--quark and its superpartners play a very
significant role. In terms of absolute value their contribution to
the upper bound on the mass of the Higgs boson remains
approximately the same as in the Minimal Supersymmetric Model.
When calculating the corrections $\Delta_{11}^{(1)}$ and
$\Delta_{11}^{(2)}$ we need to replace the parameter $\mu$ by
$\lambda y/\sqrt{2}$. Studies of the Higgs sector in the
Nonminimal Supersymmetric model and the one--loop corrections to
it were reported in \cite{A24},\cite{A33},\cite{A36}-\cite{A39}.
In \cite{A6} the upper bound on the mass of the lightest Higgs
boson in the NMSSM was compared with the corresponding bounds on
$m_h$ in the Minimal Standard and Supersymmetric Models. The
possibility of a spontaneous CP--violation in the Higgs sector of
the NMSSM was studied in \cite{A39},\cite{A40}.

It follows from condition (\ref{A5}) that the upper bound on $m_h$
increases as $\lambda$ increases. Moreover, it only differs
substantially from the corresponding bound in the MSSM in the
range of small $\tan\beta$. For high values ($\tan\beta\gg 1$) the
value of $\sin 2\beta$ tends to zero and the upper bounds on the
mass of the lightest Higgs boson in the MSSM and NMSSM are almost
the same. The case of small $\tan\beta$ is only achieved for
fairly high values of the Yukawa constant of a $t$--quark $h_t$ on
the electroweak scale ($h_t(t_0)\ge 1$ where
$t_0=\ln(M_X^2/m_t^2)$), and $\tan\beta$ decreases with increasing
$h_t(t_0)$. However, an analysis of the renormalization group
equations in the NMSSM shows that an increase of the Yukawa
constants on the electroweak scale is accompanied by an increase
of $h_t(0)$ and $\lambda(0)$ on the Grand Unification scale. It
thus becomes obvious that the upper bound on the mass of the
lightest Higgs boson in the Nonminimal Supersymmetric model
reaches its maximum on the strong Yukawa coupling limit, i.e.,
when $h_t(0)\gg g_i(0)$ and $\lambda(0)\gg g_i(0)$.

\section{Renormalization of the Yukawa couplings}

From the point of view of a renormalization group analysis,
investigation of the NMSSM presents a much more complicated
problem than investigation of the minimal SUSY model. The full set
of renormalization group equations within the NMSSM can be found
in \cite{A33},\cite{A34}. Even in the one--loop approximation,
this set of equations is nonlinear and its analytic solution does
not exist. All equations forming this set can be partitioned into
two groups. The first one contains equations that describe the
evolution of gauge and Yukawa coupling constants, while the second
one includes equations for the parameters of a soft breakdown of
SUSY, which are necessary for obtaining a phenomenologically
acceptable spectrum of superpartners of observable particles.
Since boundary conditions for three Yukawa coupling constants are
unknown, it is very difficult to perform a numerical analysis of
the equations belonging to the first group and of the full set of
the equations given above. In the regime of strong Yukawa
coupling, however, solutions to the renormalization group
equations are concentrated in a narrow region of the parameter
space near the electroweak scale, and this considerably simplifies
the analysis of the set of equations being considered.

In analysing the nonlinear differential equations entering into
the first group, it is convenient to use the quantities $\rho_t$,
$\rho_\lambda$, $\rho_\varkappa$, $\rho_1$, and $\rho_2$, defined
as follows:
\[
\rho_t(t)=\frac{Y_t(t)}{\tilde{\alpha}_3(t)},\quad
\rho_{\lambda}(t)=\frac{Y_{\lambda}(t)}{\tilde{\alpha}_3(t)},\quad
\rho_{\varkappa}(t)=\frac{Y_{\varkappa}(t)}{\tilde{\alpha}_3(t)},
\]
\[
\rho_1(t)=\frac{\tilde{\alpha}_1(t)}{\tilde{\alpha}_3(t)},\quad
\rho_2(t)=\frac{\tilde{\alpha}_2(t)}{\tilde{\alpha}_3(t)},
\]
where $\tilde{\alpha}_i(t)=g^2_i(t)/(4\pi)^2$,
$Y_t(t)=h^2_t(t)/(4\pi)^2$,
$Y_{\lambda}(t)=\lambda^2(t)/(4\pi)^2$, and
$Y_{\varkappa}(t)=\varkappa^2(t)/(4\pi)^2$.

\begin{figure}
\centering
\includegraphics[width=120mm]{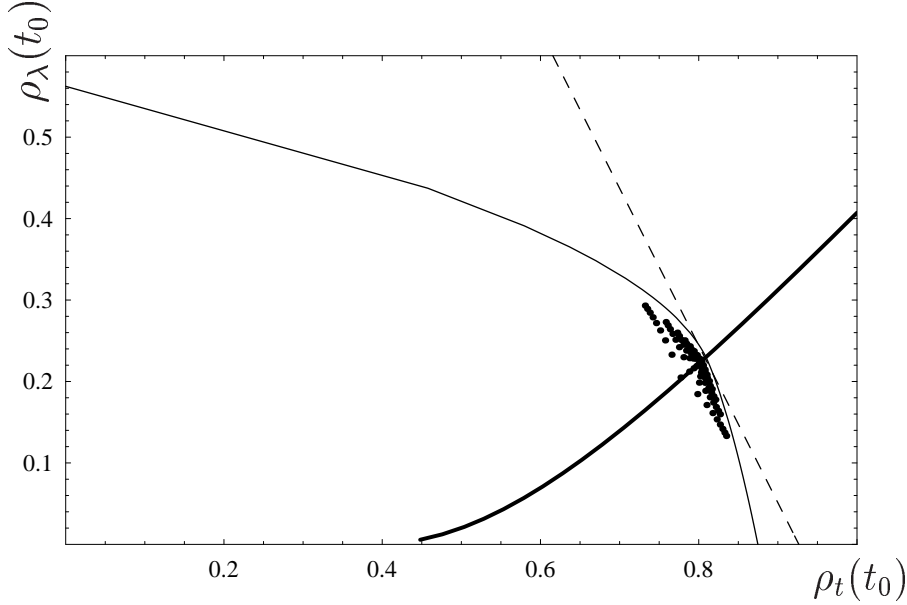}
\caption{%
The values of the Yukawa couplings at the
electroweak scale corresponding to the initial values at the GUT
scale uniformly distributed in a square $2\le
h_t^2(0),\lambda^2(0)\le 10$. The thick and thin curves represent,
respectively, the invariant and the Hill line. The dashed line is
a fit of the values $(\rho_t(t_0),\rho_\lambda(t_0))$ for $20\le
h_t^2(0),\lambda^2(0)\le 100$.}
\label{ntt-fig1}
\end{figure}

\begin{figure}
\centering
\includegraphics[width=120mm]{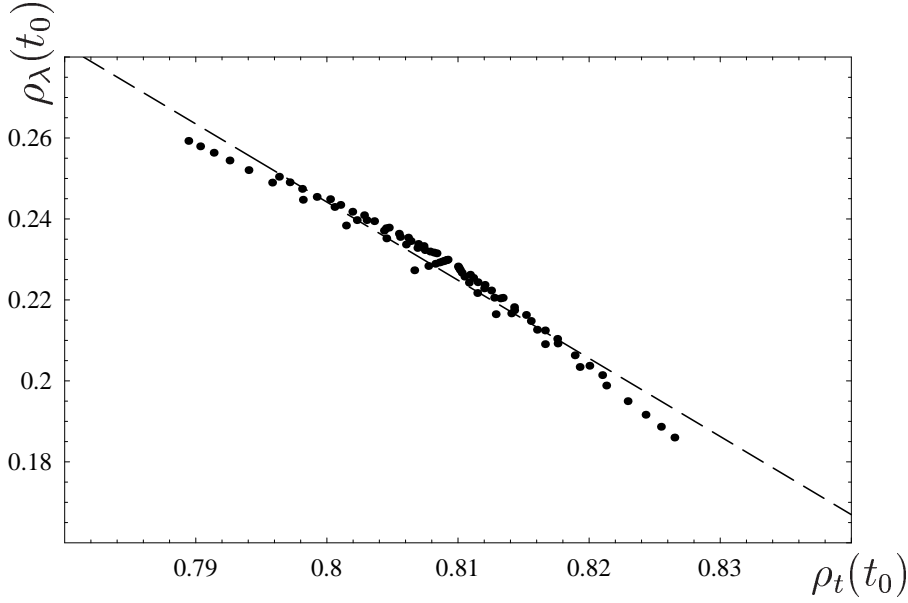}
\caption{%
The values of the Yukawa couplings at the
electroweak scale corresponding to the initial values at the GUT
scale uniformly distributed in a square $20\le
h_t^2(0),\lambda^2(0)\le 100$. The dashed line is a fit of the
values $(\rho_t(t_0),\rho_\lambda(t_0))$ for $20\le
h_t^2(0),\lambda^2(0)\le 100$.}
\label{ntt-fig2}
\end{figure}

Let us first consider the simplest case of $\varkappa=0$. The
growth of the Yukawa coupling constant $\lambda(t_0)$ at a fixed
value of $h_t(t_0)$ results in that the Landau pole in solutions
to the renormalization group equations approaches the Grand
Unification scale from above. At a specific value
$\lambda(t_0)=\lambda_{\text{max}}$, perturbation theory at $q\sim
M_X$ cease to be applicable. With increasing (decreasing) Yukawa
coupling constant for the $b$--quark, $\lambda_{\text{max}}$
decreases (increases). In the $(\rho_t,\rho_\lambda)$ plane, the
dependence $\lambda^2_{\text{max}}(h_t^2)$ is represented by a
curve bounding the region of admissible values of the parameters
$\rho_t(t_0)$ and $\rho_\lambda(t_0)$. At $\rho_\lambda=0$, this
curve intersects the abscissa at the point
$\rho_t=\rho_t^{\text{QFP}}(t_0)$. This is the way in which there
arises, in the $(\rho_t,\rho_\lambda)$ plane, the quasi--fixed (or
Hill) line near which solutions to the renormalization group
equations are grouped (see Fig. 1). With increasing $\lambda^2(0)$
and $h_t^2(0)$, the region where the solutions in questions are
concentrated sharply shrinks, and for rather large initial values
of the Yukawa coupling constants they are grouped in a narrow
stripe near the straight line
\begin{equation}
\rho_t(t_0)+0.506\rho_{\lambda}(t_0)=0.91, \label{B7}
\end{equation}
which can be obtained by fitting the results of numerical
calculations (these results are presented in Fig. 2). Moreover,
the combination $h_t^2(t_0)+0.506\lambda^2(t_0)$ of the Yukawa
coupling constants depends much more weakly on $\lambda^2(0)$ and
$h_t^2(0)$ than $\lambda^2(t_0)$ and $h_t^2(t_0)$ individually
\cite{NTB}. In other words, a decrease in $\lambda^2(t_0)$
compensates for an increase in $h_t^2(t_0)$, and vice versa. The
results in Fig. 3, which illustrate the evolution of the above
combinations of the Yukawa coupling constants, also confirm that
this combination is virtually independent of the initial
conditions.
\begin{figure}
\begin{center}
\includegraphics[width=120mm]{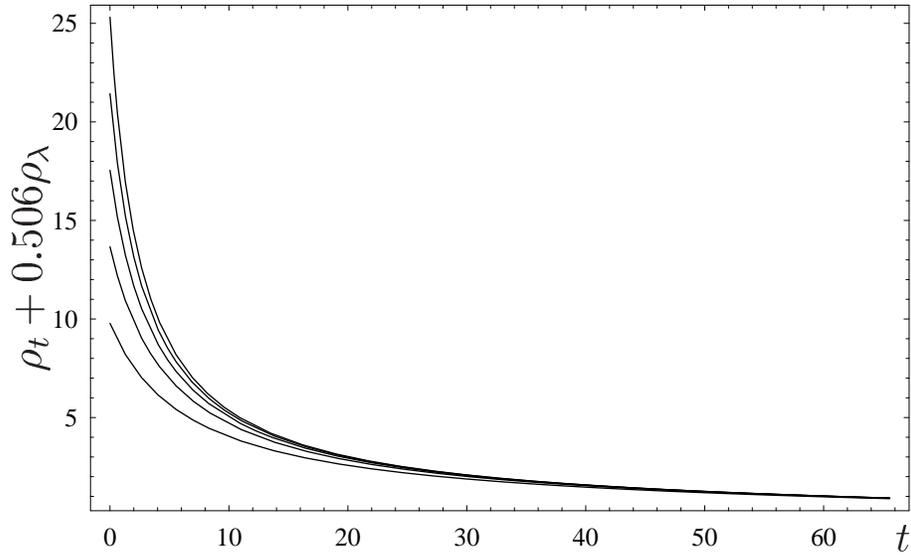}
\smallskip\centerline{(a)}\smallskip
\end{center}
\begin{center}
\includegraphics[width=120mm]{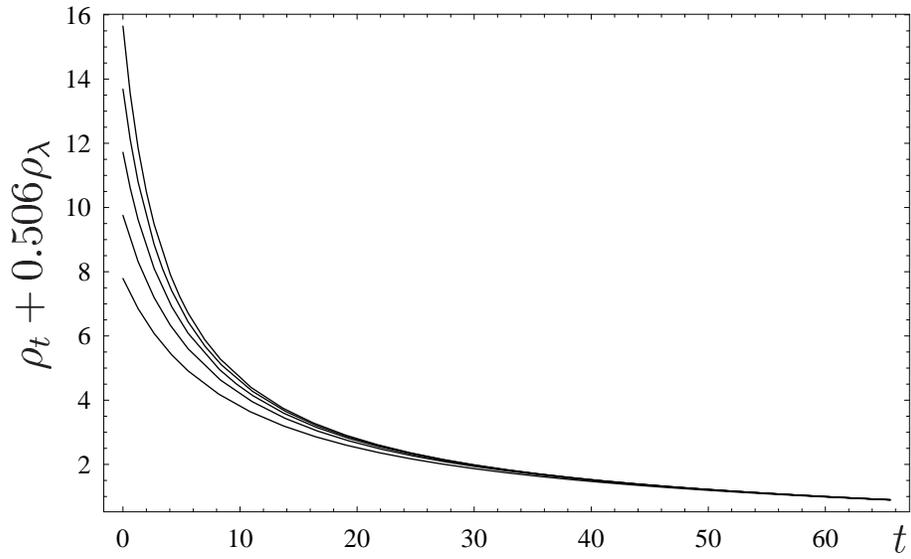}
\smallskip\centerline{(b)}\smallskip
\end{center}
\caption{%
Evolution of the combination
$\rho_t(t)+0.506\rho_\lambda(t)$ of the Yukawa couplings from the
GUT scale ($t=0$) to the electroweak scale ($t=t_0$) for
$\varkappa^2=0$ and for various initial values $h_t^2(0)$ -- Fig.
3a, $\lambda^2(0)$ -- Fig. 3b.}
\label{ntt-fig3}
\end{figure}

In analysing the results of numerical calculations, our attention
is engaged by a pronounced nonuniformity in the distribution of
solutions to the renormalization group equations along the
infrared quasi--fixed line. The main reason for this is that, in
the regime of strong Yukawa coupling, the solutions in question
are attracted not only to the quasi--fixed but also to the
infrared fixed (or invariant) line. The latter connects two fixed
points. Of these, one is an infrared fixed point of the set of
renormalization group equations within the NMSSM ($\rho_t=7/18$,
$\rho_{\lambda}=0$, $\rho_1=0$, $\rho_2=0$) \cite{B6}, while the
other fixed point $(\rho_\lambda/\rho_t=1)$ corresponds to values
of the Yukawa coupling constants in the region
$Y_t,Y_\lambda\gg\tilde{\alpha}_i$, in which case the gauge
coupling constants on the right--hand sides of the renormalization
group equations can be disregarded \cite{B35}. For the asymptotic
behaviour of the infrared fixed line at $\rho_t,\rho_\lambda\gg 1$
we have
\[
\rho_{\lambda}=\rho_t-\frac{8}{15}-\frac{2}{75}\rho_1,
\]
while in the vicinity of the point $\rho_t=7/18$, $\rho_\lambda=0$
we have
\[
\rho_\lambda\sim(\rho_t-7/18)^{25/14}.
\]

The infrared fixed line is invariant under renormalization group
transformations -- that is, it is independent of the scale at
which the boundary values $Y_t(0)$ and $Y_\lambda(0)$ are
specified and of the boundary values themselves. If the boundary
conditions are such that $Y_t(0)$ and $Y_\lambda(0)$ belong to the
fixed line, the evolution of the Yukawa coupling constants
proceeds further along this line toward the infrared fixed point
of the set of renormalization group equations within the NMSSM.
With increasing $t$, all other solutions to the renormalization
group equations are attracted to the infrared fixed line and, for
$t/(4\pi)\gg 1$, approach the stable infrared fixed point. From
the data in Figs. 1 and 2, it follows that, with increasing
$Y_t(0)$ and $Y_\lambda(0)$, all solutions to the renormalization
group equations are concentrated in the vicinity of the point of
intersection of the infrared fixed and the quasi--fixed line:
\[
\rho^{\text{QFP}}_t(t_0)=0.803,\qquad
\rho^{\text{QFP}}_{\lambda}(t_0)=0.224.
\]
Hence, this point can be considered as the quasi--fixed point of
the set of renormalization group equations within the NMSSM at
$\varkappa=0$.

In a more complicated case where all three Yukawa coupling
constants in the NMSSM are nonzero, analysis of the set of
renormalization group equations presents a much more difficult
problem. In particular, invariant (infrared fixed) and Hill
surfaces come to the fore instead of the infrared fixed and
quasi--fixed lines. For each fixed set of values of the coupling
constants $Y_t(t_0)$ and $Y_\varkappa(t_0)$, an upper limit on
$Y_\lambda(t_0)$ can be obtained from the requirement that
perturbation theory be applicable up to the Grand Unification
scale $M_X$. A change in the values of the Yukawa coupling
constants $h_t$ and $\varkappa$ at the electroweak scale leads to
a growth or a reduction of the upper limit on $Y_\lambda(t_0)$.
The resulting surface in the
$(\rho_t,\rho_\varkappa,\rho_\lambda)$ space is shown in Fig. 4.
In the regime of strong Yukawa coupling, solutions to the
renormalization group equations are concentrated near this
surface. In just the same way as in the case of $Y_\varkappa=0$, a
specific linear combination of $Y_t$, $Y_\lambda$, and
$Y_\varkappa$ is virtually independent of the initial conditions
for $Y_i(0)\to\infty$:
\begin{equation}
\rho_t(t_0)+0.72\rho_{\lambda}(t_0)+0.33\rho_{\varkappa}(t_0)=0.98.
\label{B14}
\end{equation}
The evolution of this combination of Yukawa couplings at various
initial values of the Yukawa coupling constants is illustrated in
Fig. 5.

\begin{figure}
\begin{center}
\includegraphics[width=100mm]{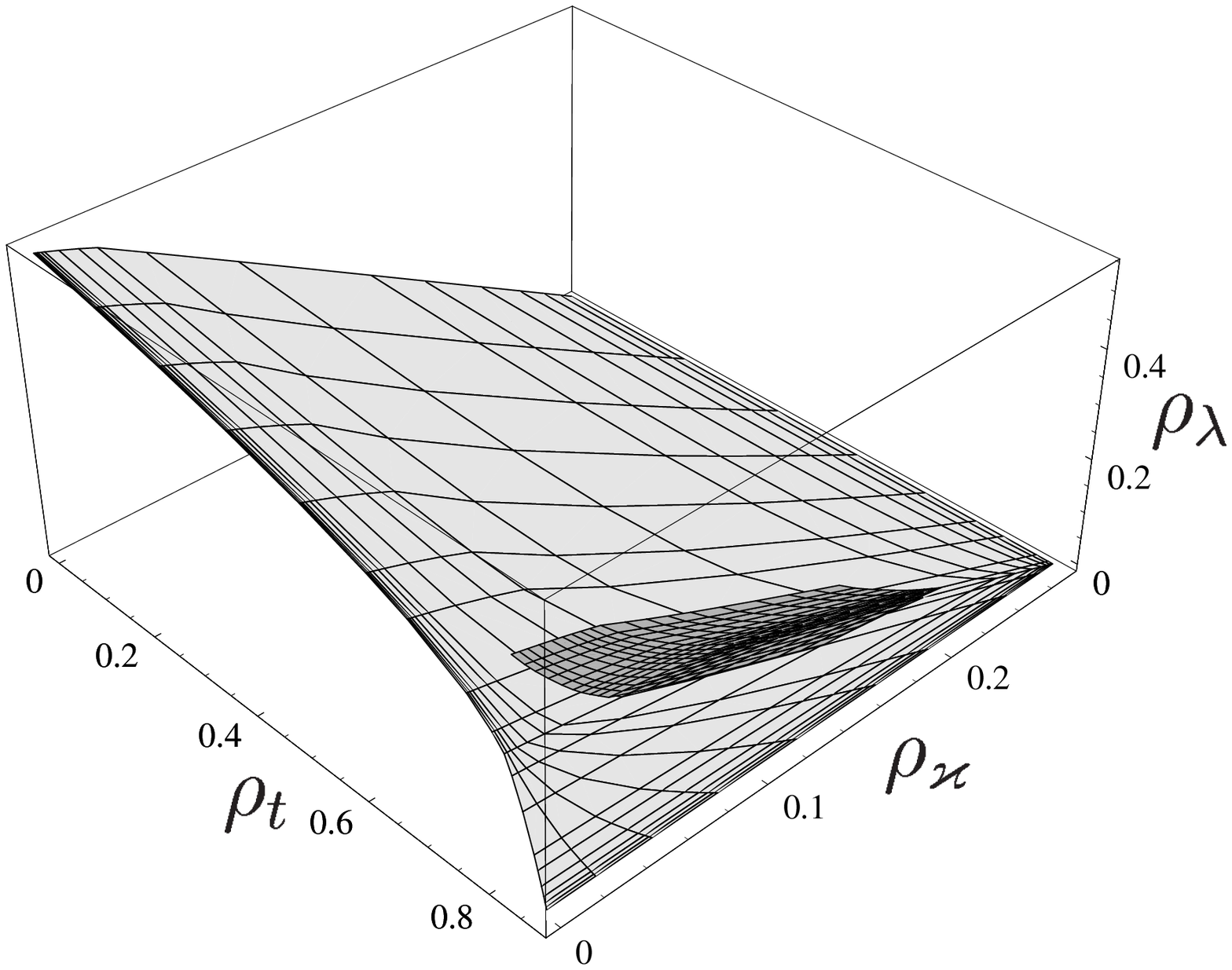}
\smallskip\centerline{(a)}\smallskip
\end{center}
\begin{center}
\includegraphics[width=100mm]{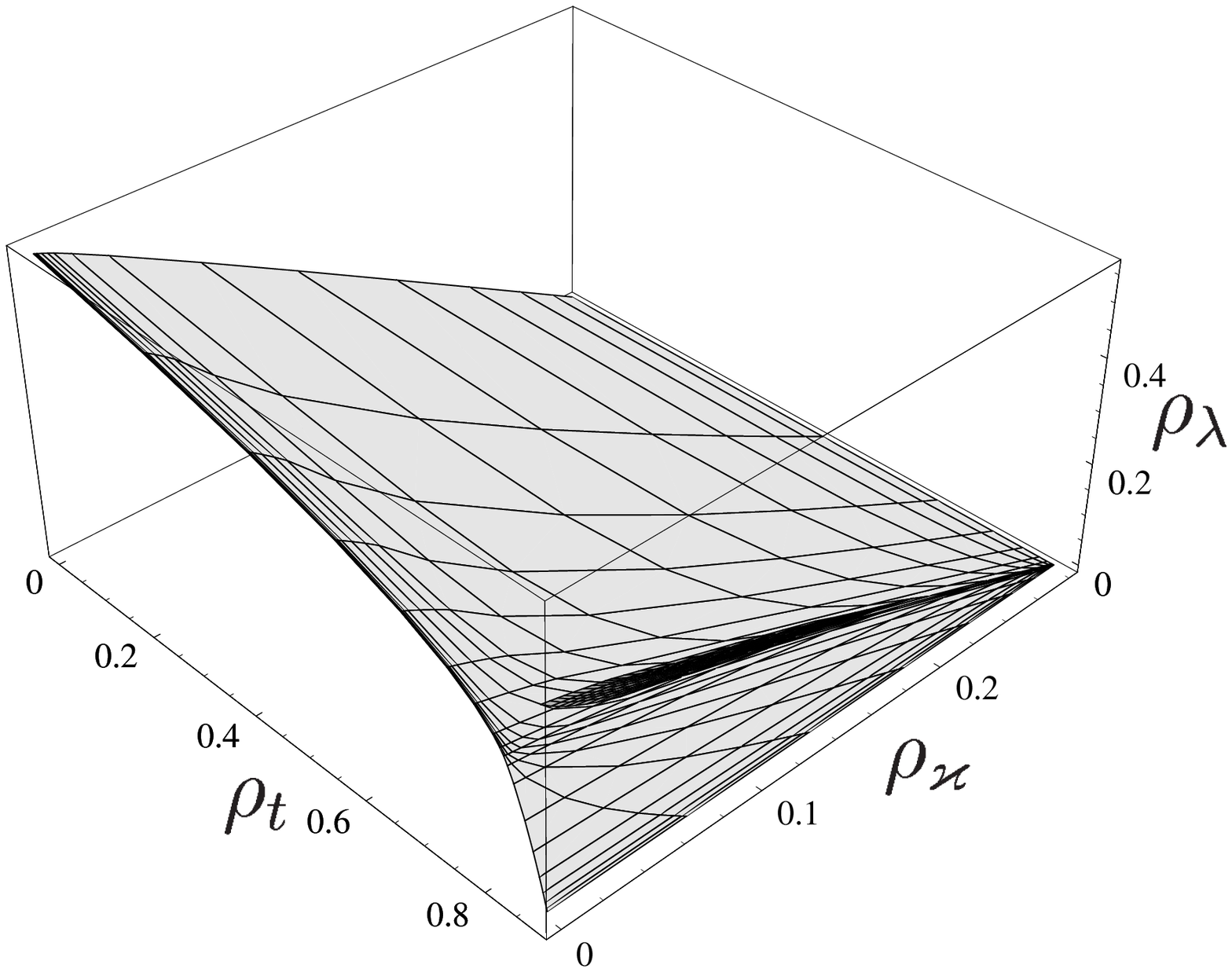}
\smallskip\centerline{(b)}\smallskip
\end{center}
\caption{%
Quasi--fixed surface in the
$(\rho_t,\rho_\varkappa,\rho_\lambda)$ space. The shaded part of
the surface represents the region near which the solutions
corresponding to the initial values $2\le
h_t^2(0),\varkappa^2(0),\lambda^2(0)\le 10$ -- Fig. 4a, $20\le
h_t^2(0),\varkappa^2(0),\lambda^2(0)\le 100$ -- Fig. 4b are
concentrated.}
\label{ntt-fig4}
\end{figure}

\begin{figure}
\begin{center}
\includegraphics[width=90mm]{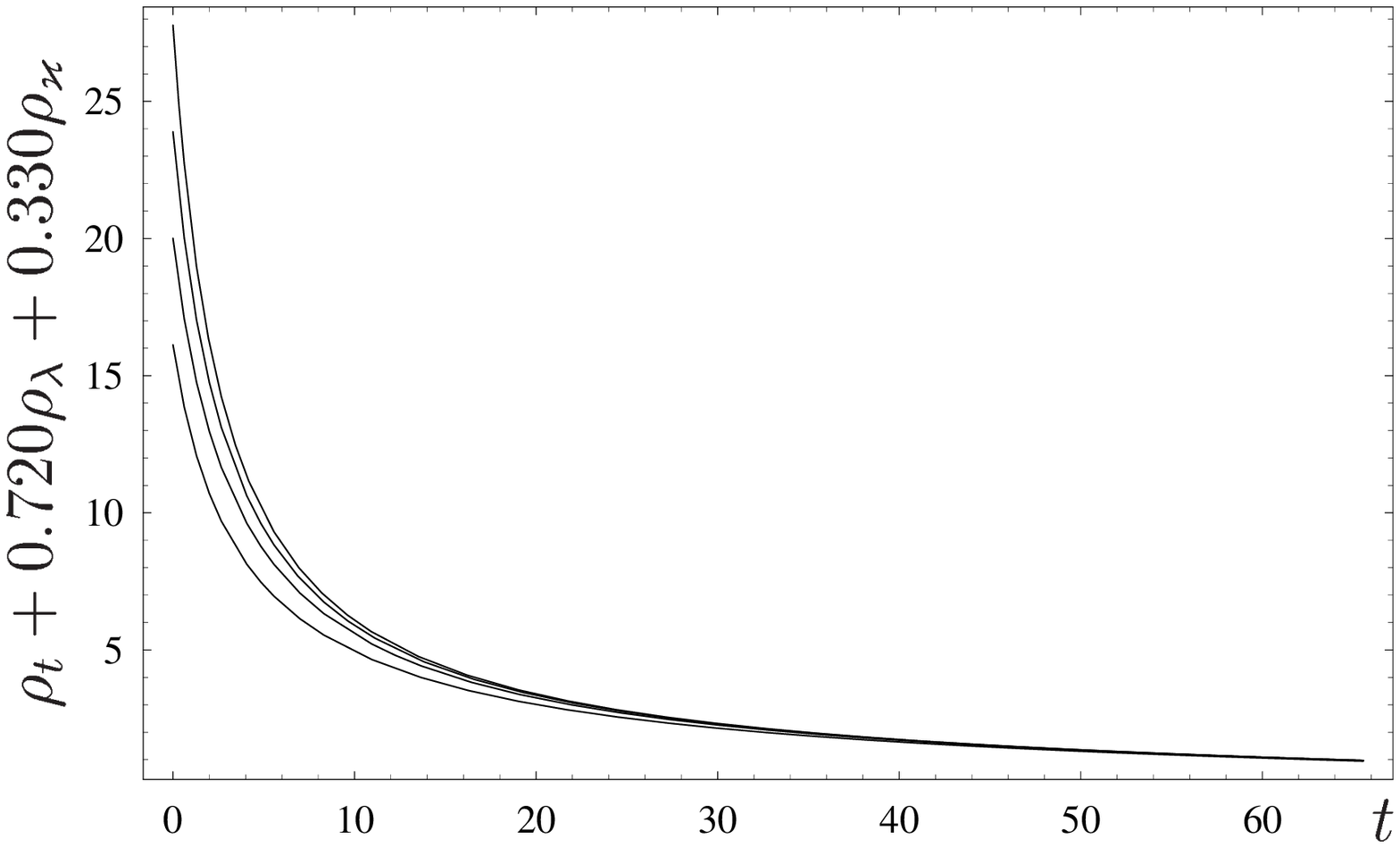}
\smallskip\centerline{(a)}\smallskip
\end{center}
\begin{center}
\includegraphics[width=90mm]{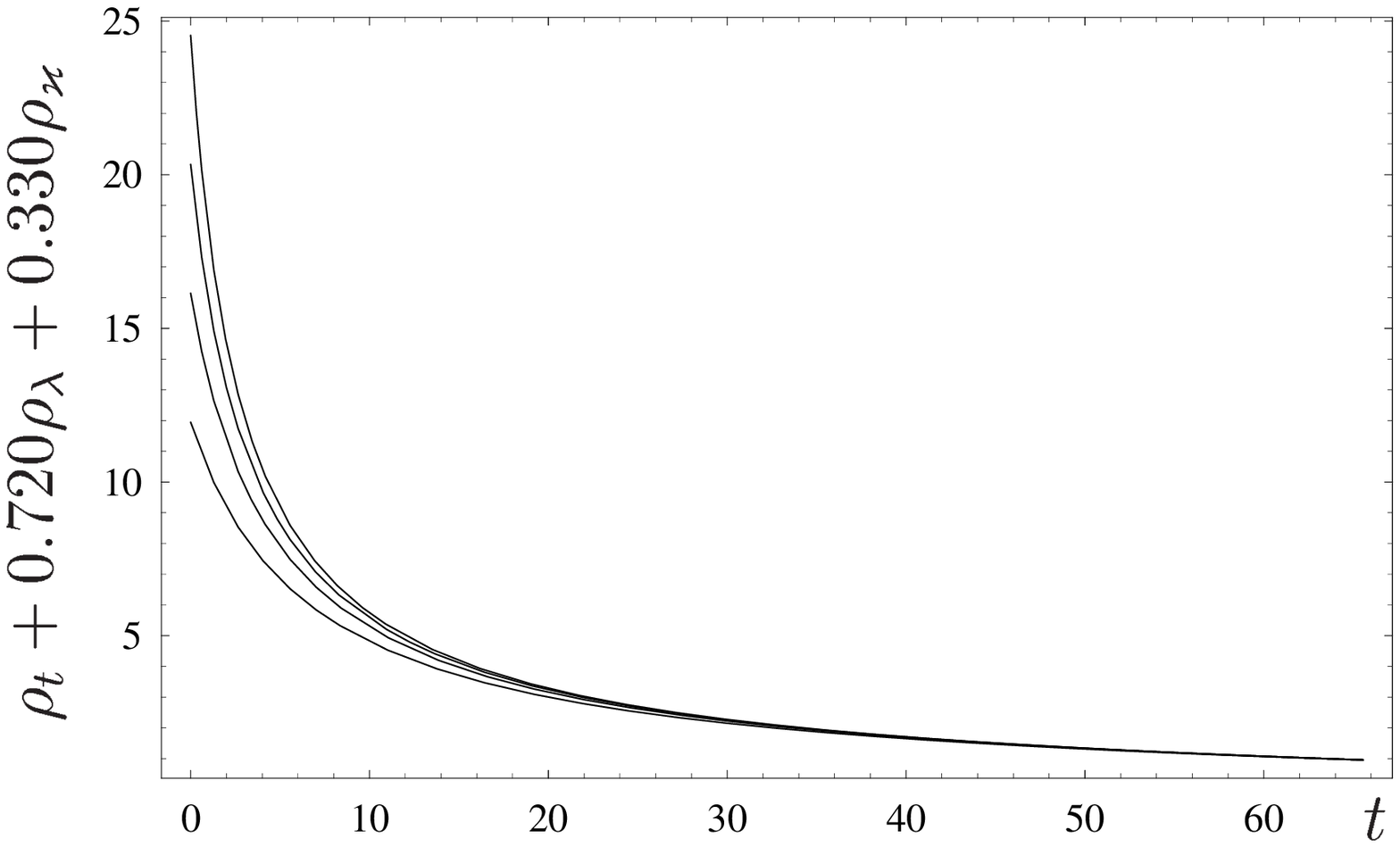}
\smallskip\centerline{(b)}\smallskip
\end{center}
\begin{center}
\includegraphics[width=90mm]{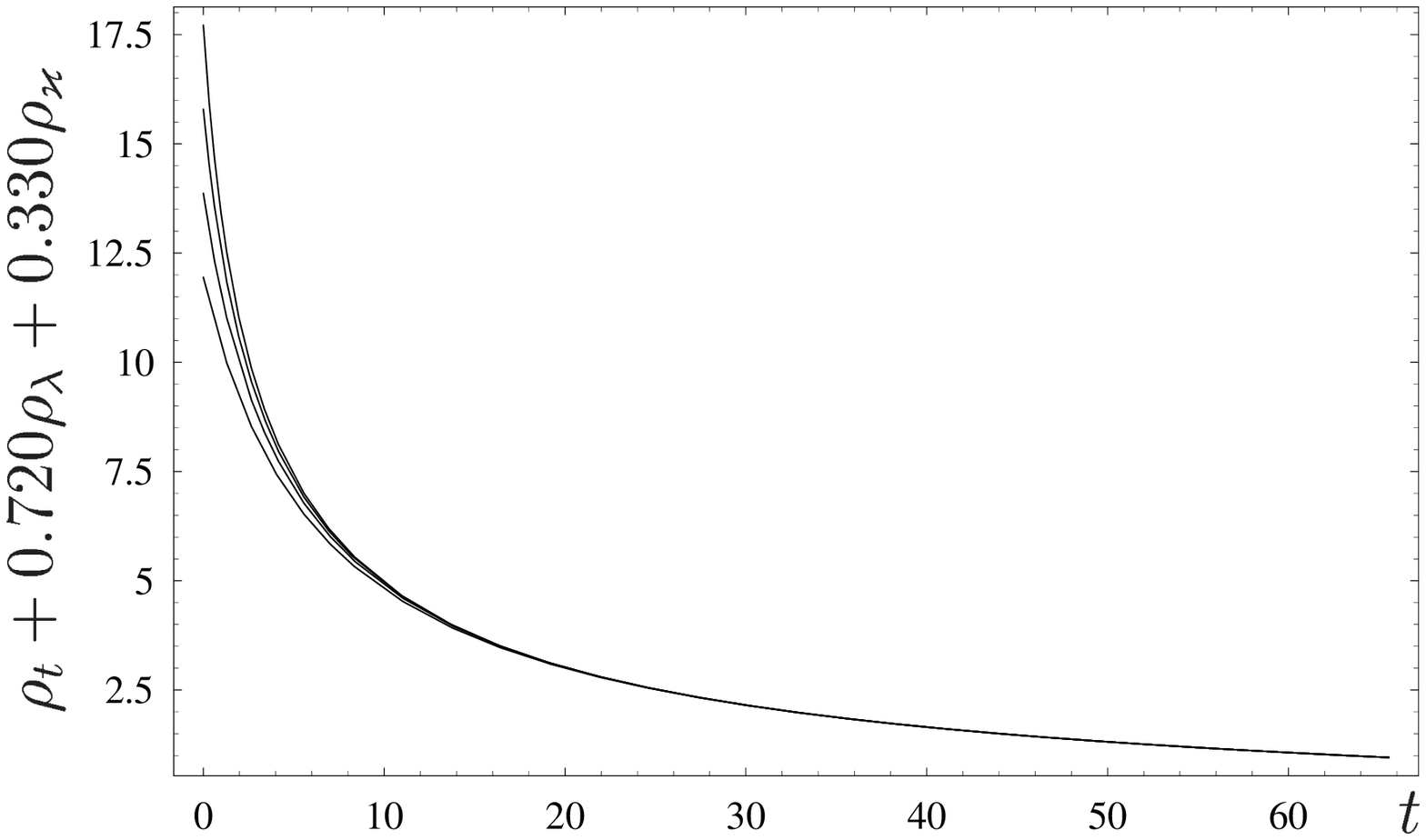}
\smallskip\centerline{(c)}\smallskip
\end{center}
\caption{%
Evolution of the combination
$\rho_t+0.720\rho_\lambda+0.3330\rho_\varkappa$ of the Yukawa
couplings from the GUT scale ($t=0$) to the electroweak scale
($t=t_0$) for various initial values $h_t^2(0)$ -- Fig. 5a,
$\lambda^2(0)$ -- Fig. 5b, $\varkappa^2(0)$ -- Fig. 5c.}
\label{ntt-fig5}
\end{figure}

On the Hill surface, the region that is depicted in Fig. 4 and
near which the solutions in question are grouped shrinks in one
direction with increasing initial values of the Yukawa coupling
constants, with the result that, at $Y_t(0)$, $Y_\varkappa(0)$,
and $Y_\lambda(0)\sim 1$, all solutions are grouped around the
line that appears as the result of intersection of the
quasi--fixed surface and the infrared fixed surface, which
includes the invariant lines lying in the $\rho_\varkappa=0$ and
$\rho_\lambda=0$ planes and connecting the stable infrared point
with, respectively, the fixed point $\rho_\lambda/\rho_t=1$ and
the fixed point $\rho_\varkappa/\rho_t=1$ in the regime of strong
Yukawa coupling. In the limit $\rho_t, \rho_\lambda,
\rho_\varkappa\gg 1$, in which case the gauge coupling constants
can be disregarded, the fixed points
$\rho_\lambda/\rho_t=1,~\rho_\varkappa/\rho_t=0$ and
$\rho_\varkappa\rho_t=1,~\rho_\lambda/\rho_t=0$ cease to be
stable. Instead of them, the stable fixed point
$R_\lambda=3/4,~R_\varkappa=3/8$ \cite{B35} appears in the
$(R_\lambda,R_\varkappa)$ plane, where
$R_\lambda=\rho_\lambda/\rho_t$ and
$R_\varkappa=\rho_\varkappa/\rho_t$. In order to investigate the
behaviour of the solutions to the renormalization group equations
within the NMSSM, it is necessary to linearise the set of these
equations in its vicinity and set $\alpha_i=0$. As a result, we
obtain
\begin{equation}
\begin{split}
R_{\lambda}(t)=&\frac{3}{4}+\left(\frac{1}{2}R_{\lambda 0}+
\frac{1}{\sqrt{5}}R_{\varkappa 0}-\frac{3(\sqrt{5}+1)}{8\sqrt{5}}
\right)\left(\frac{\rho_t(t)}{\rho_{t0}}\right)^{\lambda_1} \\
&{}+\left(\frac{1}{2}R_{\lambda 0}-\frac{1}{\sqrt{5}} R_{\varkappa
0}-\frac{3(\sqrt{5}-1)}{8\sqrt{5}}\right)
\left(\frac{\rho_t(t)}{\rho_{t0}}\right)^{\lambda_2}, \\
R_{\varkappa}(t)=&\frac{3}{8}+\frac{\sqrt{5}}{2}
\left(\frac{1}{2}R_{\lambda 0}+\frac{1}{\sqrt{5}} R_{\varkappa
0}-\frac{3(\sqrt{5}+1)}{8\sqrt{5}}\right)
\left(\frac{\rho_t(t)}{\rho_{t0}}\right)^{\lambda_1} \\
&{}-\frac{\sqrt{5}}{2}\left(\frac{1}{2}R_{\lambda 0}-
\frac{1}{\sqrt{5}}R_{\varkappa 0}-\frac{3(\sqrt{5}-1)}{8\sqrt{5}}
\right)\left(\frac{\rho_t(t)}{\rho_{t0}}\right)^{\lambda_2},\\
\end{split}
\label{B15}
\end{equation}
where $R_{\lambda 0}=R_{\lambda}(0)$, $R_{\varkappa
0}=R_{\varkappa}(0)$, $\rho_{t0}=\rho_t(0)$,
$\lambda_1=\dfrac{3+\sqrt{5}}{9}$,
$\lambda_2=\dfrac{3-\sqrt{5}}{9}$, and
$\rho_t(t)=\dfrac{\rho_{t0}}{1+7\rho_{t0}t}$. From (\ref{B15}), it
follows that the fixed point $R_\lambda=3/4,~R_\varkappa=3/8$
arises as the result of intersection of two fixed lines in the
$(R_\lambda,R_\varkappa)$ plane. The solutions are attracted most
strongly to the line
$\dfrac{1}{2}R_\lambda+\dfrac{1}{\sqrt{5}}R_\varkappa=\dfrac{3}{8}\left(1+\dfrac{1}{\sqrt{5}}\right)$,
since $\lambda_1\gg \lambda_2$. This line passes through three
fixed points in the $(R_\lambda,R_\varkappa)$ plane: $(1,0)$,
$(3/4,3/8)$, and $(0,1)$. In the regime of strong Yukawa coupling,
the fixed line that corresponds, in the
$(\rho_t,\rho_\varkappa,\rho_\lambda)$ space, to the line
mentioned immediately above is that which lies on the invariant
surface containing a stable infrared fixed point. The line of
intersection of the Hill and the invariant surface can be obtained
by mapping this fixed line into the quasi--fixed surface with the
aid of the set of renormalization group equations. For the
boundary conditions, one must than use the values $\lambda^2(0)$,
$\varkappa^2(0)$, and $h_t^2(0)\gg 1$ belonging to the
aforementioned fixed line.

In just the same way as infrared fixed lines, the infrared fixed
surface is invariant under renormalization group transformations.
In the evolution process, solutions to the set of renormalization
group equations within the NMSSM are attracted to this surface. If
boundary conditions are specified n the fixed surface, the ensuing
evolution of the coupling constants proceeds within this surface.
To add further details, we not that, near the surface being
studied and on it, the solutions are attracted to the invariant
line connecting the stable fixed point
$(\rho_\lambda/\rho_t=3/4,~\rho_\varkappa/\rho_t=3/8)$ in the
regime of strong Yukawa coupling with the stable infrared fixed
point within the NMSSM. In the limit
$\rho_t,\rho_\varkappa,\rho_\lambda\gg 1$, the equation for this
line has the form
\begin{equation}
\begin{split}
\rho_{\lambda}&=\frac{3}{4}\rho_t-\frac{176}{417}+
\frac{3}{139}\rho_2-\frac{7} {417}\rho_1, \\
\rho_{\varkappa}&=\frac{3}{8}\rho_t-\frac{56}{417}-
\frac{18}{139}\rho_2-\frac{68}{2085}\rho_1.
\end{split}
\label{B16}
\end{equation}
As one approaches the infrared fixed point, the quantities
$\rho_\lambda$ and $\rho_\varkappa$ tend to zero:
$\rho_\lambda\sim(\rho_t-7/18)^{25/14}$ and
$\rho_\varkappa\sim(\rho_t-7/18)^{9/7}$. This line intersects the
quasi--fixed surface at the point
\[
\rho^{\text{QFP}}_t(t_0)=0.82,\quad
\rho^{\text{QFP}}_{\varkappa}(t_0)=0.087,\quad
\rho^{\text{QFP}}_{\lambda}(t_0)=0.178.
\]
Since all solutions are concentrated in the vicinity of this point
for $Y_t(0), Y_\lambda(0)$, $Y_\varkappa(0)\to\infty$, it should be
considered as a quasi--fixed point for the set of renormalization
group equations within the NMSSM. We note, however, that the
solutions are attracted to the invariant line (\ref{B16}) and to
the quasi--fixed line on the Hill surface. This conclusion can be
drawn from the an analysis of the behaviour of the solutions near
the fixed point $(R_\lambda=3/4,~R_\varkappa=3/8)$ (see
(\ref{B15})). Once the solutions have approached the invariant
line
$\dfrac{1}{2}R_\lambda+\dfrac{1}{\sqrt{5}}R_\varkappa=\dfrac{3}{8}\left(1+\dfrac{1}{\sqrt{5}}\right)$,
their evolution is governed by the expression
$(\epsilon(t))^{0.085}$, where $\epsilon(t)=\rho_t(t)/\rho_{t0}$.
This means that the solutions begin to be attracted to the
quasi--fixed point and to the invariant line (\ref{B16}) with a
sizable strength only when $Y_i(0)$ reaches a value of $10^2$, at
which perturbation theory is obviously inapplicable. Thus, it is
not the infrared quasi--fixed point but the quasi--fixed line on
the Hill surface (see Fig. 4) that, within the NMSSM, plays a key
role in analysing the behaviour of the solutions to the
renormalization group equations in the regime of strong Yukawa
coupling, where all $Y_i(0)$ are much greater than
$\tilde{\alpha}_0$.

\section{Renormalization of the soft SUSY breaking parameters}

If the evolution of gauge and Yukawa coupling constants is known,
the remaining subset of renormalization group equations within the
MNSSM can be treated as a set of linear differential equations for
the parameters of a soft breakdown of supersymmetry. For universal
boundary conditions, a general solution for the trilinear coupling
constants $A_i(t)$ and for the masses of scalar fields $m_i^2(t)$
has the form
\begin{gather}
A_i(t)=e_i(t)A+f_i(t)M_{1/2}\, , \label{C8} \\
m_i^2(t)=a_i(t)m_0^2+b_i(t)M_{1/2}^2+c_i(t)AM_{1/2}+d_i(t)A^2\, .
\label{C9}
\end{gather}
The functions $e_i(t)$, $f_i(t)$, $a_i(t)$, $b_i(t)$, $c_i(t)$,
and $d_i(t)$, which determine the evolution of $A_i(t)$ and
$m_i^2(t)$, remain unknown, since an analytic solution to the full
set of renormalization group equations within the NMSSM is
unavailable. These functions greatly depend on the choice of
values for the Yukawa coupling constants at the Grand Unification
scale $M_X$. At the electroweak scale $t=t_0$, relations
(\ref{C8}) and (\ref{C9}) specify the parameters $A_i^2(t_0)$ and
$m_i^2(t_0)$ of a soft breaking of supersymmetry as functions of
their initial values at the Grand Unification scale.

The results of our numerical analysis indicate that, with
increasing $Y_i(0)$, where $Y_t(t)=\dfrac{h_t^2(t)}{(4\pi)^2}$,
$Y_\lambda(t)=\dfrac{\lambda^2(t)}{(4\pi)^2}$, and
$Y_\varkappa(t)=\dfrac{\varkappa^2(t)}{(4\pi)^2}$, the functions
$e_i(t_0)$, $c_i(t_0)$, and $d_i(t_0)$ decrease and tend to zero
in the limit $Y_i(0)\to\infty$, relations (\ref{C8}) and
(\ref{C9}) becoming much simpler in this limit. Instead of the
squares of the scalar particle masses, it is convenient to
consider their linear combinations
\begin{equation}
\begin{split}
\mathfrak{M}_t^2(t)&=m_2^2(t)+m_Q^2(t)+m_U^2(t),\\
\mathfrak{M}_{\lambda}^2(t)&=m_1^2(t)+m_2^2(t)+m_y^2(t),\\
\mathfrak{M}_{\varkappa}^2(t)&=3m_y^2(t)
\end{split}
\label{C10}
\end{equation}
in analysing the set of renormalization group equations. In the
case of universal boundary conditions, the solutions to the
differential equations for $\mathfrak{M}_i^2(t)$ can be
represented in the same form as the solutions for $m_i^2(t)$ (see
(\ref{C9})); that is
\begin{equation}
\mathfrak{M}_i^2(t)=3\tilde{a}_i(t)m_0^2+\tilde{b}_i(t)M_{1/2}^2+
\tilde{c}_i(t)A M_{1/2}+\tilde{d}_i(t)A^2. \label{C11}
\end{equation}
Since the homogeneous equations for $A_i(t)$ and
$\mathfrak{M}_i^2(t)$ have the same form, the functions
$\tilde{a}_i(t)$ and $e_i(t)$ coincide; in the limit of strong
Yukawa coupling, the $m_0^2$ dependence disappears in the
combinations (\ref{C10}) of the scalar particle masses as the
solutions to the renormalization group equations for the Yukawa
coupling constants approach quasi--fixed points. This behaviour of
the solutions implies that $A_i(t)$ and $\mathfrak{M}_i^2(t)$
corresponding to $Y_i(0)\gg\tilde{\alpha}_i(0)$ also approach
quasi--fixed points. As we see in the previous section, two
quasi--fixed points of the renormalization group equations within
the NMSSM are of greatest interest from the physical point of
view. Of these, one corresponds to the boundary conditions
$Y_t(0)=Y_\lambda(0)\gg\tilde{\alpha}_i(0)$ and $Y_\varkappa(0)=0$
for the Yukawa coupling constants. The fixed points calculated for
the parameters of a soft breaking of supersymmetry by using these
values of the Yukawa coupling constants are
\begin{equation}
\begin{aligned} \rho_{A_t}^{\text{QFP}}(t_0)&\approx 1.77,
&\rho_{\mathfrak{M}^2_t}^{\text{QFP}}(t_0)&\approx 6.09,\\
\rho_{A_{\lambda}}^{\text{QFP}}(t_0)&\approx -0.42,\qquad
&\rho_{\mathfrak{M}^2_{\lambda}}^{\text{QFP}}(t_0)&\approx -2.28,
\end{aligned}
\label{C12}
\end{equation}
where $\rho_{A_i}(t)=A_i(t)/M_{1/2}$ and
$\rho_{\mathfrak{M}_i^2}(t)=\mathfrak{M}_i^2/M_{1/2}^2$. Since the
coupling constant $\varkappa$ for the self--interaction of neutral
scalar fields is small in the case being considered,
$A_\varkappa(t)$ and $\mathfrak{M}_\varkappa^2(t)$ do not approach
the quasi--fixed point. Nonetheless, the spectrum of SUSY
particles is virtually independent of the trilinear coupling
constant $A_\varkappa$ since $\varkappa\to 0$.

In just the same way, one can determine the position of the other
quasi--fixed point for $A_i(t)$ and $\mathfrak{M}_i^2(t)$, that
which corresponds to $R_{\lambda 0}=3/4,~R_{\varkappa 0}=3/8$. The
results are
\begin{equation}
\begin{aligned} \rho_{A_t}^{\text{QFP}}(t_0)&\approx 1.73,
&\rho_{A_{\lambda}}^{\text{QFP}}(t_0)&\approx -0.43,
&\rho_{A_{\varkappa}}^{\text{QFP}}(t_0)&\approx 0.033,\\
\rho_{\mathfrak{M}^2_t}^{\text{QFP}}(t_0)&\approx 6.02,\quad
&\rho_{\mathfrak{M}^2_{\lambda}}^{\text{QFP}}(t_0)&\approx
-2.34,\quad
&\rho_{\mathfrak{M}^2_{\varkappa}}^{\text{QFP}}(t_0)&\approx 0.29,
\end{aligned}
\label{C13}
\end{equation}
where $R_{\lambda 0}=Y_\lambda(0)/Y_t(0)$ and $R_{\varkappa
0}=Y_\varkappa(0)/Y_t(0)$. It should be noted that, in the
vicinities of quasi--fixed points, we have
$\rho_{\mathfrak{M}^2_{\lambda}}^{\text{QFP}}(t_0)<0$. Negative
values of $\mathfrak{M}_\lambda^2(t_0)$ lead to a negative value
of the parameter $m_2^2(t_0)$ in the potential of interaction of
Higgs fields. In other words, an elegant mechanism that is
responsible for a radiative violation of $SU(2)\otimes U(1)$
symmetry and which does not require introducing tachyons in the
spectrum of the theory from the outset survives in the regime of
strong Yukawa coupling within the NMSSM. This mechanism of gauge
symmetry breaking was first discussed in \cite{C30} by considering
the example of the minimal SUSY model.

By using the fact that $\mathfrak{M}_i^2(t)$ as determined for the
case of universal boundary conditions is virtually independent of
$m_0^2$, we can predict $a_i(t_0)$ values near the quasi--fixed
points (see \cite{NTC}). The results are
\begin{equation}
\begin{split}
1)~&R_{\lambda 0}=1,~R_{\varkappa 0}=0,\\
&a_y(t_0)=a_u(t_0)=\frac{1}{7},~a_1(t_0)=a_q(t_0)=\frac{4}{7},~
a_2(t_0)=-\frac{5}{7}\, ;\\ 2)~&R_{\lambda 0}=3/4,~R_{\varkappa
0}=3/8,\\ &a_y(t_0)=0,~
a_1(t_0)=-a_2(t_0)=\frac{2}{3},~a_q(t_0)=\frac{5}{9},~
a_u(t_0)=\frac{1}{9}\, .
\end{split}
\label{C14}
\end{equation}
To do this, it was necessary to consider specific combinations of
the scalar particle masses, such as $m_U^2-2m_Q^2$,
$m_Q^2+m_U^2-m_2^2+m_1^2$, and $m_y^2-2m_1^2$ (at $\varkappa=0$),
that are not renormalized by Yukawa interactions. As a result, the
dependence of the above combinations of the scalar particle masses
on $m_0^2$ at the electroweak scale is identical to that at the
Grand Unification scale. The predictions in (\ref{C14}) agree
fairly well with the results of numerical calculations.

\begin{figure}
\centering
\includegraphics[width=120mm]{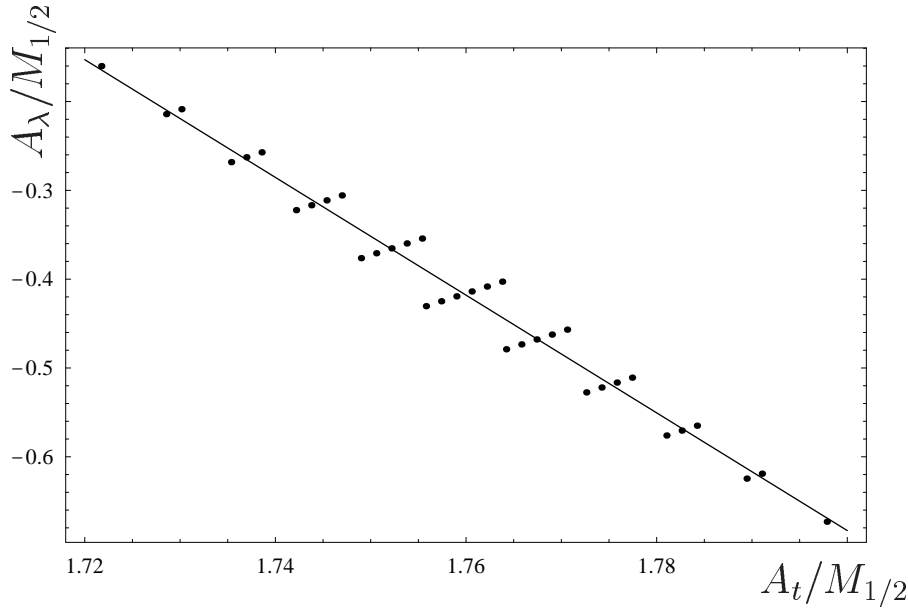}
\caption{%
The values of the trilinear couplings $A_t$
and $A_\lambda$ at the electroweak scale corresponding to the
initial values uniformly distributed in the $(A_t,A_\lambda)$
plane, calculated at $\varkappa^2=0$ and
$h_t^2(0)=\lambda^2(0)=20$. The straight line is a fit of the
values $(A_t(t_0),A_\lambda(t_0))$.}
\label{ntt-fig6}
\end{figure}

\begin{figure}
\centering
\includegraphics[width=120mm]{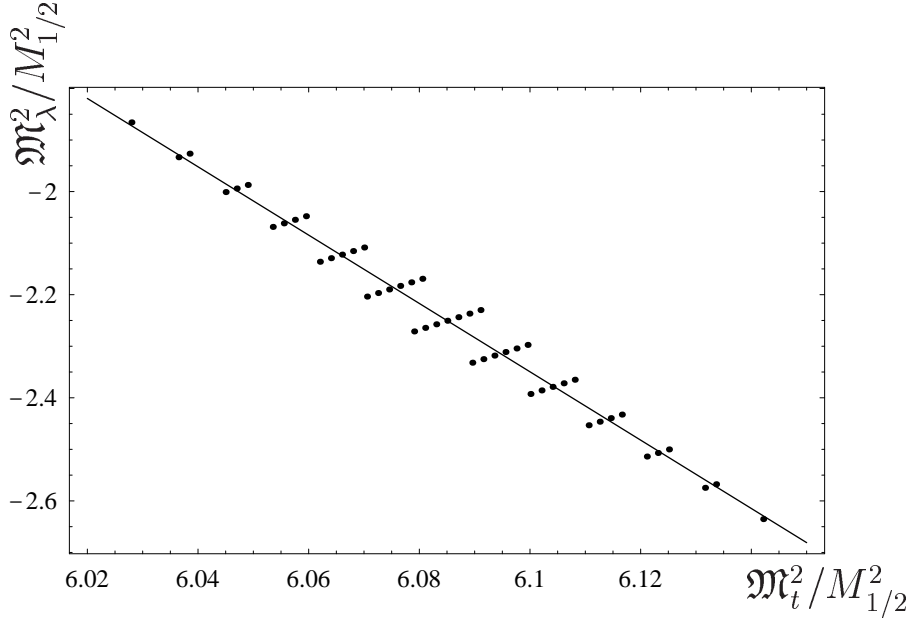}
\caption{%
The values of the combinations of masses
$\mathfrak{M}_t^2$ and $\mathfrak{M}_\lambda^2$ at the electroweak
scale corresponding to the initial values uniformly distributed in
the
$(\mathfrak{M}_t^2/M_{1/2}^2,\mathfrak{M}_\lambda^2/M_{1/2}^2)$
plane, calculated at $\varkappa^2=0$, $h_t^2(0)=\lambda^2(0)=20$,
and $A_t(0)=A_\lambda(0)=0$. The straight line is a fit of the
values $(\mathfrak{M}_t^2(t_0),\mathfrak{M}_\lambda^2(t_0))$.}
\label{ntt-fig7}
\end{figure}

Let us now consider the case of nonuniversal boundary conditions
for the soft SUSY breaking parameters. The results of our
numerical analysis, which are illustrated in Figs. 6 and 7,
indicate that, in the vicinity of the infrared fixed point at
$Y_\varkappa=0$, solutions to the renormalization group equations
at the electroweak scale are concentrated near some straight lines
for the case where the simulation was performed by using boundary
conditions uniformly distributed in the $(A_t,A_\lambda)$ and the
$(\mathfrak{M}_t^2,\mathfrak{M}_\lambda^2)$ plane. The strength
with which these solutions are attracted to them grows with
increasing $Y_i(0)$. The equations for the lines being considered
can be obtained by fitting the numerical results displayed in
Figs. 6 and 7. This yields
\begin{equation}
\begin{gathered}
A_t+0.147 A_{\lambda}=1.70M_{1/2},\\ \mathfrak{M}^2_t+0.147
\mathfrak{M}^2_{\lambda}=5.76 M_{1/2}^2 .
\end{gathered}
\label{C18}
\end{equation}
For $Y_\varkappa(0)\gg\tilde{\alpha}_0$ solutions to the
renormalization group equations are grouped near planes in the
space of the parameters of a soft breaking of supersymmetry
$(A_t,A_\lambda,A_\varkappa)$ and
$(\mathfrak{M}_t^2,\mathfrak{M}_\lambda^2,\mathfrak{M}_\varkappa^2)$
(see Figs. 8-10):
\begin{equation}
\begin{gathered}
A_t+0.128 A_{\lambda}+0.022 A_{\varkappa}=1.68 M_{1/2} ,\\
\mathfrak{M}^2_t+0.128 \mathfrak{M}^2_{\lambda}+0.022
\mathfrak{M}^2_{\varkappa}=5.77 M_{1/2}^2 .
\end{gathered}
\label{C19}
\end{equation}
It can be seen from Figs. 8 and 9 that, as the values of the
Yukawa coupling constants at the Grand Unification scale are
increased, the areas of the surfaces near which the solutions
$A_i(t)$ and $\mathfrak{M}_i^2(t)$ are concentrated shrink in one
of the directions, with the result that, at $Y_i(0)\sim 1$, the
solutions to the renormalization group equations are attracted to
one of the straight lines belonging to these surfaces.

\begin{figure}
\begin{center}
\includegraphics[width=100mm]{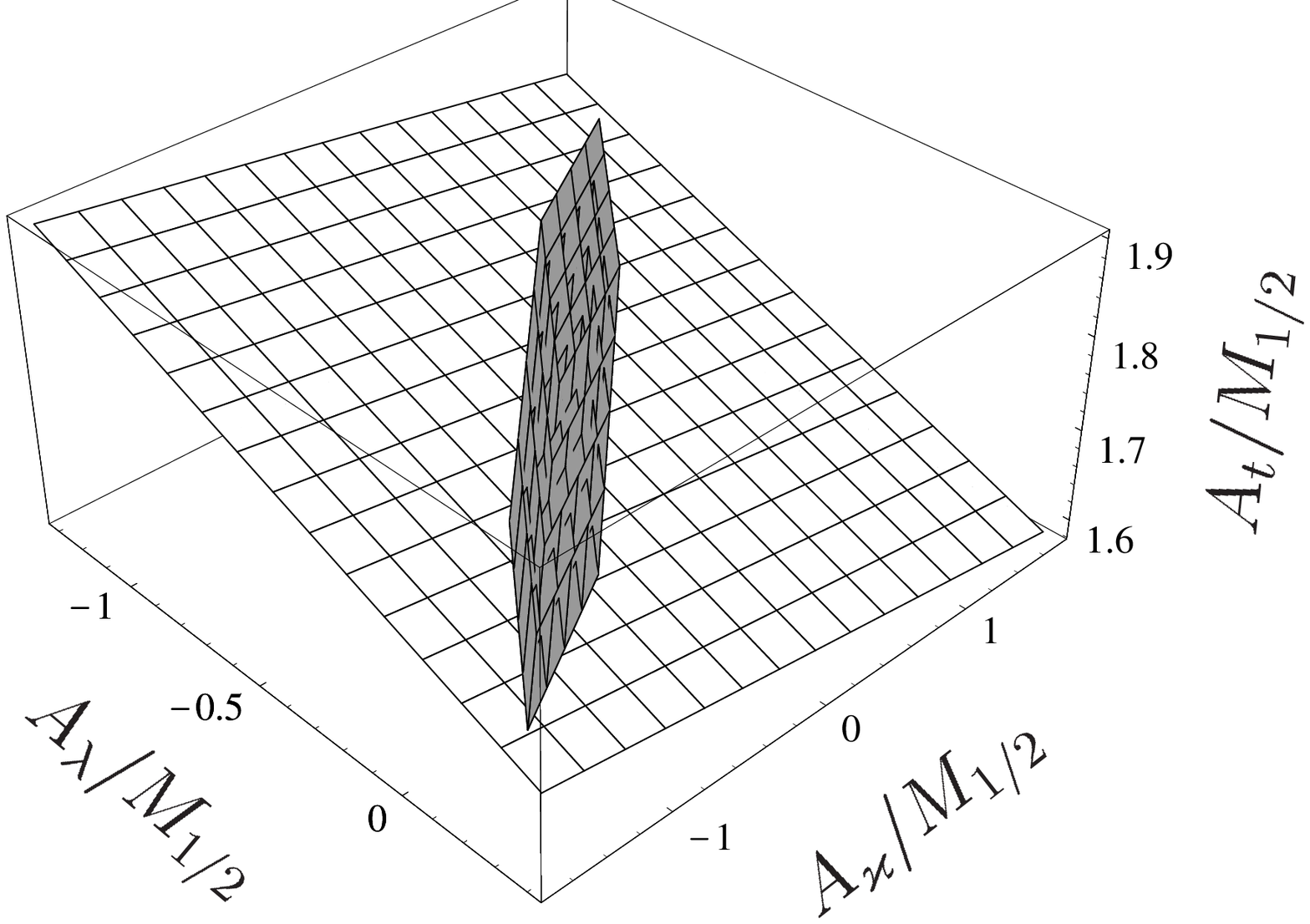}
\smallskip\centerline{(a)}\smallskip
\end{center}
\begin{center}
\includegraphics[width=100mm]{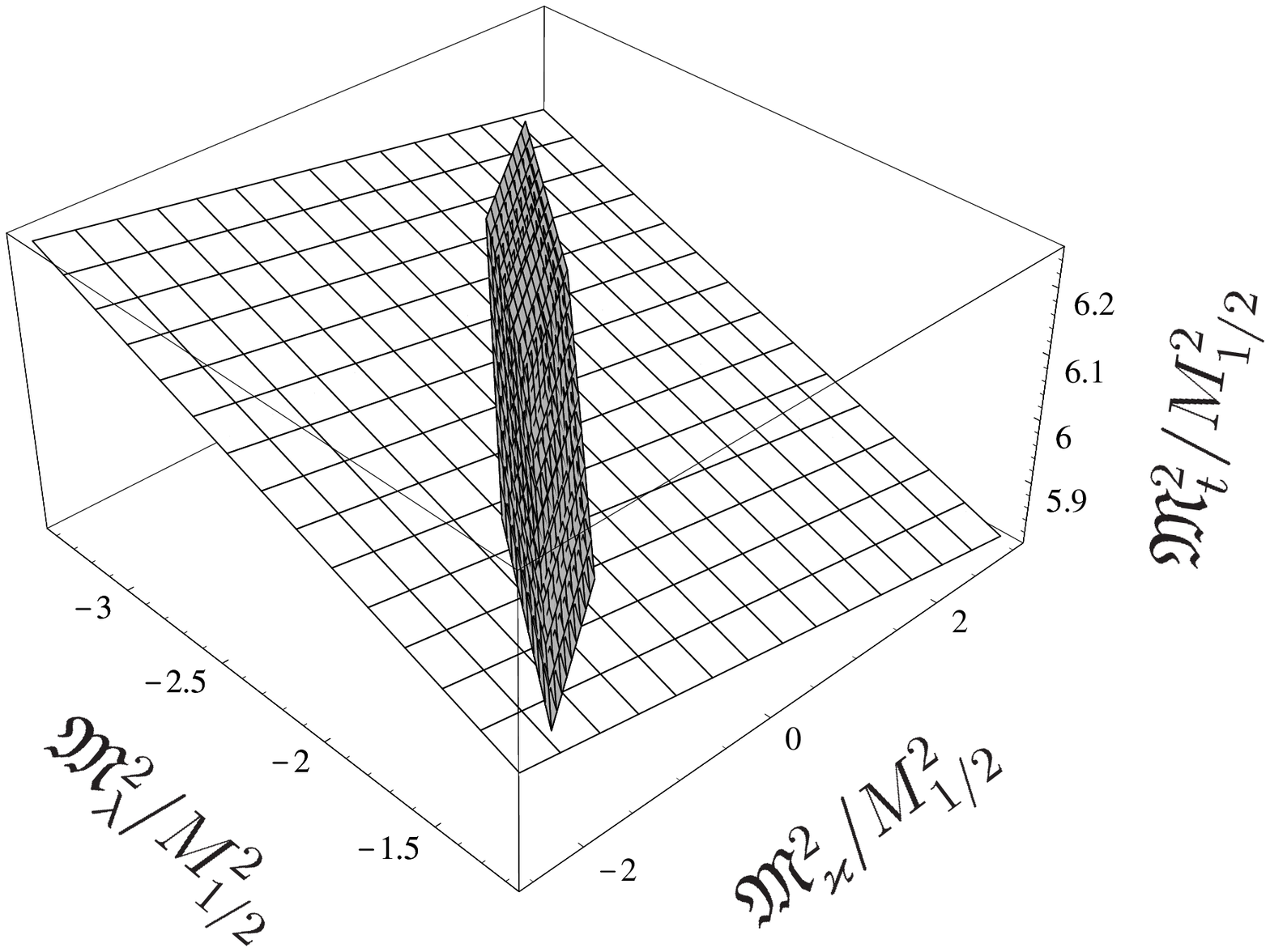}
\smallskip\centerline{(b)}\smallskip
\end{center}
\caption{%
Planes in the parameter spaces
$(A_t/M_{1/2},A_\lambda/M_{1/2},A_\varkappa/M_{1/2})$ -- Fig. 8a,
and
$(\mathfrak{M}_t^2/M_{1/2}^2,\mathfrak{M}_\lambda^2/M_{1/2}^2,\mathfrak{M}_\varkappa^2/M_{1/2}^2)$
-- Fig. 8b. The shaded parts of the planes correspond to the
regions near which the solutions at $h_t^2(0)=16$,
$\lambda^2(0)=12$, and $\varkappa^2(0)=6$ are concentrated. The
initial values $A_i(0)$ and $\mathfrak{M}_i^2(0)$ vary in the
ranges $-M_{1/2}\le A\le M_{1/2}$ and $0\le\mathfrak{M}_i^2(0)\le
3M_{1/2}^2$, respectively.}
\label{ntt-fig8}
\end{figure}

\begin{figure}
\begin{center}
\includegraphics[width=100mm]{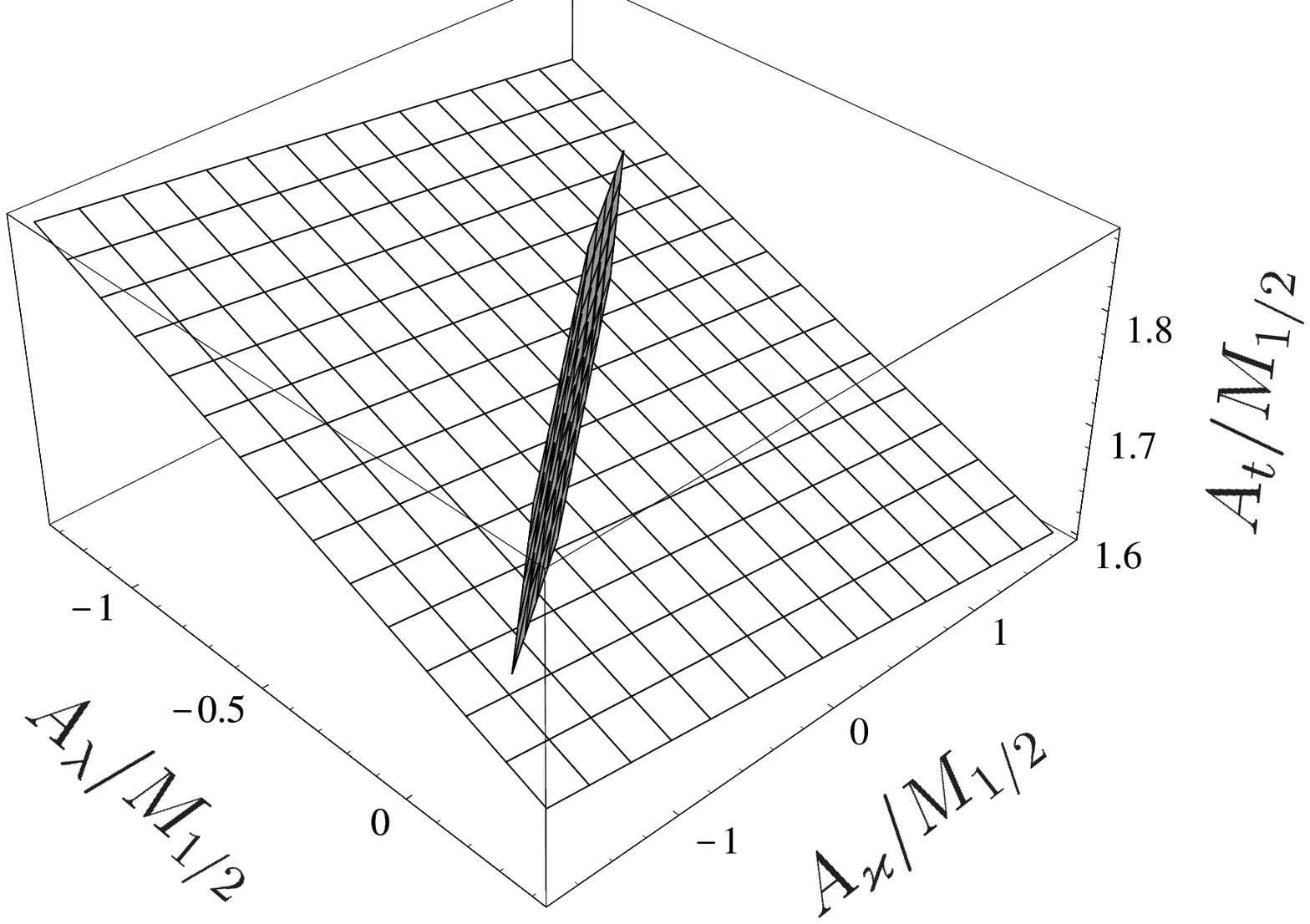}
\smallskip\centerline{(a)}\smallskip
\end{center}
\begin{center}
\includegraphics[width=100mm]{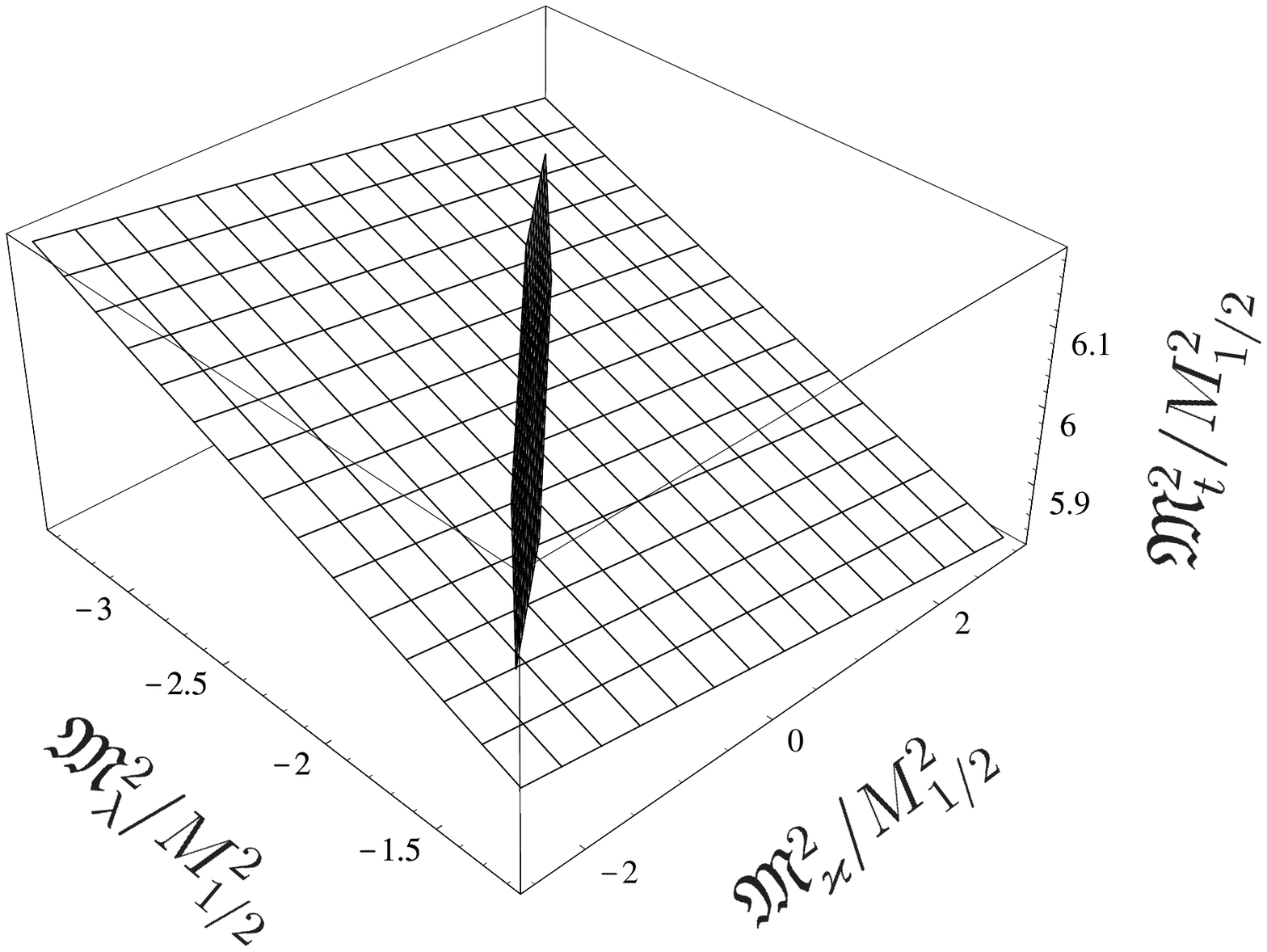}
\smallskip\centerline{(b)}\smallskip
\end{center}
\caption{%
Planes in the parameter spaces
$(A_t/M_{1/2},A_\lambda/M_{1/2},A_\varkappa/M_{1/2})$ -- Fig. 9a,
and
$(\mathfrak{M}_t^2/M_{1/2}^2,\mathfrak{M}_\lambda^2/M_{1/2}^2,\mathfrak{M}_\varkappa^2/M_{1/2}^2)$
-- Fig. 9b. The shaded parts of the planes correspond to the
regions near which the solutions at $h_t^2(0)=32$,
$\lambda^2(0)=24$, and $\varkappa^2(0)=12$ are concentrated. The
initial values $A_i(0)$ and $\mathfrak{M}_i^2(0)$ vary in the
ranges $-M_{1/2}\le A\le M_{1/2}$ and $0\le\mathfrak{M}_i^2(0)\le
3M_{1/2}^2$, respectively.}
\label{ntt-fig9}
\end{figure}

The numerical calculations also showed that, with increasing
$Y_i(0)$, only in the regime of infrared quasi--fixed points (that
is, at $R_{\lambda 0}=1,~R_{\varkappa 0}=0$ or at $R_{\lambda
0}=3/4,~R_{\varkappa 0}=3/8$) $e_i(t_0)$ and $\tilde{a}_i(t_0)$
decrease quite fast, in proportion to $1/Y_i(0)$. Otherwise, the
dependence on $A$ and $m_0^2$ disappears much more slowly with
increasing values of the Yukawa coupling constants at the Grand
Unification scale -- specifically, in proportion to
$(Y_i(0))^{-\delta}$, where $\delta<1$ (for example,
$\delta=0.35-0.40$ at $\varkappa=0$). In the case of nonuniversal
boundary conditions, only when solutions to the renormalization
group equations approach quasi--fixed points are these solutions
attracted to the fixed lines and surfaces in the space of the
parameters of a soft breaking of supersymmetry, and in the limit
$Y_i(0)\to\infty$, the parameters $A_i(t)$ and $\mathfrak{M}_i
^2(t)$ cease to be dependent on the boundary conditions.

\begin{figure}
\begin{center}
\includegraphics[width=120mm]{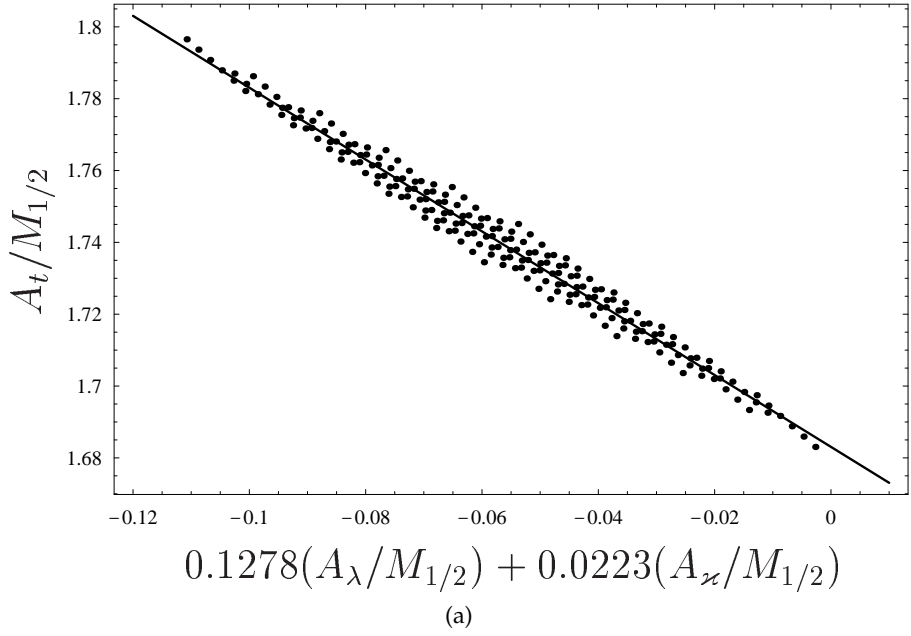}
\smallskip\centerline{(a)}\smallskip
\end{center}
\begin{center}
\includegraphics[width=120mm]{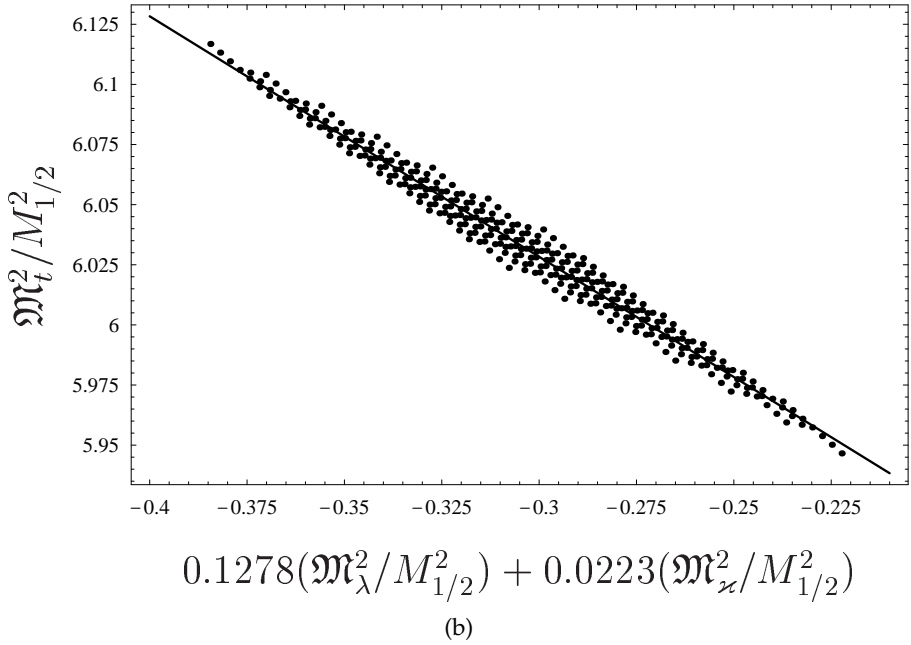}
\smallskip\centerline{(b)}\smallskip
\end{center}
\caption{%
Set of points in planes
$(0.0223(A_{\varkappa}/M_{1/2})+0.1278(A_{\lambda}/M_{1/2}),~A_t/M_{1/2})$
-- Fig. 10a, and
$(0.0223(\mathfrak{M}^2_{\varkappa}/M_{1/2}^2)+0.1278(\mathfrak{M}^2_{\lambda}/M_{1/2}^2),
~\mathfrak{M}^2_t/M^2_{1/2})$ -- Fig. 10b, corresponding to the
values of parameters of soft SUSY breaking for $h_t^2(0)=32$,
$\lambda^2(0)=24$, $\varkappa^2(0)=12$, and for a uniform
distribution of the boundary conditions in the parameter spaces
$(A_t,A_\lambda,A_\varkappa)$ and $(\mathfrak{M}^2_t,
\mathfrak{M}^2_{\lambda},\mathfrak{M}^2_{\varkappa})$. The initial
values $A_i(0)$ and $\mathfrak{M}_i^2(0)$ vary in the ranges
$-M_{1/2}\le A\le M_{1/2}$ and $0\le\mathfrak{M}_i^2(0)\le
3M_{1/2}^2$, respectively. The straight lines in Figs. 10a and 10b
correspond to the planes in Figs. 9a and 9b, respectively.}
\label{ntt-fig10}
\end{figure}

For the solutions of the renormalization group equations for the
soft SUSY breaking parameters near the electroweak scale in the
strong Yukawa coupling regime one can construct an expansion in
powers of the small parameter $\epsilon_t(t)=Y_t(t)/Y_t(0)$:
\begin{equation}
\binom{A_i(t)}{\mathfrak{M}^2_i(t)}=\sum_k u_{ik}v_{ik}(t)
\binom{\alpha_k}{\beta_k}(\epsilon_t(t))^{\lambda_k}+\dots,
\label{C25}
\end{equation}
where $\alpha_i$ and $\beta_i$ are constants of integration that
can be expressed in terms of $A_i(0)$ and $\mathfrak{M}^2_i(0)$.
The functions $v_{ij}(t)$ are weakly dependent on the Yukawa
coupling constants at the scale $M_X$, and $v_{ij}(0)=1$. They
appear upon renormalizing the parameters of a soft breaking of
supersymmetry from $q\sim 10^{12}-10^{13}\text{~GeV}$ to $q\sim
m_t$. In equations (\ref{C25}), we have omitted terms proportional
to $M_{1/2}$, $M_{1/2}^2$, $A_i(0)M_{1/2}$, and $A_i(0)A_j(0)$.

At $\varkappa=0$, we have two eigenvalues and two corresponding
eigenvectors:
\[
\lambda=\begin{pmatrix} 1 \\ 3/7 \end{pmatrix},\qquad
u=\begin{pmatrix} 1 & 1
\\ 1 & -3
\end{pmatrix},
\]
whose components specify $(A_t,A_\lambda)$ and
$(\mathfrak{M}_t^2,\mathfrak{M}_\lambda^2)$. With increasing
$Y_t(0)\simeq Y_\lambda(0)$, the dependence on $\alpha_0$ and
$\beta_0$ becomes weaker and the solutions at $t=t_0$ are
concentrated near the straight lines
$(A_t(\alpha_1),A_\lambda(\alpha_1))$ and
$(\mathfrak{M}_t^2(\beta_1),\mathfrak{M}_\lambda^2(\beta_1))$. In
order to obtain the equations for these straight lines, it is
necessary to set $A_\lambda(0)=-3A_t(0)$ and
$\mathfrak{M}_\lambda^2(0)=-3\mathfrak{M}_t^2(0)$ at the Grand
Unification scale. At the electroweak scale, there then arise a
relation between $A_t(t_0)$ and $A_\lambda(t_0)$ and a relation
between $\mathfrak{M}_t^2(t_0)$ and $\mathfrak{M}_\lambda^2(t_0)$:
\begin{equation}
\begin{gathered}
A_t+0.137 A_{\lambda}=1.70 M_{1/2},\\ \mathfrak{M}^2_t+ 0.137
\mathfrak{M}^2_{\lambda}=5.76 M^2_{1/2}.
\end{gathered}
\label{C26}
\end{equation}
These relations agree well with the equations deduced for the
straight lines at $Y_i(0)\sim 1$ by fitting the results of the
numerical calculations (\ref{C18}).

When the Yukawa coupling constant $\varkappa$ is nonzero, we have
three eigenvalues and three corresponding eigenvectors:
\[
\lambda=\begin{pmatrix} 1 \\ \frac{3+\sqrt{5}}{9} \\
\frac{3-\sqrt{5}}{9}
\end{pmatrix},\qquad
u=\begin{pmatrix} 1 & -\frac{1+\sqrt{5}}{24} &
\frac{\sqrt{5}-1}{24} \\ 1 & \frac{\sqrt{5}}{6} &
-\frac{\sqrt{5}}{6} \\ 1 & 1 & 1
\end{pmatrix},
\]
whose components specify $(A_t,A_\lambda,A_\varkappa)$ and
$(\mathfrak{M}_t^2,\mathfrak{M}_\lambda^2,\mathfrak{M}_\varkappa^2)$.
An increase in $Y_\lambda(0)\simeq 2Y_\varkappa(0)\simeq
\dfrac{3}{4}Y_t(0)$ leads to the following: first, the dependence
of $A_i(t)$ and $\mathfrak{M}_i^2(t)$ on $\alpha_0$ and $\beta_0$
disappears, which leads to the emergence of planes in the space
spanned by the parameters of a soft breaking of supersymmetry:
\begin{equation}
\begin{gathered}
A_t+0.103 A_{\lambda}+0.0124 A_{\varkappa}=1.69 M_{1/2},\\
\mathfrak{M}^2_t+0.103 \mathfrak{M}^2_{\lambda}+0.0124
\mathfrak{M}^2_{\varkappa}=5.78 M^2_{1/2}.
\end{gathered}
\label{C28}
\end{equation}
After that, the dependence on $\alpha_1$ and $\beta_1$ becomes
weaker at $Y_i(0)\sim 1$. This means that, with increasing initial
values of the Yukawa coupling constants, solutions to the
renormalization group equations are grouped near some straight
lines and we can indeed see precisely this pattern in Figs. 8-10.
All equations presented here for the straight lines and planes in
the $\mathfrak{M}_i^2$ space were obtained at $A_i(0)=0$.

From relations (\ref{C26}) and (\ref{C28}), it follows that
$A_t(t_0)$ and $\mathfrak{M}_t^2(t_0)$ are virtually independent
of the initial conditions; that is, the straight lines and planes
are orthogonal to the $A_t$ and $\mathfrak{M}_t^2$ axes. On the
other hand, the $A_\varkappa(t_0)$ and
$\mathfrak{M}_\varkappa^2(t_0)$ values that correspond to the
Yukawa self--interaction constant $Y_\varkappa$ for the neutral
fields are fully determined by the boundary conditions for the
parameters of a soft breaking of supersymmetry. For this reason,
the planes in the $(A_t,A_\lambda,A_\varkappa)$ and
$(\mathfrak{M}_t^2,\mathfrak{M}_\lambda^2,\mathfrak{M}_\varkappa^2)$
spaces are virtually parallel to the $A_\varkappa$ and
$\mathfrak{M}_\varkappa^2$ axes.

\section{Conclusions}

In the strong Yukawa coupling regime in the NMSSM, solutions to
the renormalization group equations for $Y_i(t)$ are attracted to
quasi--fixed lines and surfaces in the space of Yukawa coupling
constants and specific combinations of $\rho_i(t)$ are virtually
independent of their initial values at the Grand Unification
scale. For $Y_i(0)\to\infty$, all solutions to the renormalization
group equations are concentrated near quasi--fixed points. These
points emerge as the result of intersection of Hill lines or
surfaces with the invariant line that connects the stable fixed
point for $Y_i\gg\tilde{\alpha}_i$ with the stable infrared fixed
point. For the renormalization group equations within the NMSSM,
we have listed all the most important invariant lines and surfaces
and studied their asymptotic behaviour for
$Y_i\gg\tilde{\alpha}_i$ and in the vicinity of the infrared fixed
point.

With increasing $Y_i(0)$, the solutions in question approach
quasi--fixed points quite slowly; that is, the deviation is
proportional to $(\epsilon_t(t))^\delta$, where
$\epsilon_t(t)=Y_t(t)/Y_t(0)$ and $\delta$ is calculated by
analysing the set of the renormalization group equations in the
regime of strong Yukawa coupling. As a rule, $\delta$ is positive
and much less than unity. By way of example, we indicate that, in
the case where all three Yukawa coupling constants differ from
zero, $\delta\approx 0.085$. Of greatest importance in analysing
the behaviour of solutions to the renormalization group equations
within the NMSSM at $Y_t(0),Y_\lambda(0),Y_\varkappa(0)\sim 1$ is
therefore not the infrared quasi--fixed point but the line lying
on the Hill surface and emerging as the intersection of the Hill
and invariant surface. This line can be obtained by mapping the
fixed points $(1,0)$, $(3/4,3/8)$, and $(0,1)$ in the
$(R_\lambda,R_\varkappa)$ plane for $Y_i\gg\tilde{\alpha}_i$ into
the quasi--fixed surface by means of renormalization group
equations.

While $Y_i(t)$ approach quasi--fixed points, the corresponding
solutions for the trilinear coupling constants $A_i(t)$
characterising scalar fields and for the combinations
$\mathfrak{M}_i^2(t)$ of the scalar particle masses (see
(\ref{C10})) cease to be dependent on their initial values at the
scale $M_X$ and, in the limit $Y_i(0)\to\infty$, also approach the
fixed points in the space spanned by the parameters of a soft
breaking of supersymmetry. Since the set of differential equations
for $A_i(t)$ and $m_i^2(t)$ is linear, the $A$, $M_{1/2}$, and
$m_0^2$ dependence of the parameters of a soft breaking of
supersymmetry at the electroweak scale can be explicitly obtained
for universal boundary conditions. It turns out that, near the
quasi--fixed points, all $A_i(t)$ and all $\mathfrak{M}_i^2(t)$
are proportional to $M_{1/2}$ and $M_{1/2}^2$, respectively. Thus,
we have shown that, in the parameter space region considered here,
the solutions to the renormalization group equations for the
trilinear coupling constants and for some combinations of the
scalar particle masses are focused in a narrow interval within the
infrared region. Since the neutral scalar field $Y$ is not
renormalized by gauge interactions, $A_\varkappa(t)$ and
$\mathfrak{M}_\varkappa^2(t)$ are concentrated near zero;
therefore they are still dependent on the initial conditions. The
parameters $A_t(t_0)$ and $\mathfrak{M}_t^2(t_0)$ show the weakest
dependence on $A$ and $m_0^2$ because these parameters are
renormalized by strong interactions. By considering that the
quantities $\mathfrak{M}_i^2(t_0)$ are virtually independent of
the boundary conditions, we have calculated, near the quasi--fixed
points, the values of the scalar particle masses at the
electroweak scale.

In the general case of nonuniversal boundary conditions, the
solutions to the renormalization group equations within the NMSSM
for $A_i(t)$ and $\mathfrak{M}_i^2(t)$ are grouped near some
straight lines and planes in the space spanned by the parameters
of a soft breaking of supersymmetry. Moving along these lines and
surfaces as $Y_i(0)$ is increased, the trilinear coupling
constants and the above combinations of the scalar particle masses
approach quasi--fixed points. However, the dependence of these
couplings on $A_i(0)$ and $\mathfrak{M}_i^2(0)$ dies out quite
slowly, in proportion to $(\epsilon_t(t))^{\lambda}$, where
$\lambda$ is a small positive number; as a rule, $\lambda\ll 1$.
For example, $\lambda=3/7$ at $Y_\varkappa=0$ and $\lambda\approx
0.0085$ at $Y_\varkappa\ne 0$. The above is invalid only for the
solutions $A_i(t)$ and $\mathfrak{M}_i^2(t)$ that correspond to
universal boundary conditions for the parameters of a soft
breaking of supersymmetry and to the initial values of $R_{\lambda
0}=1,~R_{\varkappa 0}=0$ and $R_{\lambda 0}=3/4,~R_{\varkappa
0}=3/8$ for the Yukawa coupling constants at the Grand Unification
scale. They correspond to quasi--fixed points of the
renormalization group equations for $Y_i(t)$. As the Yukawa
coupling constants are increased, such solutions are attracted to
infrared quasi--fixed points in proportion to $\epsilon_t(t)$.

Straight lines in the $(A_t,A_\lambda,A_\varkappa)$ and
$(\mathfrak{M}_t^2,\mathfrak{M}_\lambda^2,\mathfrak{M}_\varkappa^2)$
spaces play a key role in the analysis of the behaviour of
solutions for $A_i(t)$ and $\mathfrak{M}_i^2(t)$ in the case where
$Y_t(0),Y_\lambda(0),Y_\varkappa(0)\sim 1$. In the space spanned
by the parameters of a soft breaking of supersymmetry, these
straight lines lie in the planes near which $A_i(t)$ and
$\mathfrak{M}_i^2(t)$ are grouped in the regime of strong Yukawa
coupling at the electroweak scale. The straight lines and planes
that were obtained by fitting the results of numerical
calculations are nearly orthogonal to the $A_t$ and
$\mathfrak{M}_t^2$ axes. This is because the constants $A_t(t_0)$
and $\mathfrak{M}_t^2(t_0)$ are virtually independent of the
initial conditions at the scale $M_X$. On the other hand, the
parameters $A_\varkappa(t_0)$ and $\mathfrak{M}_\varkappa^2(t_0)$
are determined, to a considerable extent, by the boundary
conditions at the scale $M_X$. At $R_{\lambda 0}=3/4$ and
$R_{\varkappa 0}=3/8$, the planes in the
$(A_t,A_\lambda,A_\varkappa)$ and
$(\mathfrak{M}_t^2,\mathfrak{M}_\lambda^2,\mathfrak{M}_\varkappa^2)$
spaces are therefore parallel to the $A_\varkappa$ and
$\mathfrak{M}_\varkappa^2$ axes.

\section*{Acknowledgements}

The authors are grateful to M. I. Vysotsky, D. I. Kazakov, L. B.
Okun, and K. A. Ter--Martirosyan for stimulating questions,
enlightening discussions and comments. R. B. Nevzorov is indebted
to DESY Theory Group for hospitality extended to him.

This work was supported by the Russian Foundation for Basic
Research (RFBR), projects \#\# 00-15-96786, 00-15-96562.

\title*{Multiple Point Model and Phase Transition Couplings 
in the Two-Loop Approximation of Dual Scalar Electrodynamics}
\author{%
L.V. Laperashvili$^a$, D.A. Ryzhikh$^a$ and H.B. Nielsen$^b$}
\institute{%
$^a$ Institute of Theoretical and Experimental Physics,
B.Cheremushkinskaya 25, Moscow 117 259, Russia\\
$^b$ Niels Bohr Institute, Blegdamsvej 17-21, Copenhagen, Denmark}

\authorrunning{L.V. Laperashvili, D.A. Ryzhikh and H.B. Nielsen}
\titlerunning{Multiple Point Model and Phase Transition Couplings}
\maketitle

The simplest effective dynamics describing the
confinement mechanism in the pure gauge lattice U(1) theory
is the dual Abelian Higgs model of scalar monopoles [1-3].

In the previous papers [4-6] the calculations of the U(1)
phase transition (critical) coupling constant were connected with the
existence of artifact monopoles in the lattice gauge theory and also
in the Wilson loop action model \cite{6}.
In Ref.\cite{6} we (L.V.L. and H.B.N.) have put forward the speculations
of Refs.[4,5] suggesting that the modifications of the form of
the lattice action might not change too much the phase transition value of the
effective continuum coupling constant.
In \cite{6} the Wilson loop action was considered in the
approximation of circular loops of radii $R\ge a$. It was shown that the
phase transition coupling constant is indeed approximately independent
of the regularization method: ${\alpha}_{crit}\approx{0.204}$,
in correspondence with the Monte Carlo simulation result on lattice \cite{7}:
${\alpha}_{crit}\approx{0.20\pm 0.015}$.

But in Refs.[2,3] instead of using the lattice or Wilson loop
cut--off we have considered the Higgs Monopole Model (HMM) approximating
the lattice artifact monopoles as fundamental pointlike particles described
by the Higgs scalar field.

\section{The Coleman-Weinberg effective potential for the Higgs 
monopole model}

The dual Abelian Higgs model of scalar monopoles (shortly HMM),
describing the dynamics of confinement in lattice
theories, was first suggested in Ref.\cite{1}, and considers the
following Lagrangian:
\begin{equation}
    L = - \frac{1}{4g^2} F_{\mu\nu}^2(B) + \frac{1}{2} |(\partial_{\mu} -
           iB_{\mu})\Phi|^2 - U(\Phi),\quad              \label{5y}
{\mbox{where}}\quad
 U(\Phi) = \frac{1}{2}\mu^2 {|\Phi|}^2 + \frac{\lambda}{4}{|\Phi|}^4
\end{equation}
is the Higgs potential of scalar monopoles with magnetic charge $g$, and
$B_{\mu}$ is the dual gauge (photon) field interacting with the scalar
monopole field $\Phi$. In this model $\lambda$ is the self--interaction
constant of scalar fields, and the mass parameter $\mu^2$ is negative.
In Eq.(\ref{5y}) the complex scalar field $\Phi$ contains
the Higgs ($\phi$) and Goldstone ($\chi$) boson fields:
\begin{equation}
          \Phi = \phi + i\chi.             \label{7y}
\end{equation}
The effective potential in the Higgs Scalar ElectroDynamics (HSED)
was first calculated by Coleman and Weinberg \cite{9} in the one--loop
approximation. The general method of its calculation is given in the
review \cite{10}. Using this method, we can construct the effective potential
for HMM. In this case the total field system of the gauge ($B_{\mu}$)
and magnetically charged ($\Phi$) fields is described by
the partition function which has the following form in Euclidean space:
\begin{equation}
      Z = \int [DB][D\Phi][D\Phi^{+}]\,e^{-S},     \label{8y}
\end{equation}
where the action $S = \int d^4x L(x) + S_{gf}$ contains the Lagrangian
(\ref{5y}) written in Euclidean space and gauge fixing action $S_{gf}$.
Let us consider now a shift:
$ \Phi (x) = \Phi_b + {\hat \Phi}(x)$
with $\Phi_b$ as a background field and calculate the
following expression for the partition function in the one-loop
approximation:
$$
  Z = \int [DB][D\hat \Phi][D{\hat \Phi}^{+}]
   \exp\{ - S(B,\Phi_b)
   - \int d^4x [\frac{\delta S(\Phi)}{\delta \Phi(x)}|_{\Phi=
   \Phi_b}{\hat \Phi}(x) + h.c. ]\}\\
$$
\begin{equation}
    =\exp\{ - F(\Phi_b, g^2, \mu^2, \lambda)\}.      \label{10y}
\end{equation}
Using the representation (\ref{7y}), we obtain the effective potential:
\begin{equation}
  V_{eff} = F(\phi_b, g^2, \mu^2, \lambda)        \label{11y}
\end{equation}
given by the function $F$ of Eq.(\ref{10y}) for the constant background
field $ \Phi_b = \phi_b = \mbox{const}$. In this case the one--loop
effective potential for monopoles coincides with the expression of the
effective potential calculated by the authors of Ref.\cite{9} for scalar
electrodynamics and extended to the massive theory (see review \cite{10}).
As it was shown in Ref.\cite{9}, the effective potential
can be improved by consideration of the renormalization
group equation (RGE).

\section{Renormalization group equations in the Higgs monopole model}

The RGE for the effective potential means that the potential cannot
depend on a change in the arbitrary parameter --- renormalization scale $M$:
\begin{equation}
         \frac {dV_{eff}}{dM} = 0.             \label{17y}
\end{equation}
The effects of changing it are absorbed into
changes in the coupling constants, masses and fields, giving so--called
running quantities.

Considering the RG improvement of the effective potential [8,9]
and choosing the evolution variable as
\begin{equation}
                   t = \log(\phi^2/M^2),     \label{18y}
\end{equation}
we have the following RGE for the improved $V_{eff}(\phi^2)$
with $\phi^2\equiv \phi^2_b$ \cite{11}:
\begin{equation}
   (M^2\frac{\partial}{\partial M^2} + \beta_{\lambda}\frac{\partial}
{\partial\lambda} + \beta_g\frac{\partial}{\partial g^2} +
\beta_{(\mu^2)}\mu^2\frac{\partial}{\partial \mu^2} - \gamma\phi^2
\frac{\partial}{\partial \phi^2})V_{eff}(\phi^2) = 0,    \label{19y}
\end{equation}
where $\gamma$ is the anomalous dimension and $\beta_{(\mu^2)}$,
$\beta_{\lambda}$ and $\beta_g$ are the RG $\beta$--functions for mass,
scalar and gauge couplings, respectively. RGE (\ref{19y}) leads to the
following form of the improved effective potential \cite{9}:
\begin{equation}
     V_{eff} = \frac{1}{2}\mu^2_{run}(t)G^2(t)\phi^2 +
                 \frac{1}{4}\lambda_{run}(t)G^4(t)\phi^4.  \label{20y}
\end{equation}
In our case:
\begin{equation}
 G(t) = \exp[-\frac{1}{2}\int_0^t dt'\,\gamma\left(g_{run}(t'),
         \lambda_{run}(t')\right)].                         \label{21y}
\end{equation}
A set of ordinary differential equations (RGE) corresponds to Eq.(\ref{19y}):
\begin{equation}
    \frac{d\lambda_{run}}{dt} = \beta_{\lambda}\left(g_{run}(t),\,
                    \lambda_{run}(t)\right),      \label{22y}
\end{equation}
\begin{equation}
    \frac{d\mu^2_{run}}{dt} = \mu^2_{run}(t)\beta_{(\mu^2)}
             \left(g_{run}(t),\,\lambda_{run}(t)\right),
                                                  \label{23y}
\end{equation}
\begin{equation}
    \frac{dg^2_{run}}{dt} = \beta_g\left(g_{run}(t),\,\lambda_{run}(t)\right).
                                           \label{24y}
\end{equation}
So far as the mathematical structure of HMM is equivalent
to HSED, we can use all results of the scalar electrodynamics
in our calculations, replacing the electric charge $e$ and photon
field $A_{\mu}$ by magnetic charge $g$ and dual gauge field $B_{\mu}$.

The one--loop results for $\beta_{\lambda}^{(1)}$, $\beta_{\mu^2}^{(1)}$ and
$\gamma$ are given in
Ref.\cite{9} for scalar field with electric charge $e$, but it is easy to
rewrite them for monopoles with charge $g=g_{run}$:
\begin{equation}
\gamma^{(1)} = - \frac{3g_{run}^2}{16\pi^2},   \label{30y}
\end{equation}
\begin{equation}
 \frac{d\lambda_{run}}{dt}\approx \beta_{\lambda}^{(1)} = \frac
1{16\pi^2} ( 3g^4_{run} +10 \lambda^2_{run} - 6\lambda_{run}g^2_{run}),
                                     \label{31y}
\end{equation}
\begin{equation}
\frac{d\mu^2_{run}}{dt}\approx \beta_{(\mu^2)}^{(1)}
= \frac{\mu^2_{run}}{16\pi^2}( 4\lambda_{run} - 3g^2_{run} ),
                                                \label{32y}
\end{equation}
\begin{equation}
    \frac{dg^2_{run}}{dt}\approx
     \beta_g^{(1)} = \frac{g^4_{run}}{48\pi^2}.  \label{33y}
\end{equation}

The RG $\beta$--functions for different renormalizable gauge theories with
semisimple group have been calculated in the two--loop approximation
and even beyond. But in this paper we made use the
results of Refs.\cite{12} and \cite{13} for calculation of $\beta$--functions
and anomalous dimension in the two--loop approximation, applied to the
HMM with scalar monopole fields. The higher approximations essentially
depend on the renormalization scheme.
Thus, on the level of two--loop approximation we have for all
$\beta$--functions:
\begin{equation}
  \beta = \beta^{(1)} + \beta^{(2)},           \label{34y}
\end{equation}
where
\begin{equation}
  \beta_{\lambda}^{(2)} = \frac{1}{(16\pi^2)^2}( - 25\lambda^3 +
   \frac{15}{2}g^2{\lambda}^2 - \frac{229}{12}g^4\lambda - \frac{59}{6}g^6),
                                                    \label{35y}
\end{equation}
and
\begin{equation}
\beta_{(\mu^2)}^{(2)} = \frac{1}{(16\pi^2)^2}(\frac{31}{12}g^4 + 3\lambda^2).
                                             \label{36y}
\end{equation}
The gauge coupling $\beta_g^{(2)}$--function is given by Ref.\cite{12}:
\begin{equation}
     \beta_g^{(2)} = \frac{g^6}{(16\pi^2)^2}.  \label{37y}
\end{equation}
Anomalous dimension follows from calculations made in Ref.\cite{13}:
\begin{equation}
    \gamma^{(2)} = \frac{1}{(16\pi^2)^2}\frac{31}{12}g^4.
                                                   \label{38y}
\end{equation}
In Eqs.(\ref{34y})--(\ref{38y}) and below, for simplicity, we have used the
following notations: $\lambda\equiv \lambda_{run}$, $g\equiv g_{run}$ and
$\mu\equiv \mu_{run}$.

\section{The phase diagram in the Higgs monopole model}

Now we  want to apply the effective potential calculation as a
technique for the getting phase diagram information for the condensation
of monopoles in HMM.
If the first local minimum occurs
at $\phi = 0$ and $V_{eff}(0) = 0$, it corresponds to the Coulomb--like phase.
In the case when the effective potential has the second local minimum at
$\phi = \phi_{min} \neq 0\,$ with $\,V_{eff}^{min}(\phi_{min}^2) < 0$,
we have the confinement phase. The phase transition between the
Coulomb--like and confinement phases is given by the condition when
the first local minimum at $\phi = 0$ is degenerate with the second minimum
at $\phi = \phi_0$.
These degenerate minima are shown in Fig.1 by the curve 1. They correspond
to the different vacua arising in this model. And the dashed curve 2
describes the appearance of two minima corresponding to the confinement
phases.

\begin{figure*}
\begin{center}
\noindent\includegraphics[width=120mm, height=107mm]{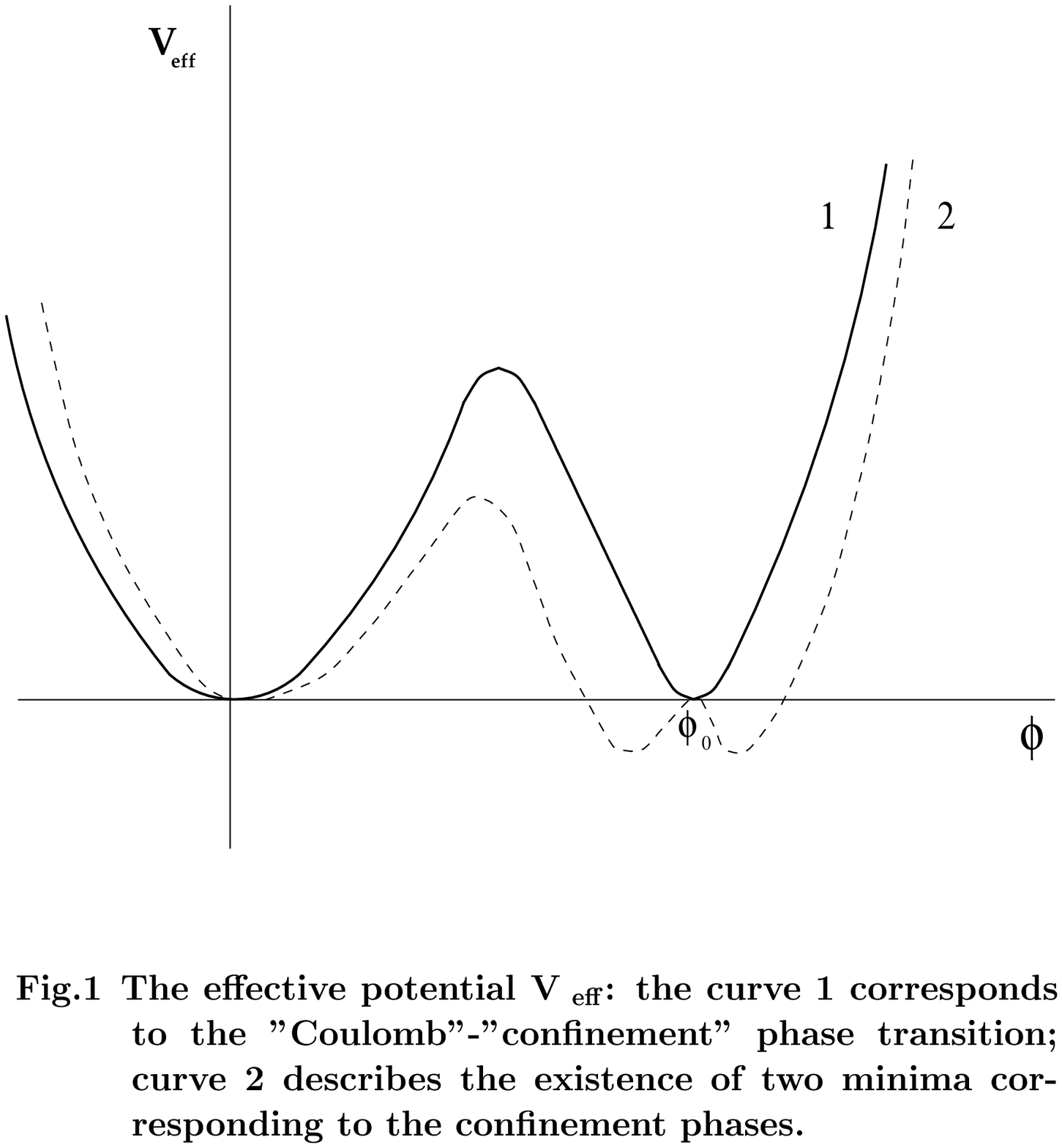}
\end{center}
\end{figure*}

The conditions of the existence of degenerate vacua are given by the
following equations:
\begin{equation}
           V_{eff}(0) = V_{eff}(\phi_0^2) = 0,     \label{39y}
\end{equation}
\begin{equation}
    \frac{\partial V_{eff}}{\partial \phi}|_{\phi=0} =
    \frac{\partial V_{eff}}{\partial \phi}|_{\phi=\phi_0} = 0,
\quad{\mbox{or}}\quad V'_{eff}(\phi_0^2)\equiv
\frac{\partial V_{eff}}{\partial \phi^2}|_{\phi=\phi_0} = 0,
                                                    \label{40y}
\end{equation}
and inequalities
\begin{equation}
    \frac{\partial^2 V_{eff}}{\partial \phi^2}|_{\phi=0} > 0, \qquad
    \frac{\partial^2 V_{eff}}{\partial \phi^2}|_{\phi=\phi_0} > 0.
                                               \label{41y}
\end{equation}
The first equation (\ref{39y}) applied to Eq.(\ref{20y}) gives:
\begin{equation}
    \mu^2_{run} = - \frac{1}{2} \lambda_{run}(t_0)\,\phi_0^2\, G^2(t_0),
\quad{\mbox{where}}\quad t_0 = \log(\phi_0^2/M^2).
                                    \label{42y}
\end{equation}
It is easy to find the joint solution of equations
\begin{equation}
      V_{eff}(\phi_0^2) = V'_{eff}(\phi_0^2) = 0.       \label{47y}
\end{equation}
Using RGE (\ref{22y}), (\ref{23y}) and Eqs.(\ref{40y})--(\ref{47y}),
we obtain:
\begin{equation}
 V'_{eff}(\phi_0^2) =\frac{1}{4}( - \lambda_{run}\beta_{(\mu^2)} +
\lambda_{run} + \beta_{\lambda} - \gamma \lambda_{run})G^4(t_0)\phi_0^2 = 0,
                                                    \label{48y}
\end{equation}
or
\begin{equation}
    \beta_{\lambda} + \lambda_{run}(1 - \gamma - \beta_{(\mu^2)}) = 0.
                                            \label{49y}
\end{equation}
Substituting in Eq.(\ref{49y}) the functions
$\beta_{\lambda}^{(1)},\,\beta_{(\mu^2)}^{(1)}$ and $\gamma^{(1)}$
given by Eqs.(\ref{30y})---(\ref{33y}), we obtain in the one--loop
approximation the following equation for the phase transition border:
\begin{equation}
     g^4_{PT} = - 2\lambda_{run}(\frac{8\pi^2}3 + \lambda_{run}).
                                                 \label{50y}
\end{equation}
The curve (\ref{50y}) is represented on the phase diagram
$(\lambda_{run}; g^2_{run})$ of Fig.2 by the curve "1" which describes
the border between the Coulomb--like phase with $V_{eff} \ge 0$
and the confinement one with $V_{eff}^{min} < 0$. This border corresponds to
the one--loop approximation.

\begin{figure*}
\begin{center}
\noindent\includegraphics[width=120mm,height=110mm]{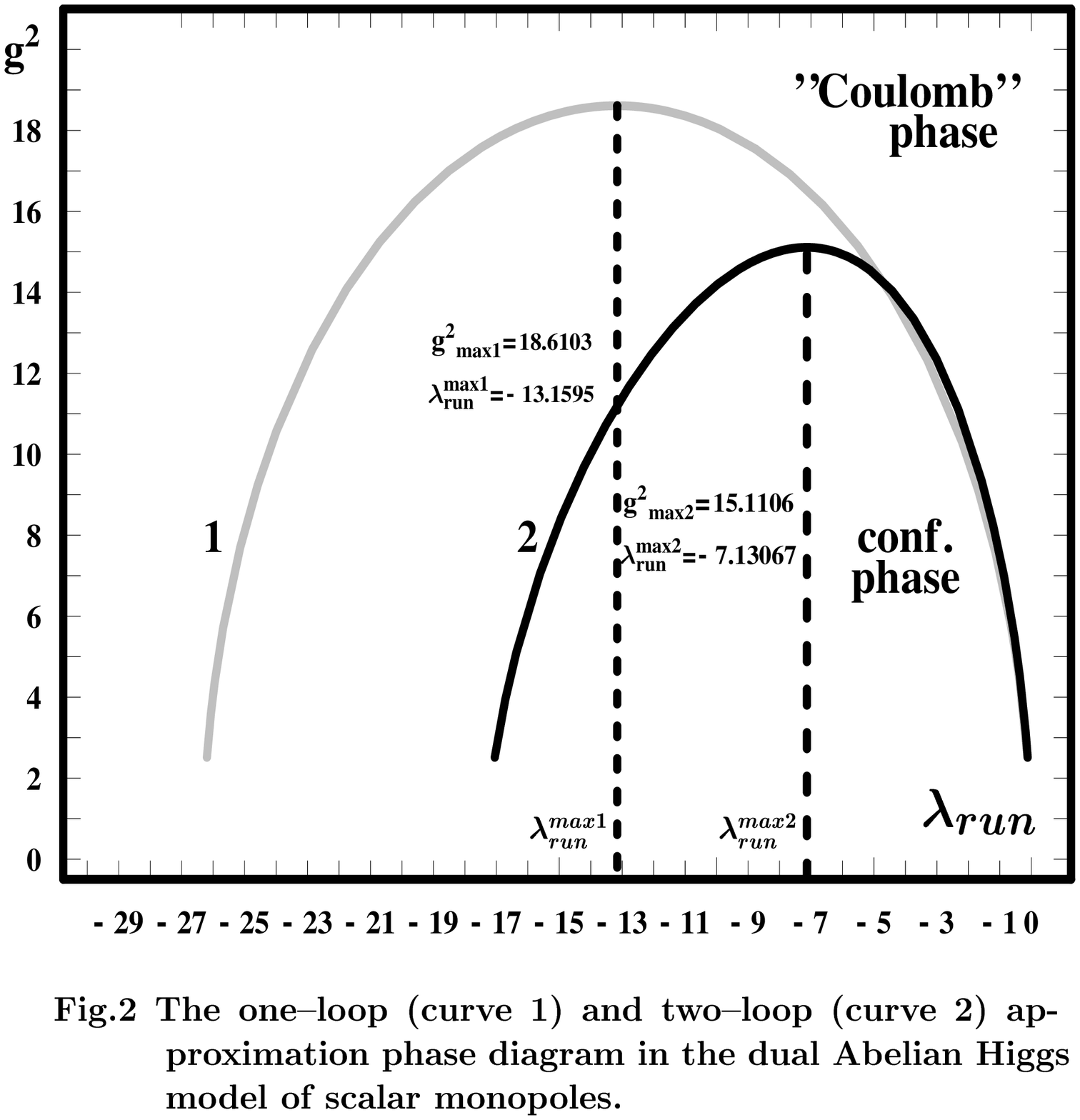}
\end{center}
\end{figure*}

Using Eqs.(\ref{30y})-(\ref{38y}), we are able to construct
the phase transition border in the two--loop approximation.
Substituting these equations into Eq.(\ref{49y}), we obtain the following
phase transition border curve equation in the two--loop approximation:
\begin{equation}
 3y^2 - 16\pi^2 + 6x^2 + \frac{1}{16\pi^2}(28x^3 + \frac{15}{2}x^2y +
  \frac{97}{4}xy^2 - \frac{59}{6}y^3) = 0,            \label{51y}
\end{equation}
where $x = - \lambda_{PT}$ and $y = g^2_{PT}$ are the phase transition
values of $ - \lambda_{run}$ and $g^2_{run}$.
Choosing the physical branch corresponding to $g^2 \ge 0$ and $g^2\to 0$,
when $\lambda \to 0$, we have received curve 2 on the phase diagram
$(\lambda_{run}; g^2_{run})$ shown in Fig.2. This curve
corresponds to the two--loop approximation and can be compared with
the  curve 1 of Fig.2, which describes the same phase transition border
calculated in the one--loop approximation.
It is easy to see that the accuracy of the 1--loop
approximation is not excellent and can commit errors of order 30\%.

According to the phase diagram drawn in Fig.2, the confinement phase
begins at $g^2 = g^2_{max}$ and exists under the phase transition border line
in the region $g^2 \le g^2_{max}$, where $e^2$ is large:
$e^2\ge (2\pi/g_{max})^2$ due to the Dirac relation:
\begin{equation}
   eg = 2\pi, \quad{\mbox{or}}\quad \alpha \tilde \alpha = \frac{1}{4}.
                                             \label{54y}
\end{equation}
Therefore, we have:
\begin{itemize}
\item in the one--loop approximation:
$$
g^2_{crit} = g^2_{max1}\approx 18.61,
\quad
   \tilde \alpha_{crit} = \frac {g^2_{crit}}{4\pi}\approx 1.48,
\quad
  \alpha_{crit} = \frac{1}{4{\tilde \alpha}_{crit}}\approx 0.17
$$
\item in the two--loop approximation:
\begin{equation}
   g^2_{crit} = g^2_{max2}\approx
  15.11, \quad
   \tilde \alpha_{crit} = \frac {g^2_{crit}}{4\pi}\approx 1.20,\quad
        \alpha_{crit} = \frac{1}{4{\tilde \alpha}_{crit}}\approx 0.208
\label{55y}
\end{equation}
\end{itemize}
Comparing these results, we obtain the accuracy of
deviation between them of order 20\%.

The last result (\ref{55y}) coincides with the lattice values
obtained for the compact QED by Monte Carlo method \cite{7}:
\begin{equation}
{\alpha}_{crit}\approx{0.20\pm 0.015},\quad\quad
{\tilde \alpha}_{crit}\approx{1.25\pm 0.10}.      \label{56}
\end{equation}
Writing Eq.(\ref{24y}) with $\beta_g$ function given by Eqs.(\ref{33y}),
(\ref{34y}), and (\ref{37y}), we have the following RGE for the monopole
charge in the two--loop approximation:
\begin{equation}
  \frac{dg^2_{run}}{dt}\approx \frac{g^4_{run}}{48\pi^2} +
    \frac{g^6_{run}}{(16\pi^2)^2},
\quad {\mbox{or}}\quad
   \frac{d\log{\tilde \alpha}}{dt}\approx \frac{\tilde \alpha}{12\pi}
            (1 + 3\frac{\tilde \alpha}{4\pi}).    \label{56ay}
\end{equation}
The values (\ref{55y})  for $g^2_{crit} = g^2_{{max}1,2}$ indicate
that the contribution of two loops described by the second term of
Eq.(\ref{56ay}) is about 0.3, confirming the validity of perturbation theory.

In general, we are able to estimate the validity of two--loop approximation
for all $\beta$--functions and $\gamma$, calculating the corresponding
ratios of two--loop contributions to one--loop contributions
at the maxima of curves 1 and 2:
\begin{equation}
\begin{array}{|l|l|}
\hline %
&\\[-0.2cm]%
\lambda_{crit} = \lambda_{run}^{max1} \approx{-13.16}&\lambda_{crit} =
\lambda_{run}^{max2}\approx{-7.13}\\[0.5cm]
g^2_{crit} = g^2_{max1}\approx{18.61}& g^2_{crit} = g^2_{max2}
\approx{15.11}\\[0.5cm]
\frac{\displaystyle\gamma^{(2)}}{\displaystyle\gamma^{(1)}}\approx{-0.0080}&\frac{\displaystyle
\gamma^{(2)}}{\displaystyle\gamma^{(1)}}\approx{-0.0065}\\[0.5cm]
\frac{\displaystyle\beta_{\mu^2}^{(2)}}{\displaystyle\beta_{\mu^2}^{(1)}}\approx{-0.0826}
&\frac{\displaystyle\beta_{\mu^2}^{(2)}}{\displaystyle\beta_{\mu^2}^{(1)}}
\approx{-0.0637}\\[0.8cm]
\frac{\displaystyle\beta_{\lambda}^{(2)}}{\displaystyle\beta_{\lambda}^{(1)}}\approx{0.1564}
&\frac{\displaystyle\beta_{\lambda}^{(2)}}
{\displaystyle\beta_{\lambda}^{(1)}}\approx{0.0412}\\[0.8cm]
\frac{\displaystyle\beta_g^{(2)}}{\displaystyle\beta_g^{(1)}}\approx{0.3536}&\frac{\displaystyle
\beta_g^{(2)}}{\displaystyle\beta_g^{(1)}}\approx{0.2871}\\[0.5cm]
\hline
\end{array}
                                         \label{57y}
\end{equation}
Here we see that all ratios are sufficiently small, i.e. all
two--loop contributions are small in comparison with one--loop contributions,
confirming the validity of perturbation theory in the 2--loop
approximation. The accuracy of deviation is worse
($\sim 30\%$) for $\beta_g$--function. But it is necessary to emphasize
that calculating the border curves 1 and 2 of Fig.2, we have not used
RGE (\ref{24y}) for monopole charge: $\beta_g$--function is absent in
Eq.(\ref{49y}). Therefore, the calculation of $g^2_{crit}$ according to
Eq.(\ref{51y}) does not depend on the approximation of $\beta_g$ function.
The above--mentioned $\beta_g$--function appears only in the second order
derivative of $V_{eff}$ which is related with the monopole mass $m$
(see Refs.[2,3]).

Eqs.(\ref{55y}) give the following result:
\begin{equation}
 \alpha_{crit}^{-1}\approx 5,      \label{56y}
\end{equation}
which is important for the phase transition at the Planck scale
predicted by the Multiple Point Model (MPM).

\section{Multiple Point Model and Critical Values
of the U(1) and SU(N) Fine Structure Constants}

Investigating the phase transition in HMM,
we had pursued two objects: from one side, we had an aim to
explain the lattice results, from the other side, we were interested
in the predictions of MPM.

\subsection{Anti-grand unification theory}

Most efforts to explain the Standard Model (SM) describing well all
experimental results known today are devoted to Grand Unification
Theories (GUTs). The supersymmetric extension of the SM consists of taking the
SM and adding the corresponding supersymmetric partners.
Unfortunately, at present time experiment does not indicate any manifestation
of the supersymmetry. In this connection, the Anti--Grand Unification
Theory (AGUT) was developed in Refs.[13-17, 4] as a realistic
alternative to SUSY GUTs.  According to this theory, supersymmetry does not
come into the existence up to the Planck energy scale:
$M_{Pl}=1.22\cdot 10^{19}$ GeV.

The Standard Model (SM) is based on the group SMG:
\begin{equation}
SMG = SU(3)_c\times SU(2)_L\times U(1)_Y.  \label{2}
\end{equation}
AGUT suggests that at the energy scale $\mu_G\sim \mu_{Pl}=M_{Pl}$ there
exists the more fundamental group $G$ containing $N_{gen}$ copies of the
Standard Model Group SMG:  \begin{equation} G = SMG_1\times SMG_2\times...\times
SMG_{N_{gen}}\equiv (SMG)^{N_{gen}}, \label{76y} \end{equation} where $N_{gen}$ designates
the number of quark and lepton generations (families).

If $N_{gen}=3$ (as AGUT predicts), then the fundamental gauge group G is:
\begin{equation}
    G = (SMG)^3 = SMG_{1st\, gen.}\times SMG_{2nd\, gen.}\times SMG_{3rd\,
                                       gen.},  \label{77y}
\end{equation}
or the generalized ones:
\begin{equation}
         G_f = (SMG)^3\times U(1)_f
         , \quad {\mbox{or}}\quad
    G_{\mbox{ext}} = (SMG\times U(1)_{B-L})^3 ,      \label{78y}
\end{equation}
which were suggested by the fitting of fermion masses of the SM
(see Refs.\cite{17}), or by the see--saw mechanism with right-handed
neutrinos \cite{19}.

\subsection{Multiple Point Principle}

AGUT approach is used in conjuction with the Multiple Point
Principle proposed in Ref.\cite{4}.
According to this principle Nature seeks a special point --- the Multiple
Critical Point (MCP) --- which is a point on the phase diagram of the
fundamental regulirized gauge theory G (or $G_f$, or $G_{ext}$), where
the vacua of all fields existing in Nature are degenerate having the same
vacuum energy density.
Such a phase diagram has axes given by all coupling constants
considered in theory. Then all (or just many) numbers of phases
meet at the MCP.
MPM assumes the existence of MCP at the Planck scale,
insofar as gravity may be "critical" at the Planck scale.

The philosophy of MPM leads to the necessity
to investigate the phase transition in different gauge theories.
A lattice model of gauge theories is the most convenient formalism for the
realization of the MPM ideas. As it was mentioned above,
in the simplest case we can imagine our
space--time as a regular hypercubic (3+1)--lattice with the parameter $a$
equal to the fundamental (Planck) scale: $a = \lambda_P = 1/M_{Pl}$.

\subsection{AGUT-MPM prediction of the Planck scale values of the
U(1), SU(2) and SU(3) fine structure constants}

The usual definition of the SM coupling constants:
\begin{equation}
  \alpha_1 = \frac{5}{3}\frac{\alpha}{\cos^2\theta_{\overline{MS}}},\quad
  \alpha_2 = \frac{\alpha}{\sin^2\theta_{\overline{MS}}},\quad
  \alpha_3 \equiv \alpha_s = \frac {g^2_s}{4\pi},     \label{81y}
\end{equation}
where $\alpha$ and $\alpha_s$ are the electromagnetic and strong
fine structure constants, respectively, is given in the Modified
minimal subtraction scheme ($\overline{MS}$).
Here $\theta_{\overline{MS}}$ is the Weinberg weak angle in $\overline{MS}$ scheme.
Using RGE with experimentally
established parameters, it is possible to extrapolate the experimental
values of three inverse running constants $\alpha_i^{-1}(\mu)$
(here $\mu$ is an energy scale and i=1,2,3 correspond to U(1),
SU(2) and SU(3) groups of the SM) from the Electroweak scale to the Planck
scale. The precision of the LEP data allows to make this extrapolation
with small errors (see \cite{20}). Assuming that these RGEs for
$\alpha_i^{-1}(\mu)$ contain only the contributions of the SM particles
up to $\mu\approx \mu_{Pl}$ and doing the extrapolation with one
Higgs doublet under the assumption of a "desert", the following results
for the inverses $\alpha_{Y,2,3}^{-1}$ (here $\alpha_Y\equiv \frac{3}{5}
\alpha_1$) were obtained in Ref.\cite{4} (compare with \cite{20}):
\begin{equation}
   \alpha_Y^{-1}(\mu_{Pl})\approx 55.5; \quad
   \alpha_2^{-1}(\mu_{Pl})\approx 49.5; \quad
   \alpha_3^{-1}(\mu_{Pl})\approx 54.0.
                                                        \label{82y}
\end{equation}
The extrapolation of $\alpha_{Y,2,3}^{-1}(\mu)$ up to the point
$\mu=\mu_{Pl}$ is shown in Fig.3.

\begin{figure*}
\begin{center}
\noindent\includegraphics[width=120mm, height=110mm]{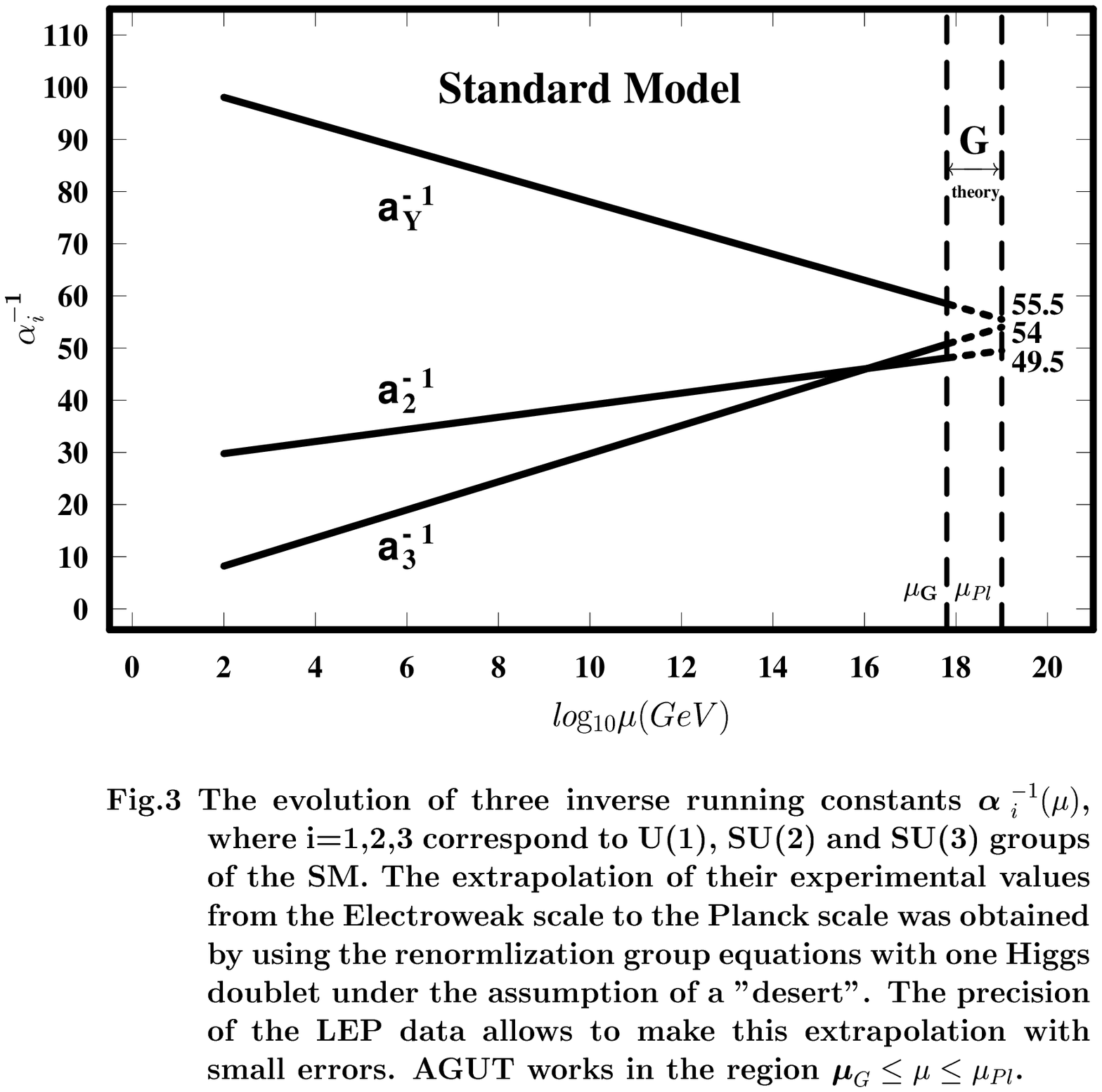}
\end{center}
\end{figure*}

According to AGUT, at some point $\mu=\mu_G < \mu_{Pl}$ (but near
$\mu_{Pl}$) the fundamental group $G$ (or $G_f$, or $G_{\mbox{ext}}$)
undergoes spontaneous breakdown to its diagonal subgroup:
\begin{equation}
      G \longrightarrow G_{diag.subgr.} = \{g,g,g || g\in SMG\},
                                                          \label{83y}
\end{equation}
which is identified with the usual (low--energy) group SMG.

The AGUT prediction of the values of $\alpha_i(\mu)$ at $\mu=\mu_{Pl}$
is based on the MPM assumptions, and gives these values
in terms of the corresponding critical couplings $\alpha_{i,crit} $
[13-15,4]:
\begin{equation}
            \alpha_i(\mu_{Pl}) = \frac {\alpha_{i,crit}}{N_{gen}}
                       = \frac{\alpha_{i,crit}}{3}
                \quad{\mbox{for}}\quad i=2,3,       \label{84y}
\end{equation}
and
\begin{equation}
\alpha_1(\mu_{Pl}) = \frac{\alpha_{1,crit}}{\frac{1}{2}N_{gen}(N_{gen} + 1)}
                   = \frac{\alpha_{1,crit}}{6} \quad{\mbox{for}}\quad U(1).
                                      \label{85y}
\end{equation}
It was assumed in Ref.\cite{4} that the MCP values
$\alpha_{i,crit}$ in Eqs.(\ref{84y}) and (\ref{85y}) coincide with
the triple point values of the effective fine structure
constants given by the lattice SU(3)--, SU(2)-- and U(1)--gauge theories.

If the point $\mu=\mu_G$ is very close to the Planck scale
$\mu=\mu_{Pl}$, then according to Eqs.(\ref{82y}) and (\ref{85y}), we have:
\begin{equation}
         \alpha_{1st\, gen.}^{-1}\approx
    \alpha_{2nd\, gen.}^{-1}\approx \alpha_{3rd\, gen.}^{-1}\approx
    \frac{\alpha_Y^{-1}(\mu_G)}{6}\approx 9,        \label{88y}
\end{equation}
what is almost equal to the value:
\begin{equation}
            \alpha_{crit.,theor}^{-1}\approx 8      \label{89y}
\end{equation}
obtained theoretically by Parisi improvement method
for the Coulomb-like phase [4,6]. The critical value (\ref{89y})
is close to the lattice and HMM ones: see Eq.(\ref{56y}).
This means that in the U(1) sector of AGUT we have $\alpha $ near
the critical point, and we can expect the existence of MCP
at the Planck scale.

\def\AGUT{{}\;\;\raisebox{.9ex}{$\times$}\raisebox{-.5ex}%
{$\!\!\!\!\!\!\!\!\scriptscriptstyle i=1,2,3$} \,(SMG_i \times %
U(1)_{\scriptscriptstyle B-L,i})}
\def\sVEV#1{\left\langle #1\right\rangle}
\def\abs#1{\left| #1\right|}
\def\sleq{\raisebox{-.6ex}{${\textstyle\stackrel{<}{\sim}}$}}
\def\sgeq{\raisebox{-.6ex}{${\textstyle\stackrel{>}{\sim}}$}}
%
\title*{Family Replicated Fit of All Quark and Lepton Masses and Mixings}
\author{H.~B.~Nielsen\thanks{E-mail: hbech@mail.desy.de} 
and Y.~Takanishi\thanks{E-mail: yasutaka@mail.desy.de}}
\institute{%
Deutsches Elektronen-Synchrotron DESY, 
Notkestra{\ss}e 85, D-22603 Hamburg, Germany \\
The Niels Bohr Institute, 
Blegdamsvej 17, DK-2100 Copenhagen {\O}, Denmark}

%
\titlerunning{Family Replicated Fit of All Quark and Lepton Masses and Mixings}
\authorrunning{H.~B.~Nielsen and Y.~Takanishi}
\maketitle

\begin{abstract}
We review our recent development of family replicated 
gauge group model, which generates the Large Mixing Angle MSW 
solution. The model is based on each family of quarks and leptons 
having its own set of gauge fields, each containing a replica of 
the Standard Model gauge fields plus a $(B-L)$-coupled gauge 
field. A fit of all the seventeen quark-lepton mass and mixing 
angle observables, using just six new Higgs field vacuum 
expectation values, agrees with the experimental data 
order of magnitudewise. However, this model can not
predict the baryogenesis in right order, therefore, we
discuss further modification of our model and present
a preliminary result of baryon number to entropy ratio.
\end{abstract}

\section{Introduction}

We have previously attempted to fit all the fermion masses and their 
mixing angles~\cite{FNT,NT1} including 
baryogenesis~\cite{NT2} in a model without
supersymmetry or grand unification. 
This model has the 
maximum number of gauge fields consistent with maintaining 
the irreduciblity of the usual Standard Model fermion 
representations, added three right-handed neutrinos. 
The predictions of this previous model are in order 
of magnitude agreement with all existing experimental data, 
however, only provided we use the Small Mixing Angle MSW~\cite{MSW} (SMA-MSW) 
solution. But, for 
the reasons given below, the SMA-MSW solution is now 
disfavoured by experiments. So here we review a modified 
version of the previous model, which 
manages to accommodate the Large Mixing Angle MSW (LMA-MSW) solution 
for solar neutrino oscillations using 6 additional Higgs fields (relative 
to the Standard Model) vacuum 
expectation values (VEVs) as adjustable parameters.

A neutrino oscillation solution to the solar neutrino problem
and a favouring of the LMA-MSW solution 
is supported by SNO results~\cite{SNO}: The measurement of the $^8$B and $hep$ 
solar neutrino fluxes shows no significant energy dependence 
of the electron neutrino survival probability in the
Super-Kamiokande and SNO energy ranges. 

Moreover, the important result which also supports LMA-MSW solution 
on the solar neutrino problem, 
reported by the Super-Kamiokande collaboration~\cite{SKDN}, 
that the day-night asymmetry data disfavour the 
SMA-MSW solution at the $95\%$ C.L..

In fact, global analyses~\cite{fogli,cc1,goswami,smirnov} of all
solar neutrino data have confirmed that the LMA-MSW solution gives the best 
fit to the data and that the SMA-MSW solution is very strongly
disfavoured and only acceptable at the $3\sigma$ level. Typical best fit
values of the mass squared difference and mixing angle parameters 
in the two flavour LMA-MSW solution are 
$\Delta m^2_\odot\approx4.5\times 10^{-5}~\mbox{\rm eV}^2$ 
and $\tan^2\theta_{\odot}\approx0.35$.

This paper is organised as follows: In the next section, we 
present our gauge group -- the family replicated gauge group -- 
and the quantum numbers of fermion and Higgs fields. 
Then, in section $3$ we discuss our philosophy of all 
gauge- and Yukawa couplings
at Planck scale being of order unity. In section 
$4$ we address how 
the family replicated gauge group breaks down to Standard Model gauge
group, and we add a small review of see-saw mechanism.
The mass matrices of all sectors are presented in section $5$,
the renormalisation group equations -- renormalisable and also 
5 dimensional non-renormalisable ones -- are shown in section $6$.
The calculation 
is described in section $7$ and the results are presented 
in section $8$. We discuss further modification of our model 
and present a preliminary results of baryon number to entropy ratio
in section $9$. Finally, section $10$ contains our conclusion.

\section{Quantum numbers of model}

Our model has, as its back-bone, the property that there are generations 
(or families) not only for fermions but also 
for the gauge bosons, \mbox{{\it i.e.}}, we have a generation (family) 
replicated gauge group namely 
\begin{equation}
  \label{eq:frg}
  \AGUT \hspace{2mm},
\end{equation}
where $SMG$ denotes the Standard Model gauge group 
$\equiv SU(3)\times SU(2)\times U(1)$, $\times$ denotes the 
Cartesian product and $i$ runs through the generations, $i=1,2,3$.

Note that this family replicated gauge group, eq.~(\ref{eq:frg}), 
is the maximal gauge group under the following assumptions:
\begin{itemize}
\item It should only contain transformations which change the known 
45 (= 3 generations of 15 Weyl particles each) Weyl fermions of the Standard Model 
and the additional three heavy see-saw (right-handed) neutrinos. 
That is our gauge group is assumed to be a subgroup of $U(48)$.
\item We avoid any new gauge transformation that would transform a 
Weyl state from one irreducible representation of the Standard Model 
group into another irreducible representation: 
there is no gauge coupling unification.
\item  The gauge group does not contain any anomalies in the gauge 
symmetry -- neither gauge nor mixed anomalies even without using the 
Green-Schwarz anomaly cancelation mechanism.
\item It should be as big as possible under the foregoing assumptions.
\end{itemize}

\begin{table}[!t]
\caption{All $U(1)$ quantum charges in the family replicated model. 
The symbols for the fermions shall be considered to mean
``proto''-particles. Non-abelian representations are given by a rule 
from the abelian ones (see Eq.~(\ref{eq:mod})).}
\vspace{3mm}
\label{Table1}
\begin{center}
\begin{tabular}{|c||c|c|c|c|c|c|} \hline
& $SMG_1$& $SMG_2$ & $SMG_3$ & $U_{\scriptscriptstyle B-L,1}$ & 
$U_{\scriptscriptstyle B-L,2}$ & $U_{\scriptscriptstyle B-L,3}$ \\ \hline\hline
$u_L,d_L$ &  $\frac{1}{6}$ & $0$ & $0$ & $\frac{1}{3}$ & $0$ & $0$ \\
$u_R$ &  $\frac{2}{3}$ & $0$ & $0$ & $\frac{1}{3}$ & $0$ & $0$ \\
$d_R$ & $-\frac{1}{3}$ & $0$ & $0$ & $\frac{1}{3}$ & $0$ & $0$ \\
$e_L, \nu_{e_{\scriptscriptstyle L}}$ & $-\frac{1}{2}$ & $0$ & $0$ & $-1$ & $0$ 
& $0$ \\
$e_R$ & $-1$ & $0$ & $0$ & $-1$ & $0$ & $0$ \\
$\nu_{e_{\scriptscriptstyle R}}$ &  $0$ & $0$ & $0$ & $-1$ & $0$ & $0$ \\ \hline
$c_L,s_L$ & $0$ & $\frac{1}{6}$ & $0$ & $0$ & $\frac{1}{3}$ & $0$ \\
$c_R$ &  $0$ & $\frac{2}{3}$ & $0$ & $0$ & $\frac{1}{3}$ & $0$ \\
$s_R$ & $0$ & $-\frac{1}{3}$ & $0$ & $0$ & $\frac{1}{3}$ & $0$\\
$\mu_L, \nu_{\mu_{\scriptscriptstyle L}}$ & $0$ & $-\frac{1}{2}$ & $0$ & $0$ & 
$-1$ & $0$\\
$\mu_R$ & $0$ & $-1$ & $0$ & $0$  & $-1$ & $0$ \\
$\nu_{\mu_{\scriptscriptstyle R}}$ &  $0$ & $0$ & $0$ & $0$ & $-1$ & $0$ \\ \hline
$t_L,b_L$ & $0$ & $0$ & $\frac{1}{6}$ & $0$ & $0$ & $\frac{1}{3}$ \\
$t_R$ &  $0$ & $0$ & $\frac{2}{3}$ & $0$ & $0$ & $\frac{1}{3}$ \\
$b_R$ & $0$ & $0$ & $-\frac{1}{3}$ & $0$ & $0$ & $\frac{1}{3}$\\
$\tau_L, \nu_{\tau_{\scriptscriptstyle L}}$ & $0$ & $0$ & $-\frac{1}{2}$ & $0$ & 
$0$ & $-1$\\
$\tau_R$ & $0$ & $0$ & $-1$ & $0$ & $0$ & $-1$\\
$\nu_{\tau_{\scriptscriptstyle R}}$ &  $0$ & $0$ & $0$ & $0$ & $0$ & $-1$ \\ 
\hline \hline
$\phi_{\scriptscriptstyle WS}$ & $0$ & $\frac{2}{3}$ & $-\frac{1}{6}$ & $0$ & 
$\frac{1}{3}$ & $-\frac{1}{3}$ \\
$\omega$ & $\frac{1}{6}$ & $-\frac{1}{6}$ & $0$ & $0$ & $0$ & $0$\\
$\rho$ & $0$ & $0$ & $0$ & $-\frac{1}{3}$ & $\frac{1}{3}$ & $0$\\
$W$ & $0$ & $-\frac{1}{2}$ & $\frac{1}{2}$ & $0$ & $-\frac{1}{3}$ & $\frac{1}{3}$ \\
$T$ & $0$ & $-\frac{1}{6}$ & $\frac{1}{6}$ & $0$ & $0$ & $0$\\
$\chi$ & $0$ & $0$ & $0$ & $0$ & $-1$ & $1$ \\
$\phi_{\scriptscriptstyle B-L}$ & $0$ & $0$ & $0$ & $0$ & $0$ & $2$ \\ 
\hline
\end{tabular}
\end{center}
\end{table}

The quantum numbers of the particles/fields in our model are found in table 1
and use of the following procedure: In table 1 one finds the charges 
under the six $U(1)$ groups in the gauge 
group \ref{eq:frg}.
Then for each particle one should take the representation under the $SU(2)_i$
and $SU(3)_i$ groups ($i=1,2,3$) with
lowest dimension matching to $y_i/2$ according to the requirement 
\begin{equation}
  \label{eq:mod}
  \frac{t_i}{3}+\frac{d_i}{2} + \frac{y_i}{2} = 0~~{\rm (mod~1)}\hspace{2mm},
\end{equation}
where $t_i$ and $d_i$ are the triality and duality for 
the $i$'th proto-generation gauge groups $SU(3)_i$ and $SU(2)_i$ 
respectively.

\section{The philosophy of all couplings being order unity}
\label{philo}

Any realistic model and at least certainly our model tends to 
get far more fundamental couplings than we have parameters 
in the Standard Model and thus pieces of data to fit. This 
is especially so for our model based on many $U(1)$ charges~\cite{FN} 
because we take it to have 
practically any not mass protected particles one may propose 
at the fundamental mass scale, taken to be the Planck mass. Especially 
we assume the existence of Dirac fermions with order of fundamental 
scale masses needed to allow the quark and lepton Weyl particles to 
take up successively gauge charges from the Higgs fields VEVs.
So unless we make assumptions about the many coupling constants 
and fundamental masses we have no chance to predict anything. Almost 
the only chance of making an assumption about all these couplings, 
which is not very model dependent, is to assume that they are {\em all of 
order unity} in the fundamental unit. This is the same type of assumption 
that is really behind use of dimensional arguments to estimate sizes of 
quantities. A procedure very often used successfully. If we really 
assumed every coupling and mass of order unity we would get the 
effective Yukawa couplings of the quarks and leptons to the 
Weinberg-Salam Higgs field to be also of order unity what is 
phenomenologically not true. To avoid this prediction we then 
blame the smallness of all but the top-Yukawa coupling on smallness 
in fundamental Higgs VEVs. That is to say we assume that the 
VEVs of the Higgs fields in Table 1, $\rho$, $\omega$, $T$, 
$W$, $\chi$,
$\phi_{B-L}$ and $\phi_{WS}$ are (possibly) very small 
compared to the fundamental/Planck unit, and these are the 
quantities we have to fit.

Technically we implement these unknown -- but of order unity according to 
our assumption -- couplings and masses by a Monte Carlo technique: we put 
them equal to random numbers with a distribution dominated by numbers 
of order unity and then perform the calculation of the observable 
quantities such as quark or lepton masses and mixing
angles again and again. At the end we average the logarithmic of 
these quantities and exponentiate them. In this way we expect to get the
typical order of magnitude predicted for the observable quantities.
In praxis we do not have to put random numbers in for all the many 
couplings in the fundamental model, but can instead just provide each 
mass matrix element with a single random number factors.

After all a product of several of order unity factors is just an order 
unity factor again. To resume our model philosophy:
{\em Only Higgs field VEVs are not of order unity. We must be
 satisfied with order of magnitude results.}

\section{Breaking of the family replicated gauge group to the Standard Model}
\indent\ The family replicated gauge group broken down to its diagonal
subgroup at scales about one or half order of magnitude under the Planck 
scale by Higgs fields -- $W$, $T$, $\omega$, $\rho$ 
and $\chi$ (in Table~\ref{Table1}):
\begin{equation}
  \label{eq:subagut}
\AGUT \rightarrow SMG\times U(1)_{B-L}\,.
\end{equation}
This diagonal subgroup is further broken down by yet two more 
Higgs fields --- the Weinberg-Salam Higgs field $\phi_{WS}$ and 
another Higgs field $\phi_{B-L}$ --- to 
$SU(3)\times U(1)_{em}$. 

\subsection{See-saw mechanism}
\indent\ See-saw mechanics is build into our model to fit the scale 
of the neutrino oscillations, \mbox{\it i.e.},  we use the 
right-handed neutrinos with heavy Majorana masses ($10^{11}$ GeV).

In order to mass-protect the right-handed neutrino from getting Planck scale
masses, we have to introduce $\phi_{\scriptscriptstyle B-L}$ 
which breaks the $B-L$ quantum charge spontaneously, and using this
new Higgs filed we are able to deal the neutrino oscillations,
\mbox{\it i.e.}, to fit the scale of the see-saw particle masses.
However, due to mass-protection by the Standard Model gauge symmetry, the 
left-handed Majorana mass terms should be negligible in our model.
Then, naturally, the light neutrino mass matrix -- effective left-left 
transition Majorana mass matrix -- can be obtained via the see-saw 
mechanism~\cite{seesaw}:
\begin{equation}
  \label{eq:meff}
  M_{\rm eff} \! \approx \! M^D_\nu\,M_R^{-1}\,(M^D_\nu)^T\,.
\end{equation}

\section{Mass matrices}

Using the $U(1)$ fermion quantum charges and Higgs field 
(presented in Table~\ref{Table1}) we can calculate the degrees of suppressions
of the left-right transition -- Dirac mass -- matrices and 
also Majorana mass matrix (right-right transition).

Note that the random complex order of unity 
numbers which are supposed to multiply all the mass matrix elements 
are not represented in following matrices:
\noindent
the up-type quarks:
\begin{eqnarray}
M_{\scriptscriptstyle U} \simeq \frac{\sVEV{(\phi_{\scriptscriptstyle\rm WS})^\dagger}}{\sqrt{2}}
\hspace{-0.1cm}
\left(\!\begin{array}{ccc}
        (\omega^\dagger)^3 W^\dagger T^2
        & \omega \rho^\dagger W^\dagger T^2
        & \omega \rho^\dagger (W^\dagger)^2 T\\
        (\omega^\dagger)^4 \rho W^\dagger T^2
        &  W^\dagger T^2
        & (W^\dagger)^2 T\\
        (\omega^\dagger)^4 \rho
        & 1
        & W^\dagger T^\dagger
\end{array} \!\right)\label{M_U}
\end{eqnarray}  
\noindent
the down-type quarks:
\begin{eqnarray}
M_{\scriptscriptstyle D} \simeq \frac{\sVEV{\phi_{\scriptscriptstyle\rm WS}}}
{\sqrt{2}}\hspace{-0.1cm}
\left (\!\begin{array}{ccc}
        \omega^3 W (T^\dagger)^2
      & \omega \rho^\dagger W (T^\dagger)^2
      & \omega \rho^\dagger T^3 \\
        \omega^2 \rho W (T^\dagger)^2
      & W (T^\dagger)^2
      & T^3 \\
        \omega^2 \rho W^2 (T^\dagger)^4
      & W^2 (T^\dagger)^4
      & W T
                        \end{array} \!\right) \label{M_D}
\end{eqnarray}
\noindent %
the charged leptons:
\begin{eqnarray}        
M_{\scriptscriptstyle E} \simeq \frac{\sVEV{\phi_{\scriptscriptstyle\rm WS}}}
{\sqrt{2}}\hspace{-0.1cm}
\left(\hspace{-0.1 cm}\begin{array}{ccc}
    \omega^3 W (T^\dagger)^2
  & (\omega^\dagger)^3 \rho^3 W (T^\dagger)^2 
  & (\omega^\dagger)^3 \rho^3 W T^4 \chi \\
    \omega^6 (\rho^\dagger)^3  W (T^\dagger)^2 
  &   W (T^\dagger)^2 
  &  W T^4 \chi\\
    \omega^6 (\rho^\dagger)^3  (W^\dagger)^2 T^4 
  & (W^\dagger)^2 T^4
  & WT
\end{array} \hspace{-0.1cm}\right) \label{M_E}
\end{eqnarray}
\noindent
the Dirac neutrinos:
\begin{eqnarray}
M^D_\nu \simeq \frac{\sVEV{(\phi_{\scriptscriptstyle\rm WS})^\dagger}}{\sqrt{2}}
\hspace{-0.1cm}
\left(\hspace{-0.1cm}\begin{array}{ccc}
        (\omega^\dagger)^3 W^\dagger T^2
        & (\omega^\dagger)^3 \rho^3 W^\dagger T^2
        & (\omega^\dagger)^3 \rho^3 W^\dagger  T^2 \chi\\
        (\rho^\dagger)^3 W^\dagger T^2
        &  W^\dagger T^2
        & W^\dagger T^2 \chi\\
        (\rho^\dagger)^3 W^\dagger T^\dagger \chi^\dagger
        &  W^\dagger T^\dagger \chi^\dagger
        & W^\dagger T^\dagger
\end{array} \hspace{-0.1 cm}\right)\label{Mdirac}
\end{eqnarray} 
\noindent %
and the Majorana (right-handed) neutrinos:
\begin{eqnarray}    
M_R \simeq \sVEV{\phi_{\scriptscriptstyle\rm B-L}}\hspace{-0.1cm}
\left (\hspace{-0.1 cm}\begin{array}{ccc}
(\rho^\dagger)^6 (\chi^\dagger)^2
& (\rho^\dagger)^3 (\chi^\dagger)^2
& (\rho^\dagger)^3 \chi^\dagger \\
(\rho^\dagger)^3 (\chi^\dagger)^2
& (\chi^\dagger)^2 & \chi^\dagger \\
(\rho^\dagger)^3 \chi^\dagger & \chi^\dagger
& 1
\end{array} \hspace{-0.1 cm}\right ) \label{Mmajo}
\end{eqnarray}

\section{Renormalisation group equations from Planck scale to week scale 
via see-saw scale}
\indent\ It should be kept in mind that the effective Yukawa couplings for 
the Weinberg-Salam Higgs field, which 
are given by the Higgs field factors in the above mass matrices 
multiplied by order unity factors, 
are the running Yukawa couplings at a scale {\em near the Planck 
scale}. In this way, we had to 
use the renormalisation group (one-loop)
$\beta$-functions to run these couplings down to the experimentally 
observable scale which we took for the charged fermion masses to be 
compared to ``measurements'' 
at the scale of $1~\mbox{\rm GeV}$, except for the 
top quark mass prediction. We define the top quark pole mass:
\begin{equation}
M_t = m_t(M)\left(1+\frac{4}{3}\frac{\alpha_s(M)}{\pi}\right)\hspace{2mm},
\end{equation}
where we put $M=180~\mbox{\rm GeV}$ for simplicity.

We use the one-loop $\beta$ functions for the gauge 
couplings and the charged fermion 
Yukawa matrices~\cite{pierre} as follows:
\begin{eqnarray}
  \label{eq:recha}
16 \pi^2 {d g_{1}\over d  t} &\!=\!& \frac{41}{10} \, g_1^3 \nonumber\\
16 \pi^2 {d g_{2}\over d  t} &\!=\!& - \frac{19}{16} \, g_2^3  \nonumber\\
16 \pi^2 {d g_{3}\over d  t} &\!=\!& - 7 \, g_3^3  \nonumber\\
16 \pi^2 {d Y_{\scriptscriptstyle U}\over d  t} &\!=\!& \frac{3}{2}\, 
\left( Y_{\scriptscriptstyle U} (Y_{\scriptscriptstyle U})^\dagger
-  Y_{\scriptscriptstyle D} (Y_{\scriptscriptstyle D})^\dagger\right)\, Y_{\scriptscriptstyle U} \nonumber\\
&& + \left\{\, Y_{\scriptscriptstyle S} - \left(\frac{17}{20} g_1^2 
+ \frac{9}{4} g_2^2 + 8 g_3^2 \right) \right\}\, Y_{\scriptscriptstyle U}\\
16 \pi^2 {d Y_{\scriptscriptstyle D}\over d  t} &\!=\!& \frac{3}{2}\, 
\left( Y_{\scriptscriptstyle D} (Y_{\scriptscriptstyle D})^\dagger
-  Y_{\scriptscriptstyle U} (Y_{\scriptscriptstyle U})^\dagger\right)\,Y_{\scriptscriptstyle D} \nonumber\\ 
&& + \left\{\, Y_{\scriptscriptstyle S} - \left(\frac{1}{4} g_1^2 
+ \frac{9}{4} g_2^2 + 8 g_3^2 \right) \right\}\, Y_{\scriptscriptstyle D} \nonumber\\
16 \pi^2 {d Y_{\scriptscriptstyle E}\over d  t} &\!=\!& \frac{3}{2}\, 
\left( Y_{\scriptscriptstyle E} (Y_{\scriptscriptstyle E})^\dagger \right)\,Y_{\scriptscriptstyle E} \nonumber\\
&& + \left\{\, Y_{\scriptscriptstyle S} - \left(\frac{9}{4} g_1^2 
+ \frac{9}{4} g_2^2 \right) \right\}\, Y_{\scriptscriptstyle E}\nonumber\\
Y_{\scriptscriptstyle S} &\!=\!& {{\rm Tr}{}}(\, 3\, Y_{\scriptscriptstyle U}^\dagger\, Y_{\scriptscriptstyle U} 
+  3\, Y_{\scriptscriptstyle D}^\dagger \,Y_{\scriptscriptstyle D} +  Y_{\scriptscriptstyle E}^\dagger\, Y_{\scriptscriptstyle E}\,)  \nonumber\hspace{2mm},
\end{eqnarray}
where $t=\ln\mu$.

By calculation we use the following initial values of gauge coupling constants:
\begin{eqnarray}
U(1): \quad & g_1(M_Z) = 0.462\hspace{2mm}, \quad & g_1(M_{\rm Planck}) = 0.614\\
SU(2):\quad & g_2(M_Z) = 0.651\hspace{2mm}, \quad & g_2(M_{\rm Planck}) = 0.504\\
SU(3):\quad & g_3(M_Z) = 1.22 \hspace{2mm}, \quad & g_3(M_{\rm Planck}) = 0.491
\end{eqnarray}

\subsection{The renormalisation group equations for the effective 
neutrino mass matrix}
\indent

The effective light neutrino masses are given by an irrelevant, 
nonrenormalisable (5 dimensional term) -- 
effective mass matrix $M_{\rm eff}$ -- for which the running
formula is~\cite{Meffrun}:
\begin{equation}
\label{eq:remeff}
16 \pi^2 {d M_{\rm eff} \over d  t}
= ( - 3 g_2^2 + 2 \lambda + 2 Y_{\scriptscriptstyle S} ) \,M_{\rm eff}
- {3\over 2} \left( M_{\rm eff}\, ( Y_{\scriptscriptstyle E} Y_{\scriptscriptstyle E}^\dagger )^T 
+ ( Y_{\scriptscriptstyle E} Y_{\scriptscriptstyle E}^\dagger ) \,M_{\rm eff}\right) \hspace{2mm},
\end{equation}
where $\lambda$ is the Weinberg-Salam Higgs self-coupling constant and
the mass of the Standard Model Higgs boson is given by 
$M_H^2 = \lambda \sVEV{\phi_{WS}}^2$. We just for simplicity 
take $M_H = 115~\mbox{\rm GeV}$ thereby we ignore the running of
the Higgs self-coupling and fixed as $\lambda=0.2185$.

Note that the renormalisation group equations are used to evolve 
the effective neutrino mass matrix from the see-saw 
sale, set by $\sVEV{\phi_{B-L}}$ in our model, to $1~\mbox{\rm GeV}$. 

\section{Method of numerical computation}
\label{sec:FN}
\indent\ In the philosophy of order unity numbers spelled out in
section~\ref{philo} we evaluate 
the product of mass-protecting Higgs VEVs required 
for each mass matrix element and provide it  
with a random complex number, $\lambda_{ij}$, of order one as a factor
taken to have Gaussian distribution with mean value zero. But we
hope the exact form of distribution does not matter much provided 
we have $\sVEV{\ln \abs{\lambda_{ij}}}=0$. In this way, we 
simulate a long chain of fundamental Yukawa couplings 
and propagators making the transition corresponding to an 
effective Yukawa coupling in the Standard Model and the parameters 
in neutrino sector. In the numerical 
computation we then calculate the masses and mixing angles time 
after time, using different sets of random numbers and, in the 
end, we take the logarithmic average of the calculated quantities 
according to the following formula:
\begin{equation}
  \label{eq:avarage}
  \sVEV{m}=\exp\left(\sum_{i=1}^{N} \frac{\ln m_i}{N}\right) \,. 
\end{equation}
Here $\sVEV{m}$ is what we take to be the prediction for one of the 
masses or mixing angles, $m_i$ is the result of the calculation
done with one set of random number combinations and $N$ is the total 
number of random number combinations used.

Since we only expect to make order of magnitude fits, we should of 
course not use ordinary $\chi^2$ defined form the experimental 
uncertainties by rather the $\chi^2$ that would correspond to a 
relative uncertainly -- an uncertain factor of order unity. Since
the normalisation of such a $\chi^2$ is not so easy to choose 
exactly we define instead a quantity which we call the 
goodness of fit (\mbox{\rm g.o.f.}). Since 
our model can only make predictions order of magnitudewise, this 
quantity \mbox{\rm g.o.f.} should only depend on the ratios of
the fitted masses and mixing angles to the experimentally
determined masses and mixing angles:
\begin{equation}
\label{gof}
\mbox{\rm g.o.f.}\equiv\sum \left[\ln \left(
\frac{\sVEV{m}}{m_{\rm exp}} \right) \right]^2 \,,
\end{equation}
where $m_{\rm exp}$ are the
corresponding experimental values presented in 
Table~\ref{convbestfit}.

We should emphasise that we \underline{do not} adjust
the order of one numbers by selection, \mbox{\it i.e.}, the
complex random numbers are needed for only calculational 
purposes. That means that we have only six adjustable parameters
-- VEVs of Higgs fields -- and, on the other hand, that the 
averages of the predicted quantities, $\sVEV{m}$, are just 
results of integration
over the ``dummy'' variables -- random numbers -- therefore, the
random numbers are not at all parameters!

Strictly speaking, however, one could consider the choice of the 
distribution of the random order unity numbers as parameters. But 
we hope that provided we impose on the distribution the conditions that 
the average be zero and the average of the logarithm of the numerical
value be zero, too, any reasonably smooth distribution would 
give similar results for the $\sVEV{m}$ values at the end. In our 
early work~\cite{NT1} we did see that a couple of different proposals 
did not make too much difference.

\section{Results}
\indent\ We averaged over $N=10,000$ complex order unity random 
number combinations. These complex numbers
are chosen to be a number picked from a Gaussian 
distribution, with mean value zero and standard deviation one, 
multiplied by a random phase factor. We put them as factors into 
the mass matrices (eqs.~\ref{M_U}-\ref{Mmajo}). Then 
we computed averages according to eq.~(\ref{eq:avarage}) and used 
eq.~(\ref{gof}) as a $\chi^2$ to fit the $6$ free parameters and found:
\begin{eqnarray} 
\label{eq:VEVS} 
&&\sVEV{\phi_{\scriptscriptstyle WS}}= 246~\mbox{\rm GeV}\,,  
\,\sVEV{\phi_{\scriptscriptstyle B-L}}=1.64\times10^{11}~\mbox{\rm GeV}\,, 
\,\sVEV{\omega}=0.233\,,\nonumber\\
&&\,\sVEV{\rho}=0.246\,,\,\sVEV{W}=0.134\,,
\,\sVEV{T}=0.0758\,,\,\sVEV{\chi}=0.0737\,,
\end{eqnarray}
where, except for the Weinberg-Salam Higgs field and 
$\sVEV{\phi_{\scriptscriptstyle B-L}}$, the VEVs are expressed in Planck units. 
Hereby we have considered that the Weinberg-Salam Higgs field VEV is 
already fixed by the Fermi constant.
The results of the best fit, with the VEVs in eq.~(\ref{eq:VEVS}), 
are shown in Table~\ref{convbestfit} and the fit has  
$\mbox{\rm g.o.f.}=3.63$. 

We have $11=17 - 6$ degrees of freedom -- predictions -- leaving each of 
them with a logarithmic error of
$\sqrt{3.63/11}\simeq0.57$, which is very close to the 
theoretically expected value $64\%$~\cite{FF}. This means 
that we can fit {\rm all quantities} within a factor 
$\exp\left(\sqrt{3.63/11}\right)\simeq1.78$ of the experimental values.

\begin{table}[!t]
\caption{Best fit to conventional experimental data.
All masses are running
masses at $1~\mbox{\rm GeV}$ except the top quark mass which is the pole mass.
Note that we use the square roots of the neutrino data in this 
Table, as the fitted neutrino mass and mixing parameters 
$\sVEV{m}$, in our goodness of fit ($\mbox{\rm g.o.f.}$) definition, 
eq.~(\ref{gof}).}
\begin{displaymath}
\begin{array}{|c|c|c|}
\hline\hline
 & {\rm Fitted} & {\rm Experimental} \\ \hline
m_u & 4.4~\mbox{\rm MeV} & 4~\mbox{\rm MeV} \\
m_d & 4.3~\mbox{\rm MeV} & 9~\mbox{\rm MeV} \\
m_e & 1.0~\mbox{\rm MeV} & 0.5~\mbox{\rm MeV} \\
m_c & 0.63~\mbox{\rm GeV} & 1.4~\mbox{\rm GeV} \\
m_s & 340~\mbox{\rm MeV} & 200~\mbox{\rm MeV} \\
m_{\mu} & 80~\mbox{\rm MeV} & 105~\mbox{\rm MeV} \\
M_t & 208~\mbox{\rm GeV} & 180~\mbox{\rm GeV} \\
m_b & 7.2~\mbox{\rm GeV} & 6.3~\mbox{\rm GeV} \\
m_{\tau} & 1.1~\mbox{\rm GeV} & 1.78~\mbox{\rm GeV} \\
V_{us} & 0.093 & 0.22 \\
V_{cb} & 0.027 & 0.041 \\
V_{ub} & 0.0025 & 0.0035 \\ \hline
\Delta m^2_{\odot} & 9.5 \times 10^{-5}~\mbox{\rm eV}^2 &  4.5 \times 10^{-5}~\mbox{\rm eV}^2 \\
\Delta m^2_{\rm atm} & 2.6 \times 10^{-3}~\mbox{\rm eV}^2 &  3.0 \times 10^{-3}~\mbox{\rm eV}^2\\
\tan^2\theta_{\odot} &0.23 & 0.35\\
\tan^2\theta_{\rm atm}& 0.65 & 1.0\\
\tan^2\theta_{13}  & 4.8 \times 10^{-2} & \sleq~2.6 \times 10^{-2}\\
\hline\hline
\mbox{\rm g.o.f.} &  3.63 & - \\
\hline\hline
\end{array}
\end{displaymath}
\label{convbestfit}
\end{table}

{}From the table~\ref{convbestfit} the experimental mass values are a factor
two higher than predicted for down, charm and for the Cabibbo angle
$V_{us}$ while they are smaller by a factor for strange and 
electron. Thinking only on the angles and masses (not squared) the agreement
is in other cases better than a factor two.

Experimental results for the values of neutrino mixing angles 
are often presented in terms of the function $\sin^22\theta$ 
rather than $\tan^2\theta$ (which, contrary to $\sin^22\theta$, 
does not have a maximum at $\theta=\pi/4$ and thus still varies 
in this region).
Transforming from $\tan^2\theta$ variables to $\sin^22\theta$ 
variables, our predictions for the neutrino mixing angles become:
\begin{eqnarray}
  \label{eq:sintan}
 \sin^22\theta_{\odot} &\!=\!&0.61\,,\\
 \sin^22\theta_{\rm atm} &\!=\!& 0.96\,, \\
 \sin^22\theta_{13} &\!=\!& 0.17\,.
\end{eqnarray}  
We also give here our predicted hierarchical neutrino mass 
spectrum:
\begin{eqnarray}
m_1 &\!=\!&  4.9\times10^{-4}~~\mbox{\rm eV}\,, 
\label{eq:neutrinomass1}\\
m_2 &\!=\!&  9.7\times10^{-3}~~\mbox{\rm eV}\,, 
\label{eq:neutrinomass2}\\
m_3 &\!=\!&  5.2\times10^{-2}~~\mbox{\rm eV}\,.
\label{eq:neutrinomass3} 
\end{eqnarray}

Our agreement with experiment is excellent: all of our 
order of magnitude neutrino predictions lie 
inside the $99\%$ C.L. border determined from phenomenological fits 
to the neutrino data, even including the CHOOZ upper bound.
Our prediction of the solar mass squared 
difference is about a factor of $2$ larger than the global data fit
even though the prediction is inside of the LMA-MSW region, 
giving a contribution to our goodness of fit of \mbox{\rm g.o.f.} 
$\approx 0.14$. Our CHOOZ angle also turns out to be 
about a factor of $2$ larger than the experimental limit at 
$90\%$ C.L., delivering another contribution of \mbox{\rm g.o.f.} 
$\approx 0.14$. In summary our predictions for the neutrino sector 
agree extremely well with the data, giving a contribution of only 
0.34 to \mbox{\rm g.o.f.} while the charged fermion sector contributes 
3.29 to \mbox{\rm g.o.f.}.

\subsection{$CP$ violation}
\indent\ Since we have taken our random couplings to be -- 
whenever allowed -- \underline{complex} we have order of 
unity or essentially maximal $CP$-violation so a unitary 
triangle with angles of order one is a success of our 
model. After our fitting of masses and of mixings 
we can simply predict order of magnitudewise of $CP$-violation in 
\mbox{\it e.g.} $K^0\! - \!\bar{K^0}$ decay or in CKM and MNS 
mixing matrices in general. 

The Jarlskog area $J_{\scriptscriptstyle CP}$ provides a measure of the 
amount of $CP$ violation in the quark sector~\cite{cecilia} and, 
in the approximation of setting cosines of mixing angles to unity, 
is just twice the area of the unitarity triangle:
\begin{equation}
  \label{eq:jarkskog}
  J_{\scriptscriptstyle CP}=V_{us}\,V_{cb}\,V_{ub}\,\sin \delta \,,
\end{equation}
where $\delta$ is the $CP$ violation phase in the CKM matrix.
In our model the quark mass matrix elements have random phases, 
so we expect $\delta$ (and also the three angles $\alpha$, 
$\beta$ and $\gamma$ of the unitarity triangle) to be of 
order unity and, taking an average value of 
$|\sin\delta| \approx 1/2$, the area of the 
unitarity triangle becomes
\begin{equation}
  \label{eq:jarkskog*0.5}
  J_{\scriptscriptstyle CP}\approx \frac{1}{2}\,V_{us}\,V_{cb}\,V_{ub}\,.
\end{equation}
Using the best fit values for the CKM elements from 
Table~\ref{convbestfit}, we predict 
$J_{\scriptscriptstyle CP} \approx 3.1\times10^{-6}$ to be compared with 
the experimental value $(2-3.5)\times10^{-5}$. 
Since our result for the Jarlskog area  
is the product of four quantities, we do not expect the 
usual $\pm64\%$ logarithmic uncertainty but rather 
$\pm\sqrt{4}\cdot64\%=128\%$ logarithmic
uncertainty. This means our result deviates from the 
experimental value by 
$\ln (\frac{2.7 \times 10^{-5}}{3.1 \times 10^{-6}})/1.28$ 
= 1.7 ``standard deviations''. 

The Jarlskog area has been calculated from the best fit parameters in 
Table 2, it is also possible to calculate them directly while 
making the fit. So we have calculated $J_{\scriptscriptstyle CP}$ 
for $N=10,000$ complex order unity random 
number combinations. Then we took the logarithmic average 
of these $10,000$ samples of $J_{\scriptscriptstyle CP}$
and obtained the following result:
\begin{eqnarray}
  \label{eq:jcpabsm}
   J_{\scriptscriptstyle CP}&=& 3.1\times 10^{-6} \,,
\end{eqnarray}  
in good agreement with the values given above.

\subsection{Neutrinoless double beta decay}
\indent\ Another prediction, which can also be made from this model, is 
the electron ``effective Majorana mass'' -- the parameter in  
neutrinoless beta decay -- defined by: 
\begin{equation}
\label{eq:mmajeff}
\abs{\sVEV{m}} \equiv \abs{\sum_{i=1}^{3} U_{e i}^2 \, m_i} \,,
\end{equation}
where $m_i$ are the masses of the neutrinos $\nu_i$ 
and $U_{e i}$ are the MNS mixing matrix elements for the 
electron flavour to the mass eigenstates $i$. We can 
substitute values for the neutrino masses $m_i$ from 
eqs.~(\ref{eq:neutrinomass1}-\ref{eq:neutrinomass3}) and for the 
fitted neutrino mixing angles from Table~\ref{convbestfit} into 
the left hand side of 
eq.~(\ref{eq:mmajeff}). 
As already mentioned, the $CP$ violating phases in the MNS mixing 
matrix are essentially random in our model. So we combine the 
three terms in eq.~(\ref{eq:mmajeff}) by taking the square root of 
the sum of the modulus squared of each term, which gives
our prediction:
\begin{equation}
  \label{eq:meffresult}
  \abs{\sVEV{m}} \approx 3.1\times 10^{-3}~~\mbox{\rm eV}\,.
\end{equation}

In the same way as being calculated the Jarlskog area we
can compute using $N=10,000$ complex order unity random 
number combinations to get the $\abs{\sVEV{m}}$. Then we 
took the logarithmic average 
of these $10,000$ samples of $\abs{\sVEV{m}}$ as 
usual: 
\begin{eqnarray}
    \abs{\sVEV{m}}&=& 4.4\times 10^{-3}~~\mbox{\rm eV}\,.
\end{eqnarray}  
This result does not agree with the central value of recent 
result -- ``evidence'' -- from the
Heidelberg-Moscow collaboration~\cite{evidence}.

\section{Baryogenesis via Lepton Number Violation}
\indent\ Having now a well fitted model giving orders of magnitude 
for all the Yukawa couplings and having the see-saw mechanism,
it is obvious that we ought to calculate the amount of baryons $Y_B$
relative to entropy being produced via the Fukugita-Yanagida 
mechanism~\cite{FY}. According to this mechanism the decay of the 
right-handed 
neutrinos by $CP$-violating couplings lead to an excess of the $B-L$
charge (meaning baryon number minus lepton number), the relative excess
in the decay from Majorana neutrino generation number $i$ 
being called $\epsilon_i$. This 
excess is then immediately -- and continuously back and forth -- being
converted partially to a baryon number excess, although it starts out 
as being a lepton number $L$ asymmetry, since the right-handed 
neutrinos decay to leptons and Weinberg-Salam Higgs particles.   
It is a complicated discussion to estimate to what extend the 
$B-L$ asymmetry
is washed out later in the cosmological development, but our estimates 
goes that there is not enough baryon number excess left to fit the 
Big Bang development at the stage of formation of the light elements 
primordially (nuclearsynthesis). 

Recently we have, however, developed a modified version~\cite{NTfur} 
of our model -- only deviating in the right-handed sector -- characterized 
by changing the quantum numbers assumed for the see-saw scale producing
Higgs field $\phi_{B-L}$ in such a way that the biggest matrix elements in 
the right-handed mass matrix (eq.~\ref{Mmajo}) becomes the pair of 
-- because of the symmetry -- identical off diagonal 
elements (row, column)=(2,3) and (3,2).
Thereby we obtain two almost degenerate right-handed neutrinos and that helps
for making the $B-L$ asymmetry in the decay bigger. In this modified model 
that turns out to fit the rest of our predictions approximately equally well
or even better we then get a very satisfactory baryon number relative to 
entropy prediction
\begin{equation}
Y_B \approx  2.5 \times 10^{-11}\,. 
\end{equation}   
 In the same time as making this modification of the $\phi_{B-L}$ 
quantum numbers we also made some improvements in the calculation 
by taking into account
the running of the Dirac neutrino Yukawa couplings from the Planck scale to 
the corresponding right-handed neutrino scales. Also, we calculated 
more accurate dilution factors than previous our work~\cite{NT2}. However,
foregoing work was based on the mass matrices which predicted
the SMA-MSW, so we must investigate the baryogenesis using the
present mass matrices, of course, with the modified right-handed
Majorana mass matrix.

\section{Conclusion}
\indent\ We have reviewed our model which is able to 
predict the experimentally favored LMA-MSW 
solution rather than the SMA-MSW solution for solar neutrino 
oscillations after careful choice of the $U(1)$ charges for the
Higgs fields causing transitions between 1st and 2nd 
generations. However, the fits of charged lepton quantities
become worse compare to our ``old'' model that can predict
SMA-MSW solar neutrino solution. On the other hand, we 
now can fit the neutrino quantities very well: the price 
paid for the greatly improved neutrino mass matrix fit -- the 
neutrino parameters now contribute only very little to the 
\mbox{\rm g.o.f.} -- is a slight deterioration in the fit to the 
charged fermion 
mass matrices. In particular the predicted values of the quark masses 
$m_d$ and $m_c$ and the Cabibbo angle $V_{us}$ are reduced compared to 
our previous fits. However the overall fit agrees with the seventeen 
measured quark-lepton mass and mixing angle parameters in Table 2 
within the theoretically expected uncertainty~\cite{FF} of about 
$64\%$; it is a perfect fit order of magnitudewise. 
It should be remarked that our model provides 
an order of magnitude fit{}/{}understanding of all the effective
Yukawa couplings of the Standard Model and the neutrino 
oscillation parameters in terms of only $6$ parameters -- the Higgs 
field vacuum expectation values!

\section*{Acknowledgments}
We wish to thank T.~Asaka, W.~Buchm{\"u}ller, L.~Covi and P.~Di Bari 
for useful discussions. 
H.B.N. thanks the Alexander 
von Humboldt-Stiftung for Forschungspreis. Y.T. thanks 
the Frederikke L{\o}rup f{\o}dt Helms Mindelegat for 
a travel grant to attend the EPS HEP 2001 and the 4th Bled workshop,
DESY for financial support. 


%
\def\sVEV#1{\left\langle #1\right\rangle}
\def\abs#1{\left| #1\right|}
\def\braket#1{\left\langle #1\right\rangle}
\def\ket#1{\left| #1\right\rangle} 
\def\bra#1{\left\langle #1\right|}
\newcommand{\MeV}{\mbox{\rm MeV}}
\newcommand{\GeV}{\mbox{\rm GeV}}
\newcommand{\eV}{\mbox{\rm eV}}
\def\sleq{\raisebox{-.6ex}{${\textstyle\stackrel{<}{\sim}}$}}
\def\sgeq{\raisebox{-.6ex}{${\textstyle\stackrel{>}{\sim}}$}}
\title*{Family Replicated Calculation of Baryogenesis}
\author{H.~B.~Nielsen\thanks{E-mail: hbech@mail.desy.de} 
and Y.~Takanishi\thanks{E-mail: yasutaka@mail.desy.de}}
\institute{%
Deutsches Elektronen-Synchrotron DESY,
Notkestra{\ss}e 85,
D-22603 Hamburg, Germany\\
and
The Niels Bohr Institute, 
Blegdamsvej 17, Copenhagen {\O},  Denmark}

\titlerunning{Family Replicated Calculation of Baryogenesis}
\authorrunning{H.~B.~Nielsen and Y.~Takanishi}
\maketitle

\begin{abstract}
In our model with a Standard Model gauge group extended with 
a baryon number minus lepton number charge {\em for each
family of quarks and leptons}, we calculate the baryon number 
relative to entropy produced in early Big Bang by 
the Fukugita-Yanagida mechanism. With the parameters, 
$\hbox{\it i.e.}{}$, the Higgs 
VEVs already fitted in a very successful way to quark and lepton 
masses and mixing angles we obtain the {\em order of magnitude}
pure prediction $Y_B=2.59{+17.0\atop-2.25}\times 10^{-11}$
which according to a theoretical estimate should mean in this case 
an uncertainty of the order of a factor 7 up or down (to be 
compared to $Y_B=(1.7-8.1)\times 10^{-11}$) using a relatively crude 
approximation for the dilution factor, while using another 
estimate based 
on Buchm{\"u}ller and Pl{\"u}macher a factor $500$ less, but this 
should rather be considered a lower limit. With a realistic 
uncertainty due to wash-out of a factor $100$ up or down we 
even with the low estimate only deviate by $1.5\sigma$.
\end{abstract}

\section{Introduction}
Using the model for mass matrices presented by us in an other 
contribution~\cite{Dub} at this conference we want 
to compute the amount of baryons 
produced in the early universe. This model works by having the 
mass matrix elements being suppressed by approximately 
conserved quantum numbers from a gauge group repeated for each family 
of quarks and leptons and also having a $(B-L)$
charge for each family.

The baryon number density relative to entropy density, $Y_B$, is one 
of the rather few quantities that can give us information about the 
laws of nature beyond the Standard Model and luckily we have from 
the understanding of the production of light isotopes at 
the minute scale in Big Bang fits to this quantity~\cite{KT}. The 
``experimental'' data of the ratio of 
baryon number density to the entropy density is 
\begin{equation}
  \label{eq:YBexpnor}
  Y_B~\Big|_{\rm exp}=\left(1.7-8.1\right)\times10^{-11}\hspace{2mm}.
\end{equation}
We already had a good fit of all the masses and mixings~\cite{NT1,FNT} 
for both 
quarks and leptons measured so far and agreeing with all the bounds 
such as neutrinoless beta decay and proton decay not being seen and
matching on the borderline but consistent with the accuracy of our model 
and of the experiment of CHOOZ the electron to heaviest left-handed neutrino
mixing, and that in a version of our model in which the dominant matrix 
element in the right-handed neutrino mass matrix is the diagonal one
for the ``third'' ($\hbox{\it i.e.}{}$ with same $(B-L)_i$ as the third family) 
family $\nu_{R 3}$ right-handed neutrino.
This version of our model which fits otherwise very well does 
not give sufficient 
$(B-L)$ excess, that survives, but the by now the best model in our 
series should have the right-handed mass matrix dominated by the 
off-diagonal elements $(2,3)$ and $(3,2)$, so that there appears two
almost mass degenerate see-saw neutrinos, in addition to the 
third one (first family) which is much lighter.  
  
\section{Mass matrices and results for masses and mixing angles}
Our model produces mass matrix elements -- or 
effective Yukawa couplings  -- which are suppressed from being of the 
order of the top-mass because they are forbidden by the conservation 
of the gauge charges of our model and can only become different from 
zero using the $6$ Higgs fields~\cite{FNT,NT2} which we have in addition to 
the field replacing the Weinberg-Salam one. In the neutrino sector
according to the see-saw mechanism~\cite{hbnseesaw} we have to 
calculate Dirac- and Majorana-mass matrices,
$M_{\rm eff} \! \approx \! M^D_\nu\,M_R^{-1}\,(M^D_\nu)^T$, to 
obtain the effective mass matrix $M_{\rm eff}$ for the 
left handed neutrinos we in practice can ``see''.
Here we present all mass matrices as they follow from our choice 
of quantum numbers for the $7$ Higgs fields in our model and for the 
quarks and leptons (as they can be found in the other contribution).
Only the quantum numbers for the field called $\phi_{B-L}$ 
is -- in order to get degenerate see-saw neutrinos -- changed into
having the $B-L$ quantum numbers of family $2$ and $3$ 
equal to $1$, $\hbox{\it i.e.}{}$, $(B-L)_2 = (B-L)_3 =1$, while the other family quantum 
numbers are just zero:

\noindent
the up-type quarks:
\begin{eqnarray}
M_{\scriptscriptstyle U} \simeq \frac{\sVEV{(\phi_{\scriptscriptstyle\rm WS})^\dagger}}{\sqrt{2}}
\hspace{-0.1cm}
\left(\!\begin{array}{ccc}
        (\omega^\dagger)^3 W^\dagger T^2
        & \omega \rho^\dagger W^\dagger T^2
        & \omega \rho^\dagger (W^\dagger)^2 T\\
        (\omega^\dagger)^4 \rho W^\dagger T^2
        &  W^\dagger T^2
        & (W^\dagger)^2 T\\
        (\omega^\dagger)^4 \rho
        & 1
        & W^\dagger T^\dagger
\end{array} \!\right)\label{hbnM_U}
\end{eqnarray}  
\noindent
the down-type quarks:
\begin{eqnarray}
M_{\scriptscriptstyle D} \simeq \frac{\sVEV{\phi_{\scriptscriptstyle\rm WS}}}
{\sqrt{2}}\hspace{-0.1cm}
\left (\!\begin{array}{ccc}
        \omega^3 W (T^\dagger)^2
      & \omega \rho^\dagger W (T^\dagger)^2
      & \omega \rho^\dagger T^3 \\
        \omega^2 \rho W (T^\dagger)^2
      & W (T^\dagger)^2
      & T^3 \\
        \omega^2 \rho W^2 (T^\dagger)^4
      & W^2 (T^\dagger)^4
      & W T
    \end{array} \!\right) \label{hbnM_D}
\end{eqnarray}
\noindent %
the charged leptons:
\begin{eqnarray}        
M_{\scriptscriptstyle E} \simeq \frac{\sVEV{\phi_{\scriptscriptstyle\rm WS}}}
{\sqrt{2}}\hspace{-0.1cm}
\left(\hspace{-0.1 cm}\begin{array}{ccc}
    \omega^3 W (T^\dagger)^2
  & (\omega^\dagger)^3 \rho^3 W (T^\dagger)^2 
  & (\omega^\dagger)^3 \rho^3 W T^4 \chi \\
    \omega^6 (\rho^\dagger)^3  W (T^\dagger)^2 
  &   W (T^\dagger)^2 
  &  W T^4 \chi\\
    \omega^6 (\rho^\dagger)^3  (W^\dagger)^2 T^4 
  & (W^\dagger)^2 T^4
  & WT
\end{array} \hspace{-0.1cm}\right) \label{hbnM_E}
\end{eqnarray}
\noindent
the Dirac neutrinos:
\begin{eqnarray}
M^D_\nu \simeq \frac{\sVEV{(\phi_{\scriptscriptstyle\rm WS})^\dagger}}{\sqrt{2}}
\hspace{-0.1cm}
\left(\hspace{-0.1cm}\begin{array}{ccc}
        (\omega^\dagger)^3 W^\dagger T^2
        & (\omega^\dagger)^3 \rho^3 W^\dagger T^2
        & (\omega^\dagger)^3 \rho^3 W^\dagger  T^2 \chi\\
        (\rho^\dagger)^3 W^\dagger T^2
        &  W^\dagger T^2
        & W^\dagger T^2 \chi\\
        (\rho^\dagger)^3 W^\dagger T^\dagger \chi^\dagger
        &  W^\dagger T^\dagger \chi^\dagger
        & W^\dagger T^\dagger
\end{array} \hspace{-0.1 cm}\right)\label{hbnMdirac}
\end{eqnarray} 
\noindent %
and the Majorana (right-handed) neutrinos:
\begin{eqnarray}    
M_R \simeq \sVEV{\phi_{\scriptscriptstyle\rm B-L}}\hspace{-0.1cm}
\left (\hspace{-0.1 cm}\begin{array}{ccc}
(\rho^\dagger)^6 \chi^\dagger
& (\rho^\dagger)^3 \chi^\dagger /2
& (\rho^\dagger)^3/2  \\
(\rho^\dagger)^3 \chi^\dagger /2
& \chi^\dagger & 1 \\
(\rho^\dagger)^3/2 & 1 & \chi
\end{array} \hspace{-0.1 cm}\right ) \label{Majorana}
\end{eqnarray}       
We shall remember that it is here understood that all the matrix 
elements are to be provided with order of unity factors which we
do not know and in practice have treated by 
inserting random order of unity factors over which we then average at the 
end (in a logarithmic way).

\section{Renormalisation group equations}\label{RGE}
The model for the Yukawa couplings we use gives, in principle, these couplings 
at the fundamental scale, taken to be the Planck scale, at first, and we then 
use the renormalisation group to run them down to the scales where they are
to be confronted with experiment. From the Planck scale down to 
the see-saw scale or rather from 
where our gauge group is broken down to $SMG\times U(1)_{B-L}$ we use
the one-loop renormalisation group running of the Yukawa coupling constant 
matrices and the gauge couplings~\cite{hbnNT1} in GUT notation 
including the running of Dirac neutrino Yukawa coupling:
\begin{eqnarray*}
16 \pi^2 {d g_{1}\over d  t} &\!=\!& \frac{41}{10} \, g_1^3 \hspace{2mm},\hspace{2mm}\hspace{2mm}\hspace{2mm} 16 \pi^2 {d g_{2}\over d  t} \!\!=\!\! - \frac{19}{16} \, g_2^3 \hspace{2mm}, \hspace{2mm}\hspace{2mm}\hspace{2mm} 16 \pi^2 {d g_{3}\over d  t} \!\!=\!\! - 7 \, g_3^3  \hspace{2mm},\\
16\pi^2{d Y_{\scriptscriptstyle U}\over d t}&\!=\!&\frac{3}{2}\left(Y_{\scriptscriptstyle U}(Y_{\scriptscriptstyle U})^\dagger\!-\!Y_{\scriptscriptstyle D}(Y_{\scriptscriptstyle D})^\dagger\right)Y_{\scriptscriptstyle U}\!+\!\left\{Y_{\scriptscriptstyle S}\!-\!\left(\frac{17}{20}g_1^2+\frac{9}{4}g_2^2+8g_3^2\right)\right\}Y_{\scriptscriptstyle U} ,\\
16\pi^2{d Y_{\scriptscriptstyle D}\over d t}&\!=\!&\frac{3}{2}\left(Y_{\scriptscriptstyle D}(Y_{\scriptscriptstyle D})^\dagger\!-\!Y_{\scriptscriptstyle U}(Y_{\scriptscriptstyle U})^\dagger\right)Y_{\scriptscriptstyle D}\!+\!\left\{Y_{\scriptscriptstyle S}\!-\!\left(\frac{1}{4}g_1^2+\frac{9}{4}g_2^2+8g_3^2\right)\right\}Y_{\scriptscriptstyle D} ,\\
16 \pi^2 {d Y_{\scriptscriptstyle E}\over d  t} &\!=\!& \frac{3}{2}\, 
\left( Y_{\scriptscriptstyle E} (Y_{\scriptscriptstyle E})^\dagger
-  Y_{\scriptscriptstyle \nu} (Y_{\scriptscriptstyle \nu})^\dagger\right)\,Y_{\scriptscriptstyle E} 
+ \left\{\, Y_{\scriptscriptstyle S} - \left(\frac{9}{4} g_1^2 
+ \frac{9}{4} g_2^2 \right) \right\}\, Y_{\scriptscriptstyle E} \hspace{2mm},\\
16 \pi^2 {d Y_{\scriptscriptstyle \nu}\over d  t} &\!=\!& \frac{3}{2}\, 
\left( Y_{\scriptscriptstyle \nu} (Y_{\scriptscriptstyle \nu})^\dagger -  
Y_{\scriptscriptstyle E} (Y_{\scriptscriptstyle E})^\dagger\right)\,Y_{\scriptscriptstyle \nu} 
+ \left\{\, Y_{\scriptscriptstyle S} - \left(\frac{9}{20} g_1^2 
+ \frac{9}{4} g_2^2 \right) \right\}\, Y_{\scriptscriptstyle \nu} \hspace{2mm},\\ \label{YScon} 
Y_{\scriptscriptstyle S} &\!=\!& {{\rm Tr}{}}(\, 3\, Y_{\scriptscriptstyle U}^\dagger\, Y_{\scriptscriptstyle U} 
+  3\, Y_{\scriptscriptstyle D}^\dagger \,Y_{\scriptscriptstyle D} +  Y_{\scriptscriptstyle E}^\dagger\, 
Y_{\scriptscriptstyle E} +  Y_{\scriptscriptstyle \nu}^\dagger\, Y_{\scriptscriptstyle \nu}\,) \hspace{2mm},
\end{eqnarray*}
where $t=\ln\mu$ and $\mu$ is the renormalisation point. 

In order to run the renormalisation group
equations down to $1~\mbox{\rm GeV}$, we use the following initial values:
\begin{eqnarray}
U(1):\quad & g_1(M_Z) = 0.462 \hspace{2mm},\quad & g_1(M_{\rm Planck}) = 0.614\,, \\
SU(2):\quad & g_2(M_Z) = 0.651 \hspace{2mm},\quad & g_2(M_{\rm Planck}) = 0.504\,, \\
SU(3):\quad & g_3(M_Z) = 1.22  \hspace{2mm},\quad & g_3(M_{\rm Planck}) = 0.491\,.
\end{eqnarray}
We varied the $6$ free 
parameters and found the best fit, corresponding to the lowest value 
for the quantity $\mbox{\rm g.o.f.}\equiv\sum \left[\ln \left(
\frac{\sVEV{m}_{\rm pred}}{m_{\rm exp}} \right) \right]^2=3.38$, 
with the following values for the VEVs:
\begin{eqnarray} 
\label{eq:hbnVEVS} 
&&\sVEV{\phi_{\scriptscriptstyle WS}}= 246~\mbox{\rm GeV}\hspace{2mm},  
\hspace{2mm}\sVEV{\phi_{\scriptscriptstyle B-L}}=1.23\times10^{10}~\mbox{\rm GeV}\hspace{2mm}, 
\hspace{2mm}\sVEV{\omega}=0.245\hspace{2mm},\nonumber\\
&&\hspace{2mm}\sVEV{\rho}=0.256\hspace{2mm},\hspace{2mm}\sVEV{W}=0.143\hspace{2mm},
\hspace{2mm}\sVEV{T}=0.0742\hspace{2mm},\hspace{2mm}\sVEV{\chi}=0.0408\hspace{2mm},
\end{eqnarray}
where, except for the Weinberg-Salam Higgs field and 
$\sVEV{\phi_{\scriptscriptstyle B-L}}$, the VEVs are expressed in Planck units. 
Hereby we have considered that the Weinberg-Salam Higgs field VEV is 
already fixed by the Fermi constant.
The results of the best fit, with the VEVs in eq.~(\ref{eq:hbnVEVS}), 
are shown in Table~\ref{hbnconvbestfit}.
\begin{table}[!b]
\begin{displaymath}
\begin{array}{|c|c|c|}
\hline\hline
 & {\rm Fitted} & {\rm Experimental} \\ \hline
m_u & 5.2~\MeV & 4~\MeV \\
m_d & 5.0~\MeV & 9~\MeV \\
m_e & 1.1~\MeV & 0.5~\MeV \\
m_c & 0.70~\GeV & 1.4~\GeV \\
m_s & 340~\MeV & 200~\MeV \\
m_{\mu} & 81~\MeV & 105~\MeV \\
M_t & 208~\GeV & 180~\GeV \\
m_b & 7.4~\GeV & 6.3~\GeV \\
m_{\tau} & 1.11~\GeV & 1.78~\GeV \\
V_{us} & 0.10 & 0.22 \\
V_{cb} & 0.024 & 0.041 \\
V_{ub} & 0.0025 & 0.0035 \\ \hline
\Delta m^2_{\odot} & 9.0 \times 10^{-5}~\eV^2 &  4.5 \times 10^{-5}~\eV^2 \\
\Delta m^2_{\rm atm} & 1.8 \times 10^{-3}~\eV^2 &  3.0 \times 10^{-3}~\eV^2\\
\tan^2\theta_{\odot} &0.23 & 0.35\\
\tan^2\theta_{\rm atm}& 0.83 & 1.0\\
\tan^2\theta_{\rm chooz}  & 3.3 \times 10^{-2} & \sleq~2.6 \times 10^{-2}\\
\hline\hline
\mbox{\rm g.o.f.} &  3.38 & - \\
\hline\hline
\end{array}
\end{displaymath}
\caption{Best fit to conventional experimental data.
All masses are running
masses at $1~\mbox{\rm GeV}$ except the top quark mass which is the pole mass.
Note that we use the square roots of the neutrino data in this 
Table, as the fitted neutrino mass and mixing parameters 
$\sVEV{m}$, in our goodness of fit ($\mbox{\rm g.o.f.}$) definition.}
\label{hbnconvbestfit}
\end{table}

\section{Quantities to use for baryogenesis calculation}
Since the baryogenesis in the Fukugita-Yanagida scheme~\cite{hbnFY} arises from
a negative excess of lepton number being converted by Sphalerons to
a positive baryon number excess partly and this negative excess comes
from the $CP$ violating decay of the see-saw neutrinos we shall 
introduce the parameters $\epsilon_i$ giving the measure of the relative 
asymmetry under $C$ or $CP$ in the decay of neutrino number $i$: 
\noindent\ Defining the measure $\epsilon_i$ for the $CP$ violation 
\begin{equation}
  \label{eq:epsilonCP}
 \epsilon_i \equiv\frac{\sum_{\alpha,\beta}\Gamma(N_{{\scriptscriptstyle R}\, i} \to \ell^\alpha\phi_{\scriptscriptstyle WS}^\beta)-\sum_{\alpha,\beta}\Gamma(N_{{\scriptscriptstyle R}\, i}\to \bar\ell^\alpha \phi_{\scriptscriptstyle WS}^{\beta \dagger})}{\sum_{\alpha,\beta}\Gamma(N_{{\scriptscriptstyle R}\, i}
\to \ell^\alpha\phi_{\scriptscriptstyle WS}^\beta) + \sum_{\alpha,\beta}\Gamma(N_{{\scriptscriptstyle R}\, i}\to\bar\ell^\alpha \phi_{\scriptscriptstyle WS}^{\beta \dagger})}\hspace{2mm}, 
\end{equation}
where $\Gamma$ are $N_{{\scriptscriptstyle R}\, i}$ decay rates (in the $N_{{\scriptscriptstyle R}\, i}$ 
rest frame), summed over the
neutral and charged leptons (and Weinberg-Salam Higgs fields) 
which appear as final states in the $N_{{\scriptscriptstyle R}\, i}$ decays 
one sees that the 
excess of leptons over anti-leptons produced in the decay
of one $N_{{\scriptscriptstyle R}\, i}$ is just $\epsilon_i$. The total 
decay rate at the tree level is given by
\begin{equation}
  \label{eq:LOCP}
  \Gamma_{N_i}=\Gamma_{N_i\ell}+\Gamma_{N_i\bar\ell}
  ={((\widetilde{M_\nu^D})^\dagger \widetilde{M_\nu^D)}_{ii}\over 
    4\pi \sVEV{\phi_{\scriptscriptstyle WS}}^2}\,M_i \hspace{2mm},
\end{equation}%
where $\widetilde{M_\nu^D}$ can be expressed through the 
unitary matrix diagonalising the right-handed neutrino 
mass matrix $V_R$:
\begin{eqnarray}
  \label{eq:tildemd}
 \widetilde{M_\nu^D} \!&\equiv&\! M_\nu^D\,V_R \hspace{2mm},\\
V_R^\dagger \,M_R\,M_R^\dagger\, V_R \!&=&\! 
{\rm diag} \left(\,M^2_1, M^2_2, M^2_3\,\right) \hspace{2mm}.
\end{eqnarray}
The $CP$ violation rateis computed according to~\cite{CRV,BuPlu}
\begin{equation}
\label{eq:CPepsilon}
\epsilon_i = \frac{\sum_{j\not= i} {\rm Im}[((\widetilde{M_\nu^D})^{\dagger} \widetilde{M_\nu^D})^2_{ji}] \left[\, f \left( \frac{M_j^2}{M_i^2} \right) + g \left( \frac{M_j^2}{M_i^2} \right)\,\right]}{4 \pi \sVEV{\phi_{\scriptscriptstyle WS}}^2 ((\widetilde{M_\nu^D})^{\dagger}\widetilde{M_\nu^D})_{ii}}
\end{equation}
where the function, $f(x)$, comes from the one-loop vertex contribution and
the other function, $g(x)$, comes from the self-energy contribution.
These $\epsilon$'s can be calculated in perturbation theory 
only for differences between Majorana neutrino masses which 
are sufficiently large compare to their decay widths, $\hbox{\it i.e.}{}$, the 
mass splittings satisfy the condition, 
$\abs{M_i-M_j}\gg\abs{\Gamma_i-\Gamma_j}$:
\begin{equation}
f(x)=\sqrt{x} \left[1-(1+x) \ln \frac{1+x}{x}\right]\hspace{2mm}, 
\hspace{2mm} \hspace{2mm} g(x)=\frac{\sqrt{x}}{1-x} \hspace{2mm}.
\end{equation}
We as usual~\cite{KT} introduce the dacay rate relative to 
\begin{equation}
  \label{eq:Kdrei}
K_i\equiv\frac{\Gamma_i}{2 H} \,\Big|_{ T=M_{i} } = \frac{M_{\rm
Planck}}{1.66 \sVEV{\phi_{\scriptscriptstyle WS}}^2  8 \pi 
g_{*\,i}^{1/2}}\frac{((\widetilde{M_\nu^D})^{\dagger} 
\widetilde{M_\nu^D})_{ii}}{M_{i}} \qquad
(i=1, 2, 3)\hspace{2mm}, \end{equation}%
where $\Gamma_i$ is the width of the flavour $i$ Majorana neutrino,
$M_i$ is its mass and $g_{*\,i}$ is the number of degrees of freedom
at the temperature $M_i$ (in our model $\sim100$).

In order to estimate the effective $K$ factors we first 
introduce some normalized state vectors for the decay products:
\begin{eqnarray*}
\ket{ i }\equiv\!\!&&\left(\sum_{k=1}^{3} 
\abs{\,\left[\widetilde{M_\nu^D}(M_i)\right]_{k i}}^2\right)^{-\frac{1}{2}}\nonumber\\
&&\times\left(\,\left[ \widetilde{M_\nu^D}(M_i)\right]_{1 i} 
\,, \left[\widetilde{M_\nu^D}(M_i)\right]_{2 i}
\,, \left[\widetilde{M_\nu^D}(M_i)\right]_{3 i}\,\right)\nonumber\hspace{2mm},
\end{eqnarray*}

\noindent
Then we may take an approximation for the effective $K$ factors:
\begin{eqnarray}
\label{eq:keff1}
{K_{\scriptscriptstyle\rm eff}}_1 &=& K_1(M_1)\hspace{2mm}, \\
\label{eq:keff2}
{K_{\scriptscriptstyle\rm eff}}_2 &=& K_2(M_2) + \abs{\braket{2 | 3}}^2 \,K_3(M_3) 
+ \abs{\braket{2 | 1}}^2\,K_1(M_1)\hspace{2mm},\\ 
\label{eq:keff3}
{K_{\scriptscriptstyle\rm eff}}_3 &=& K_3(M_3) + \abs{\braket{3 | 2}}^2\,K_2(M_2) + 
\abs{\braket{3 | 1}}^2\,K_1(M_1) \hspace{2mm}.
\end{eqnarray}

\section{Result for baryogenesis}
\indent\ Using  the Yukawa couplings -- as coming from the 
VEVs of our seven different
Higgs fields --  the numerical calculation of baryogenesis were 
performed using our random order unity factor method. In order 
to get baryogenesis in Fukugita-Yanagida scheme, we calculated the 
see-saw neutrino masses, ${K_{\scriptscriptstyle\rm eff}}_i$ 
factors and $CP$ violation parameters 
using $N=10,000$ random number 
combinations and logarithmic average method:
\begin{eqnarray*}
\label{eq:Metc} 
\begin{array}{l l l}
  M_1= 2.1 \times 10^{5}~\mbox{\rm GeV}\hspace{2mm}, & {K_{\scriptscriptstyle\rm eff}}_1= 31.6 \hspace{2mm}, & \abs{\epsilon_1} = 4.62\times 10^{-12}\hspace{2mm}\\*[0.2cm]
  M_2= 8.8 \times 10^{9}~\mbox{\rm GeV}\hspace{2mm}, & {K_{\scriptscriptstyle\rm eff}}_2 = 116.2\hspace{2mm},& 
\abs{\epsilon_2} = 4.00\times 10^{-6}\hspace{2mm} \\*[0.2cm]
  M_3= 9.9 \times 10^{9}~\mbox{\rm GeV}\hspace{2mm}, & {K_{\scriptscriptstyle\rm eff}}_3= 114.7\hspace{2mm}, & \abs{\epsilon_3} = 3.27\times 10^{-6}\hspace{2mm}
\end{array}
\end{eqnarray*}
The sign of $\epsilon_i$ is unpredictable due to the complex 
random number coefficients in mass matrices, therefore we are 
not in the position to say the sign of $\epsilon$'s. 
 Using the complex order unity random 
numbers being given by a Gaussian distribution we get after logarithmic
averaging using the dilution factors as presented by~\cite{KT,NT1}
\begin{equation}
  \label{eq:YB}
  Y_B = 2.59{+17.0\atop-2.25}\times 10^{-11} \hspace{2mm},
\end{equation}
where we have estimated the uncertainty  
in the natural exponent according to the ref.~\cite{hbnFF} 
to be $64~\%\cdot\sqrt{10}\approx 200~\%$. 
 
The understanding of how this baryon to entropy prediction $Y_B$ comes
about in the model may be seen from the following (analytical) estimate
\begin{equation}
Y_B \approx \frac{1}{3}\cdot \frac{\chi}{\sqrt{g_{*}} \,T^2}\cdot\frac{M_3}{M_{\rm Planck} } \approx \frac{1}{3} \cdot 10^{-9} 
\label{analyticba}
\end{equation}
where we left out for simplicity the $\ln K$ factor in the denominator of the 
dilution factor $\kappa$ and where $M_3$ is the mass of one of the 
heavy right-handed neutrinos 
in our model $M_3\approx \sVEV{\phi_{B-L}}$. Since the atmospheric mass 
square difference square root $\sqrt{\Delta m^2_{\rm atm}}
\approx 0.05~\mbox{\rm eV} \approx \sVEV{\phi_{WS}}^2 (WT)^2/M_3$ 
we see that keeping it leaves us with the dependence 
\begin{equation}
Y_B \approx\frac{\sVEV{\phi_{WS}}^2 \chi}{3 \sqrt{0.05~\mbox{\rm eV} 
\cdot \,g_* \,M_{\rm Planck}\, W^2 T^4}}\approx \frac{1}{5}\times 10^{-4} 
\cdot \frac{\chi}{\sqrt{g_*} \,W^2 T^4} 
\end{equation}

\section{Problem with wash-out effects?}
To make a better estimate of the wash-out effect we may make 
use of the calculations by~\cite{BP} by putting effective values
for the see-saw neutrino mass $M$ and $\widetilde{m}$. The most important 
wash-out is due to ``on-shell'' formation of right-handed neutrinos 
and only depends on $K$ or the thereto proportional $\widetilde{m}$, but 
next there are wash-out effects going rather than by $K$ or $\widetilde{m}$ as 
$M \widetilde{m}^2$. In the presentation of the results by \cite{BP} fixed 
ratios between right-handed neutrino masses were assumed. However, in reality
a very important wash-out comes form the off-shell inverse decay and that 
goes as 
\begin{equation}
  \label{eq:til}
  M_1 \sum_j \frac{M_j^2}{M_1^2}\,\widetilde{m}_j^2 \hspace{2mm}\hspace{2mm} {\rm with} \hspace{2mm}\hspace{2mm}
\widetilde{m}_j \equiv
\frac{[(\widetilde{M_\nu^D})^\dagger
\widetilde{M_\nu^D}]_{jj}}{M_j}
\end{equation}
Here we use the notation with $\widetilde{m}_j$ from~\cite{BP}:
$ \widetilde{m}_j \approx K_j \cdot 2.2 \cdot 10^{-3} \mbox{\rm eV}$.

Using such a term (see eq.~\ref{eq:til}) with the ansatz 
ratios used in~\cite{BP},
$M_3^2 =10^{6}~M_1^2$ and $M_2^2 = 10^3~M_1^2$ one gets for 
eq.~(\ref{eq:til})
$\approx 10^6~M_1 ~\widetilde{m}_3^2$, while we would with our mass ratios 
(eq.~\ref{eq:Metc}) $M_3^2 \approx 1/4 \cdot 10^{10} ~M_1^2$ and 
$M_2^2 \approx 1/4 \cdot 10^{10} ~M_1^2$ obtain 
correspondingly $2\cdot 10^5~\mbox{\rm GeV} \cdot 1/4 \cdot 10^{10}
~\widetilde{m}_3^2 \approx 1/2 \cdot 10^{15}~\mbox{\rm GeV}~\widetilde{m}_3^2$, which 
then being identified with $ 10^6~M_{1 \,{\rm use}}~\widetilde{m}_3^2$  
would lead to that we should effectively use 
for simulating our model the mass of the right handed neutrino -- which
is a parameter in the presentation of the dilution effects in~\cite{BP} --
$M_{1\,{\rm use}} = 1/2 \cdot 10^{15}~\mbox{\rm GeV}/10^6 = 1/2 \cdot 
10^9~\mbox{\rm GeV}$.
Inserting this $M_{1\,{\rm use}}$ value for our estimate  $\widetilde{m}_2 
\approx \widetilde{m}_3 \approx 0.1~\mbox{\rm eV}$ gives a dilution 
factor $\kappa\approx 10^{-4}$, $\hbox{\it i.e.}{}$, a factor $500$ less 
than  what we used with our 
estimate using the $K_{{\rm eff}}$'s.
(Our $\widetilde{m}_3 =\widetilde{m}_2$ are surprisingly large compared to
the $\sqrt{\Delta m^2_{\rm atm}}$ because of renormalzation running .)
Using the better calculation of ~\cite{BP} which has a very steep dependence 
-- a fourth power say -- as function of $\widetilde{m}$ our uncertainty 
should also be corrected to a factor 100 up or down. So then we have 
one and a half standard deviations of getting too little baryon number. 

\section{Conclusion}
We calculated the baryon density relative to the entropy density 
-- baryogenesis -- from our model order of magnitudewise. This model 
already fits to quark and lepton masses and mixing 
angles using {\em only 
six parameters}, vacuum expectation values. We got a result for the 
baryon number
predicting about a factor only three less than the fitting to microwave 
background fluctuations obtained by 
Buchm{\"u}ller~$\hbox{\it et al.}$~\cite{BB}, when 
we used our crude  $K_{\rm eff}$'s approximation. However, using 
the estimate extracted from the calculations of ~\cite{BP} we got
three orders of magnitude too low prediction of the baryon number. 
This estimate must though be considered a possibly too low estimate 
because there is one scattering effect that is strongly suppressed with
our masses but which were included in that calculation. But even 
the latter estimate should because of the steep dependence of the result
on the parameters be considered more uncertain and considering the 
deviation of our prediction only $1.56\sigma$ is not unreasonable.

Since we used the Fukugita-Yanagida mechanism of obtaining first 
a lepton number excess being converted (successively by Sphalerons) 
into the baryon number, our success in this prediction should be 
considered not only a victory for our model for mass matrices 
but also for this mechanism. Since our model would be hard to 
combine with supersymmetry -- it would loose much of its 
predictive power by having to double the Higgs fields -- 
we should consider it in a \underline{non} SUSY scenario and 
thus we can without problems take the energy scale to 
inflation/reheating to be so high that the plasma had 
already had time to go roughly to thermal equilibrium 
before the right-handed neutrinos go out-of-equilibrium
due to their masses. We namely simply have no problem 
with getting too many gravitinos because gravitinos do 
not exist at all in our scheme.

Another ``unusual'' feature of our model is that the 
dominant contribution to the baryogenesis comes from 
the \underline{heavier}
right-handed neutrinos. In our model 
it could be arranged without any troubles that the two 
heaviest right-handed neutrinos have masses only deviating 
by $10\%$ namely given by our VEV parameters $\chi$. This 
leads to significant enhancement of the $\epsilon_2$ and 
$\epsilon_3$ which is crucial for the success of our prediction.
There is namely a significant wash-our taking place, 
by a factor of the order of 
$\kappa=10^{-3}~\hbox{\rm to}~10^{-6} $. It is remarkable that we have 
here worked with a model that order of magnitudewise has with only
six adjustable parameters been able to fit all the masses and mixings 
angles for quarks and leptons measured so far, including the 
Jarlskog $CP$ violation area and most importantly and 
interestingly the baryogenesis in the early Universe. To 
confirm further our model we are in strong need for further 
data -- which is not already predicted by the Standard 
Model, or we would have to improve it to give in principle 
accurate results rather than only orders of magnitudes. The 
latter would, however, be against the hall mark of our model, 
which precisely makes use of that we can guess that the huge 
amount of unknown coupling constants in our scheme with lots 
of particles can be counted as being {\em of order unity}.

\section*{Acknowledgments}
We wish to thank W.~Buchm{\"u}ller, P.~Di Bari and 
M.~Hirsch for useful discussions. We thank the Alexander 
von Humbold-Stiftung and DESY for financial support. 


%
\newcommand{\vp}{\varphi}
\newcommand{\vt}{\vartheta}
\newcommand{\lan}{\langle}
\newcommand{\ran}{\rangle}

\def\la{\mathrel{\mathpalette\fun <}}
\def\ga{\mathrel{\mathpalette\fun >}}
\def\fun#1#2{\lower3.6pt\vbox{\baselineskip0pt\lineskip.9pt
\ialign{$\mathsurround=0pt#1\hfil##\hfil$\crcr#2\crcr\sim\crcr}}}

\title*{Neutrino Oscillations in Vacuum on the Large Distance: Influence of the Leptonic CP-phase.}
\author{%
D.A. Ryzhikh\thanks{ryzhikh@heron.itep.ru} and 
K.A. Ter-Martirosyan\thanks{termarti@heron.itep.ru}}
\institute{%
ITEP, Moscow, 117259, B.Cheremushkinskaya, 25} 

\authorrunning{D.A. Ryzhikh and K.A. Ter-Martirosyan}
\titlerunning{Neutrino Oscillations in Vacuum on the Large Distance}
\maketitle

\begin{abstract}
Vacuum neutrino oscillations for three generations are considered. The influence of the leptonic CP-violating phase
(similar to the quarks CP-phase) on neutrino oscillations is taken into account in the matrix of leptons mixing. The
dependence of probabilities of a transition of one kind neutrino to another kind on three mixing angles and on the
CP-phase is obtained in a general form. It is pointed that one can reconstruct the value of the leptonic CP-phase
by measuring probabilities for a transition of one kind neutrino to another kind averaging over all oscillations.
Also it is noted that the manifestation of the CP-phase in deviations of probabilities of forward neutrino
transitions from probabilities of backward neutrino transitions is an effect practically slipping from an
observation.
\end{abstract}

\section{Introduction.}

It is unclear up to now in spite of great number of papers devoted to the investigation of neutrino
oscillations,
what is the real precision of experimental values of three mixing angles and masses of neutrino from different %
generations? And consequently do neutrino really
oscillate? The central values of these angles and values of
errors obtained in different papers and given in our paper
change from author to author and from paper to paper.
Therefore these data are very suspicious. Below in this paper we gave the
value of this precision approximately because it is defined
very roughly. Nevertheless the investigations of neutrino
look rather encouraging since the set of large perspective
devices (K2K in Japan~\cite{SK2,SK4}, CERN-GRAND Sacco
(CNGS)~\cite{CHOOZ} in Europe and Fermilab-Soudan in USA)
and some small but also perspective devices in another
regions of the Earth began to work recently or will begin
to work in the near future. In particular, the precision of
defining of the values of $\nu_\tau$ and $\nu_\mu$ masses and
also the values of sines of the neutrino mixing angles will be
appreciably improved in the nearest future (in one or two
years). Furthermore, we believe that attempts to obtain the
value of the CP-phase from experimental data will be made in
the future in spite of apparent present-day hopelessness.

The present work is devoted to neutrino oscillations and, in particular, to a possible manifestation of the
leptonic CP-phase in neutrino oscillations. At the second section of this paper we consider the standard theory of the neutrino
oscillations with regard for the leptonic CP-violating phase. Then we give the formulas for the probabilities of the
conservation of the neutrino kind and for the probabilities of the neutrino transition to another neutrino kind with some
examples of the possible manifestation of the leptonic CP-phase using modern experimental data. And then we
investigate the difference between the $\nu_\alpha\to\nu_\beta$ transitions probability and the $\nu_\beta\to\nu_\alpha$ transitions
probability and the possible influence of the leptonic CP-phase on this difference.

\section{Standard theory of neutrino oscillations with regard for leptonic CP-violating phase.}
In this section we describe the standard theory of neutrino oscillations including the leptonic CP-phase.
So, neutrino $(\nu_e)_L,(\nu_\mu)_L,(\nu_\tau)_L$ which were born in the decay reactions or in collisions do not have
definite masses. They are superpositions of neutrino states $\nu_1,\nu_2,\nu_3$ with definite masses, and their
wave functions are:
\begin{equation}
\label{nu} \nu_\beta(\vec x,t)= \sum_{k=1}^3(\hat V_{ÌÍÑ}^l)_{\beta k}\nu_k(\vec x,t),\quad \beta=e,\mu,\tau; \quad
k=1,2,3. %
\end{equation} %
Here it is supposed that $\nu_k=(\nu_1,\nu_2,\nu_3)$ are the wave functions of the neutrino with definite masses moved in a
beam along the axis $OX$ with not small momentum
$|\vec p_\nu|\gg m_\nu$ and ultrarelativistic energy 
$E_k=\sqrt{\vec p_\nu^2+m_k^2}\simeq$ 
$|\vec p_\nu|+m_k^2/2p_\nu$, $k=1,2,3$. Thus their wave functions look like:
\begin{equation}
\label{nu0} \nu_k(\vec x,t)=e^{i\vec p_\nu\vec
x}e^{-iE_kt}\nu_k(0)=e^{-i\frac{m_k^2}{2p_\nu}t}\nu_k(0)
\end{equation}

Mixing of leptons, i.e. mixing of neutrino when the mass matrix of "electrons" of three-generations is diagonal,
is defined by a unitary $3\times 3$ matrix $\hat V^l=\hat V^l_{MNS}$ Maki--Nakagava--Sakata. This matrix depends
on three mixing angles of the leptons $\vt_{12},\vt_{13}$ and $\vt_{23}$. It is
similar to CKM matrix of quarks mixing and has a well-known form:
\begin{equation}
\label{Vmhc} \hat V^l= \left( \begin{array}{ccl}
c_{12}c_{13}&s_{12}c_{13}&s_{13}e^{-i\delta_l}\\
-s_{12}c_{13}-c_{12}s_{23}s_{13}e^{i\delta_l}&c_{12}c_{23}-
s_{12}s_{23}s_{13}e^{i\delta_l}&s_{23}c_{13}\\
s_{12}s_{23}-c_{12}c_{23}s_{13}e^{i\delta_l}&-c_{12}s_{23}-
s_{12}c_{23}s_{13}e^{i\delta_l}&c_{23}c_{13}\\ \end{array} %
\right) %
\end{equation} %
Note that the $\hat V^l$ can be represented in the form of the product of three matrices of rotations (or of mixing of
two generations in pairs) $\hat O_{12}, \hat O_{13}(\delta_l)$ and $\hat O_{23}$. It is easy to verify that $\hat
V^l\equiv \hat O_{12}\hat O_{13}(\delta_l)\hat O_{23}$, where:
\begin{eqnarray}
\hat O_{12} &=& \left( \begin{array} {rrc}c_{12}&s_{12}&0\\-s_{12}&c_{12}& 0\\0&0&1 \end{array} \right), \hat O_{13}(\delta_l)=
\left( \begin{array} {ccc} c_{13}&0&s_{13}
    e^{-i\delta_l}\\0&1&0\\-s_{13} e^{i\delta_l} &0&c_{13} \end{array}
\right), \nonumber \\
{   }\label{O} \\ 
 \hat O_{23} &=&
\left( \begin{array} {lrc}1&0&0\\0&c_{23}&s_{23}\\0&-s_{23}&c_{23} %
\end{array} %
\right); %
\nonumber
\end{eqnarray} %
here and in~(\ref{Vmhc}) $\delta_l$ is the leptonic CP-violating phase. Its value is not known up to now, sometimes,
for example, it is considered to be equal to $0$ whereas its analogue -- the quark CP-phase $\delta_q$ seems to be
close to $\pi/2$~\cite{CPq}. Acting by matrix~(\ref{Vmhc}) on column
$\hat\nu=\left(\begin{array}{c}\nu_1\\\nu_2\\\nu_3\end{array}\right)$ we obtain according to~(\ref{nu}):
\begin{equation}
\label{nu||}
\begin{array}{c}\left(\begin{array}{c}\nu_e\\\nu_\mu\\\nu_\tau\end{array}\right)(t)=\hat
V_l\left(\begin{array}{c}\nu_1\\ \nu_2\\\nu_3\end{array}\right)\\ \\ \left\{\begin{array}{lll}
\nu_e(t)&=&[c_{12}c_{13}\nu_1(0)+ s_{12}c_{13}
\nu_2(0)e^{-i\varphi_{21}} \\
 & & \qquad {}+ s_{13}\nu_3(0)e^{-i\varphi_{31}-i\delta_l}]e^{-i\frac{m_1^2}{2p_\nu}t}\\
\nu_{\mu}(t)&=&[-(s_{12}c_{23}+c_{12}s_{23}s_{13}
e^{i\delta_l})\nu_1(0)\\
 & & \qquad {}+(c_{12}c_{23}-s_{12}s_{23}s_{13}e^{i\delta_l})\nu_2(0)e^{-i\varphi_{21}}\\
& & \qquad {}+c_{13}s_{23}\nu_3(0)e^{-i\varphi_{31}}]e^{-i\frac{m_1^2}{2p_\nu}t}\\
\nu_{\tau}(t)&=&[(s_{12}s_{23}-c_{12}c_{23}s_{13}
e^{i\delta_l})\nu_1(0)\\
 & & \qquad {}-(c_{12}s_{23}+s_{12}c_{23}s_{13}e^{i\delta_l})\nu_2(0)e^{-i\varphi_{21}}\\
& & \qquad {}+c_{13}c_{23}\nu_3(0)e^{-i\varphi_{31}}]
e^{-i\frac{m_1^2}{2p_\nu}t} \end{array} \right. %
\end{array} %
\end{equation} %
where, using dependence~(\ref{nu0}) of neutrino states on the time $t=L/c$ we have:
\begin{equation}
\label{vp} \varphi_{ij}=\frac{(m_i^2-m_j^2)}{2p_\nu}t=1.27\frac{(m_i^2-m_j^2)\mbox{(eV$^2$)}}
{E_\nu\mbox{(MeV)}}L\mbox{(m)} %
\end{equation} %
where $E_\nu\simeq cp_\nu$ is an energy of the neutrino beam: $E_\nu\gg m_3>m_2>m_1$. Neutrino states with
definite masses are mutually orthogonal and are normalized to unity. Using these statements we
can easy obtain
expressions for probabilities of a transition in vacuum of one kind neutrino to neutrino of another kind during the time $t$.

For probabilities of conservation of $e,\mu,\tau$--neutrino kind we have, respectively:
\begin{equation}
\label{P|ee|} \left\{ \begin{array}{rrl}
P(\nu_e\nu_e)&=&|c^2_{12}c^2_{13}+s^2_{12}c^2_{13}e^{i\varphi_{21}}+
s^2_{13}e^{i\varphi_{31}}|^2\\ P(\nu_\mu
\nu_\mu)&=&||c_{13}s_{12}+c_{12}s_{13}s_{23}e^{i\delta_l}|^2+
|c_{12}c_{23}-s_{12}s_{23}s_{13}e^{i\delta_l}|^2e^{i\varphi_{21}}\\
 & &  {}+c^2_{13}s^2_{23}e^{i\varphi_{31}}|^2\\
P(\nu_\tau\nu_\tau)&=&||s_{12}s_{23}-c_{12}c_{23}s_{13}e^{i\delta_l}|^2+
|c_{12}s_{23}+s_{12}c_{23}s_{13}e^{i\delta_l}|^2e^{i\varphi_{21}}\\
 & & {}+c^2_{13}c^2_{23}e^{i\varphi_{31}}|^2\\
\end{array} \right. \end{equation} And for probabilities of transitions of
$\nu_\alpha$ neutrino to neutrino of another kind $\nu_\beta$ we
obtain:
\begin{equation}
\label{P|emu|} \left\{ \begin{array}{lll}
P(\nu_e\nu_\mu)&=&|c_{12}c_{13}(c_{13}s_{12}+c_{12}s_{23}s_{13}e^{i\delta_l})\\
 & & {}-c_{13}s_{12}(c_{12}c_{23}-s_{12}s_{23}s_{13}e^{i\delta_l})e^{i\varphi_{21}}\\
& & {}-s_{13}c_{13}s_{23}e^{i(\delta_l+\varphi_{31})}|^2\\
P(\nu_e\nu_\tau)&=&|c_{12}c_{13}(s_{23}s_{12}-c_{12}c_{23}s_{13}e^{i\delta_l})\\
 & &
 {}-c_{13}s_{12}(c_{12}s_{23}+c_{23}s_{12}s_{13}e^{i\delta_l})e^{i\varphi_{21}}\\
&& {}+s_{13}c_{13}c_{23}e^{i(\delta_l+\varphi_{31})}|^2\\
P(\nu_\mu\nu_\tau)&=&|(c_{13}s_{12}+c_{12}s_{13}s_{23}e^{i\delta_l})
(s_{12}s_{23}-c_{12}c_{23}s_{13}e^{-i\delta_l})\\
& & {}+(c_{12}c_{23}-s_{12}s_{13}s_{23}e^{i\delta})(c_{12}s_{23}+c_{23}s_{12}s_{13}e^{i\delta})e^{i\varphi_{21}}\\
& & {}-c_{23}c^2_{13}s_{23}e^{i\varphi_{31}}|^2 \end{array} \right. 
\end{equation}

\section{Probabilities of the change of the neutrino kind $\boldsymbol{1-P(\nu_\alpha\nu_\alpha)}$ and
of neutrino $\boldsymbol{\nu_\alpha}$ transition to neutrino $\boldsymbol{\nu_\beta}$ of another kind:
$\boldsymbol{P(\nu_\alpha\nu_\beta)}$.}%
After not complicated, but cumbersome transformations of the formulas~(\ref{P|ee|}),(\ref{P|emu|}) we have complete
expressions for the probability of the change of neutrino kind $1-P(\nu_\alpha\nu_\alpha)$ and for the probabilities of
transition of one kind neutrino to neutrino of another kind. But these formulas are very complex for an
analysis and because of experimental peculiarities of the neutrino registration it is more convenient to use probabilities
averaging over oscillations, i.e. over phases~(\ref{vp}) of neutrino of the continuous spectra. Therefore
we adduce these complete formulas only for references.
$$ %
\begin{array}{lll} %
1-P(\nu_e\nu_e)&=& c_{12}^2 \sin^2(2\vt_{13})\sin^2(\vp_{31}/2)+
c^4_{13}\sin^2(2\vt_{12})\sin^2(\vp_{21}/2)\\
& & {}+s_{12}^2\sin^2(2\vt_{13})\sin^2(\vp_{32}/2) %
\end{array} %
$$%
\vspace{0.5em} %
$$
\begin{array}{lll} %
1-P(\nu_\mu\nu_\mu) &=&
\{c^4_{23}\sin^2(2\vt_{12})+s^4_{12}s^2_{13}\sin^2(2\vt_{23})+s^4_{23}s^4_{13}\sin^2(2\vt_{12})\\
 & & {}+ c^4_{12}s^2_{13}\sin^2(2\vt_{23})
+\cos\delta_l\sin(4\vt_{12})\sin(2\vt_{23})(s_{13}c_{23}^2-s^3_{13}s^2_{23})\\
 & & {}-\cos^2\delta_l
s^2_{13}\sin^2(2\vt_{23})\sin^2(2\vt_{12})\} \sin^2(\vp_{21}/2)\\
 & &
 {}+\{s^2_{12}c^2_{13}\sin^2(2\vt_{23})+c^2_{12}s^4_{23}\sin^2(2\vt_{13})\\
 & & {}+
\cos\delta_ls^2_{23}c_{13}\sin(2\vt_{12})\sin(2\vt_{23})\sin(2\vt_{13})\}
\sin^2(\vp_{31}/2)\\
& &
{}+\{c^2_{12}c^2_{13}\sin^2(2\vt_{23})+s^2_{12}s^4_{23}\sin^2(2\vt_{13})\\
 & & {}-\cos\delta_l
s^2_{23}c_{13} \sin(2\vt_{12})\sin(2\vt_{23})\sin(2\vt_{13})\}
\sin^2(\vp_{32}/2) %
\end{array} %
$$ %
\vspace{0.5em} %
$$ %
\begin{array}{lll}%
1-P(\nu_\tau\nu_\tau) &=&
\{s^4_{23}\sin^2(2\vt_{12})+s^4_{12}s^2_{13}\sin^2(2\vt_{23})\\
 & & {}+ c^4_{23}s^4_{13}\sin^2(2\vt_{12})+c^4_{12}s^2_{13}\sin^2(2\vt_{23})\\
 & &
 {}+\cos\delta_l\sin(4\vt_{12})\sin(2\vt_{23})(s^3_{13}c^2_{23}-s^2_{13}s_{23}^2)\\
 & & {}-\cos^2\delta_ls^2_{13}\sin^2(2\vt_{23})\sin^2(2\vt_{12})\}
\sin^2(\vp_{21}/2)\\
& &
{}+\{s^2_{12}c^2_{13}\sin^2(2\vt_{23})+c^2_{12}c^4_{23}\sin^2(2\vt_{13})\\
& & {}-\cos\delta_lc^2_{23}c_{13}\sin(2\vt_{12})\sin(2\vt_{23})\sin(2\vt_{13})\}
\sin^2(\vp_{31}/2)\\ 
& & {}+\{c^2_{12}c^2_{13}\sin^2(2\vt_{23})+
s^2_{12}c^4_{23}\sin^2(2\vt_{13}) \\
 & & {}+\cos\delta_lc^2_{23}c_{13}
\sin(2\vt_{12})\sin(2\vt_{23})\sin(2\vt_{13})\} \sin^2(\vp_{32}/2)%
\end{array} %
$$ %
\vspace{0.5em} %
$$ %
\begin{array}{lll}%
P(\nu_e\nu_{\mu}) &=& \frac{\displaystyle 1}{\displaystyle 4}
\{\sin^2(2\vt_{13})(s^2_{23}+c^4_{12}s^2_{23}+s^4_{12}s^2_{23}) \\
 & & {}+ \frac {\displaystyle 1}{\displaystyle
2}c_{13}\sin(2\vt_{13})\sin(2\vt_{23})\sin(4\vt_{12})
\cos\delta_l)\\
 & & {}-2c^2_{13}\sin^2(2\vt_{12})(c^2_{23}-s^2_{13}s^2_{23})\cos(\vp_{21})\\
 & & {}-2s^2_{23}\sin^2(2\vt_{13})(c^2_{12}\cos(\vp_{31})+s^2_{12}
\cos(\vp_{32}))\\
 & & {}+c_{13}\sin(2\vt_{12})\sin(2\vt_{13})\sin(2\vt_{23}) \\
 & & \quad{}\cdot(s^2_{12}\cos(\delta_l+\vp_{21})-c^2_{12}\cos(\delta_l-\vp_{21}))\\
 & & {}+c_{13}\sin(2\vt_{12})\sin(2\vt_{13})\sin(2\vt_{23}) \\
 & & \quad{}\cdot(\cos(\delta_l+\vp_{32})-\cos(\delta_l-\vp_{31}))\\
 & & {}+2c^2_{13}c^2_{23}\sin^2(2\vt_{12})\} %
\end{array} %
$$%
\vspace{0.5em} %
$$ %
\begin{array}{lll} %
P(\nu_e\nu_{\tau}) &=& \frac{\displaystyle 1}{\displaystyle 4}
\{\sin^2(2\vt_{13})(c^2_{23}+c^4_{12}c^2_{23}+s^4_{12}c^2_{23}) \\
 & & {}-\frac {\displaystyle 1}{\displaystyle
2}c_{13}\sin(2\vt_{13})\sin(2\vt_{23})\sin(4\vt_{12})
\cos\delta_l)\\
 & & {}+2c^2_{13}\sin^2(2\vt_{12})(s^2_{23}-c^2_{13}s^2_{23})\cos(\vp_{21})\\
 & & \quad{}-2c^2_{23}\sin^2(2\vt_{13})(c^2_{12}\cos(\vp_{31})+s^2_{12}
\cos(\vp_{32}))\\
 & & {}+c_{13}\sin(2\vt_{12})\sin(2\vt_{13})\sin(2\vt_{23})\\
 & & \quad{}\cdot(c^2_{12}\cos(\delta_l-\vp_{21})-s^2_{12}\cos(\delta_l+\vp_{21}))\\
 & & {}+c_{13}\sin(2\vt_{12})\sin(2\vt_{13})\sin(2\vt_{23})\\
 & & \quad{}\cdot(\cos(\delta_l+\vp_{31})-\cos(\delta_l+\vp_{32}))\\
 & & {}+2c^2_{13}s^2_{23}\sin^2(2\vt_{12})\} %
\end{array} %
$$ %
$$ %
\begin{array}{lll}%
P(\nu_{\mu}\nu_{\tau}) &=& \frac{\displaystyle 1}{\displaystyle 4}
\{2s^2_{13}\sin^2(2\vt_{12})\cos^2(2\vt_{23})\\
 & & {}+(c^4_{13}+c^4_{12}+s^4_{12}+(c^4_{12}+s^4_{12})s^4_{13})
\sin^2(2\vt_{23})\\
 & & {}-[2s^2_{13}(c^4_{23}+s^4_{23})\sin^2(2\vt_{12})+
[2s^2_{13}(c^4_{12}+s^4_{12})\\
 & & \quad{}-(1+s^4_{13})\sin^2(2\vt_{12})]
\sin^2(2\vt_{23})]\cos(\vp_{21})\\
 & & {}-[2c^2_{13}(s^2_{12}+c^2_{12}s^2_{13})\sin^2(2\vt_{23})\\
 & & \quad{}- \frac
{\displaystyle 1}{\displaystyle 2}c_{13}\sin(2\vt_{12})\sin(2\vt_{13})
\sin(4\vt_{23})\cos\delta_l]\cos(\vp_{31})\\
 & & {}+[2c^2_{13}\sin^2(2\vt_{23})(s^2_{12}s^2_{13}-c^2_{12})\\
 & & \quad{}-\frac
{\displaystyle 1}{\displaystyle 2}c_{13}\sin(2\vt_{12})\sin(2\vt_{13})
\sin(4\vt_{23})\cos\delta_l]\cos(\vp_{32})\\
 & & {}+2c_{13}\sin(2\vt_{12})\sin(2\vt_{13})
\sin(2\vt_{23})\\
& & \quad{}\cdot\sin\delta_l\sin(\vp_{21}/2)
\cos(\frac{\displaystyle\vp_{31}+\vp_{32}}{\displaystyle 2})\\
 & & {}+s_{13}\sin(4\vt_{12})\sin(4\vt_{23})\cos\delta_l
[1+s^2_{13}]\sin^2(\vp_{21}/2)\\
 & & {}-c_{13}\sin(2\vt_{12})\sin(2\vt_{13})\sin(2\vt_{23})
\sin\delta_l\sin(\vp_{21})\\
 & & {}-2s_{13}^2\sin^2(2\vt_{12})\sin^2(2\vt_{23})\cos(2\delta_l)
\sin^2(\vp_{21}/2)\}\\ %
\end{array} %
$$
Note that the probability of $\nu_e\nu_e$ oscillations does not depend on the leptonic CP-phase in contrast to another
probabilities of the neutrino oscillations.
After averaging these formulas over all phases $\vp_{ij}$ and taking into account
$\langle \cos(\vp_{ij}\pm\delta_l)\rangle=0,\;\langle\sin^2\vp_{ij}\rangle=\langle \cos^2\vp_{ij}\rangle=1/2$ we
have:
\begin{equation}
\label{Pav} \left\{ \begin{array}{rll} \langle
1-P(\nu_e\nu_e)\rangle&=&A_{ee}\\ \langle
1-P(\nu_{\mu}\nu_{\mu})\rangle&=&A_{\mu\mu}+B_{\mu\mu}\cos\delta_l+
C_{\mu\mu}\cos^2\delta_l\\ \langle
1-P(\nu_{\tau}\nu_{\tau})\rangle&=&A_{\tau\tau}+B_{\tau\tau}\cos\delta_l+
C_{\tau\tau}\cos^2\delta_l\\ \langle
P(\nu_e\nu_{\mu})\rangle&=&A_{e\mu}+B_{e\mu}\cos\delta_l\\ \langle
P(\nu_e\nu_{\tau})\rangle&=&A_{e\tau}+B_{e\tau}\cos\delta_l\\
\langle
P(\nu_{\mu}\nu_{\tau})\rangle&=&A_{\mu\tau}+B_{\mu\tau}\cos\delta_l
+C_{\mu\tau}\cos(2\delta_l) \end{array} \right. \end{equation} 

\begin{equation}
\label{A} 
\left\{ 
\begin{array}{rl} A_{ee}=\frac{\displaystyle 1}{\displaystyle 2}
[c^4_{13}\sin^2(2\vt_{12})+\sin^2(2\vt_{13})] 
&\\
\vspace{0.5cm}
\begin{array}{r} A_{\mu\mu}=\frac{1}{2}
[(c^2_{13}+(c^4_{12}+s^4_{12})s^2_{13})\sin^2(2\vt_{23})\\
+(s^4_{13}\sin^2(2\vt_{12})+\sin^2(2\vt_{13}))s^4_{23}\\ +c^4_{23}\sin^2(2\vt_{12})]\\ \end{array} 
& 
\begin{array}{lll} B_{\mu\mu} &=&
  \frac{1}{2}(c^2_{23}-s^2_{23}s^2_{13}) s_{13}\\ & & {}\cdot\sin(2\vt_{23})\sin(4\vt_{12})\\
C_{\mu\mu} &=& -\frac{1}{2}s^2_{13}\sin^2(2\vt_{23})\\
 & & \quad{}\cdot \sin^2(2\vt_{12}) \end{array}\\ 
\vspace{0.5cm}
\begin{array}{r}
A_{\tau\tau}=\frac{1}{2} [(c^2_{13}+(c^4_{12}+s^4_{12})s^2_{13})\sin^2(2\vt_{23})\\
+(s^4_{13}\sin^2(2\vt_{12})+\sin^2(2\vt_{13}))c^4_{23}\\ +s^4_{23}\sin^2(2\vt_{12})]\\ \end{array} 
& 
\begin{array}{lll} B_{\tau\tau} &=& -\frac{1}{2}s_{13}\sin(2\vt_{23})\\
 & & {}\cdot(s^2_{23}-c^2_{23}s^2_{13})\sin(4\vt_{12})\\ 
C_{\tau\tau} &=& {}-\frac{1}{2}s^2_{13}\sin^2(2\vt_{23})\\
 & & \cdot\sin^2(2\vt_{12})
\end{array}\\ \begin{array}{r} A_{e\mu}=\frac{1}{4}
[(1+c^4_{12}+s^4_{12})s^2_{23}\sin^2(2\vt_{13})]\hspace{0.5cm}\\ +2c^2_{13}c^2_{23}\sin^2(2\vt_{12})] \end{array}
& 
\begin{array}{lll}
B_{e\mu} &=& \frac{1}{8}c_{13}\sin(2\vt_{13})\sin(2\vt_{23})\\
 & & {}\cdot\sin(4\vt_{12}) \end{array}
\\
\begin{array}{r} A_{e\tau}=\frac{1}{4} [(1+c^4_{12}+s^4_{12})c^2_{23}\sin^2(2\vt_{13})]\hspace{0.5cm}\\
+2c^2_{13}s^2_{23}\sin^2(2\vt_{12})] \end{array} 
& 
\begin{array}{lll}
B_{e\tau} &=& -\frac{1}{8}c_{13}\sin(2\vt_{13})\\
 & & {}\cdot\sin(2\vt_{23})\sin(4\vt_{12})
\end{array} \\ \begin{array}{r} A_{\mu\tau}=\frac{1}{4}
[2s^2_{13}\sin^2(2\vt_{12})\cos^2(2\vt_{23})\hspace{1cm}\\ +\sin^2(2\vt_{23})\{(c^4_{12}+s^4_{12})s^4_{13}\\
+c^4_{13}+c^4_{12}+s^4_{12}\}] \end{array} 
& 
\begin{array}{lll} B_{\mu\tau} &=& \frac{1}{8}
(1+s^2_{13})s_{13}\sin(4\vt_{12})\\
 & & {}\cdot\sin(4\vt_{23})\\ C_{\mu\tau} &=& {}-\frac{1}{4}
s^2_{13}\sin^2(2\vt_{12})\\
 & & {}\cdot\sin^2(2\vt_{23})\\ \end{array} \end{array} \right. %
\end{equation} %
These expressions are organized in such a way for to emphasize the influence of the leptonic CP-phase on the averaging
probabilities of the neutrino oscillations.

Note that the probabilities of the change of neutrino kind and the probabilities of transitions to another two
neutrino states obviously obey the following rules: %
$$
1-P(\nu_\alpha\nu_\alpha)=P(\nu_\alpha\nu_\beta)+P(\nu_\alpha\nu_\gamma),\mbox{ where }\alpha,
\beta,\gamma=e,\mu,\tau. %
$$
\section{Some examples of the manifestation of the leptonic CP-phase}
In this section we give some examples demonstrating a possible dependence of the probabilities~(\ref{Pav}) on CP-phase. %
But first, for the convenience we introduce new designations:
$$
b_{ik}=B_{ik}/A_{ik},\; c_{ik}=C_{ik}/A_{ik}
$$

The first set of possible values for mixing angles taken from experimental data~\cite{ispan} is:\\
{\bf example a)}
\begin{equation}%
\label{vta} \vt_{12}=(42\pm2)^\circ,\;\vt_{13}=(4.0\pm0.5)^\circ,\;\vt_{23}=(43.6\pm0.5)^\circ%
\end{equation}%
(small mixing of 1,3 generations was obtained from experimental data of 
nuclear reactor CHOOZ~\cite{CHOOZ}). Here and below in
our paper an average error is taken from tables adduced in papers~\cite{ispan,camp}. These values are preliminary and are used below
for our estimations. %
In this case all coefficients $b_{ik}$, $c_{ik}$ in formulas~(\ref{Pav})
have very small values, in particular because of smallness of mixing angle $s_{13}=\sin\vt_{13}$. The table of all
coefficients of formulas~(\ref{Pav}) for this case is:
\begin{equation}
\label{Anuma} \left\{ \begin{array}{l} A_{ee}=0.499;\\ A_{\mu\mu}=0.636,\;b_{\mu\mu}=0.0058,\;c_{\mu\mu}=-0.0038;\\
A_{\tau\tau}=0.613,\;b_{\tau\tau}=0.0055,\;c_{\tau\tau}=-0.0040;\\ A_{e\mu}=0.261,\;b_{e\mu}=0.014;\\
A_{e\tau}=0.238,\;b_{e\tau}=-0.015;\\ A_{\mu\tau}=0.373,\;b_{\mu\tau}=0.0005,\;c_{\mu\tau}=-0.0032. \end{array} \right.
\end{equation}
As we can see, the ratio of the number of the $\mu$-neutrino to the number of the $\tau$-neutrino produced in the initial
beam of electron neutrino $\nu_e$ at a large distance from the source is:
\begin{equation}
\frac{\displaystyle\langle P(\nu_e\nu_{\mu})\rangle}{\displaystyle\langle P(\nu_e\nu_{\tau})\rangle}=\frac{\displaystyle A_{e\mu}}{\displaystyle
A_{e\tau}}\cdot \frac{\displaystyle (1+b_{e\mu}\cos\delta_l)}{\displaystyle (1+b_{e\tau}\cos\delta_l)}\simeq \frac{\displaystyle A_{e\mu}}{\displaystyle
A_{e\tau}}(1+(b_{e\mu}-b_{e\tau})\cos\delta_l) %
\end{equation} %
where $b_{e\mu}-b_{e\tau}\simeq 2b_{e\mu}\simeq 2.8\%$, (as $b_{e\tau}\simeq -b_{e\mu}$), with $\frac{\displaystyle
A_{e\mu}}{\displaystyle A_{e\tau}}\simeq 1.1$. Thus, the contribution of terms containing $\cos\delta_l$ to the ratio of
probabilities $\langle P(\nu_e\nu_{\mu})\rangle$ and $\langle P(\nu_e\nu_{\tau})\rangle$ is of order of $3\%$.
Therefore an experimental observation of the CP-violating phase manifestation is very difficult for this set of mixing
angles (see Fig.1).

\begin{figure}
\begin{center}
\noindent\includegraphics[width=120mm,
keepaspectratio=true]{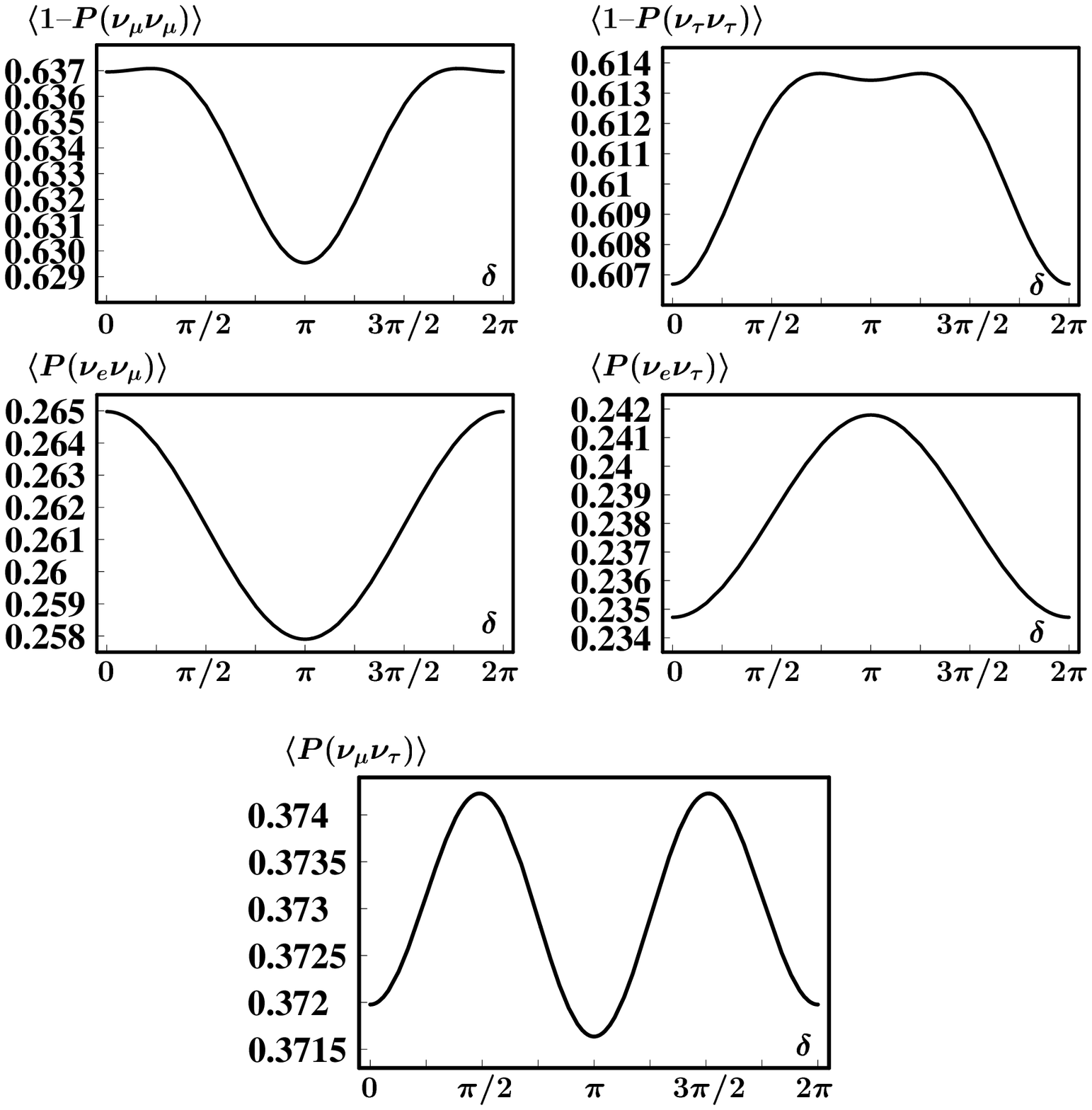}
\end{center}
\caption{%
In this figure the dependences of the probabilities averaging over neutrino oscillations on the leptonic CP-phase
are shown. These dependences are small and difference between maximal and minimal values of probabilities
equals approximately $0.7\%$ in the $\nu_\mu\nu_\mu$ and $\nu_\tau\nu_\tau$ cases, $0.8\%$ in the $\nu_e\nu_\mu$ and
$\nu_e\nu_\tau$ transitions and $0.03\%$ in the $\nu_\mu\nu_\tau$
transitions.}
\end{figure}

The second set of possible values for mixing angles taken from experimental data is:\\
{\bf example b)}
\begin{equation}
\label{vtb} \vt_{12}=(42.0\pm2.0)^\circ,\;\vt_{13}=(14.0\pm1.0)^\circ,\;\vt_{23}=(43.6\pm0.5)^\circ, %
\end{equation} %
(this example is not in a good agreement with experimental data~\cite{CHOOZ} because, although it gives appropriate values
for $\vt_{12}$ and $\vt_{23}$~(\ref{vta}), the value of $\vt_{13}$ is rather large here~\cite{yap}).

In this case, i.e. at $\sin\vt_{13}=0.24$, coefficients $b_{ik},\, c_{ik}$ have larger values in comparison with
the previous case. The values of these coefficients are of the order of several percent, in
particular,
$b_{e\mu}-b_{e\tau}\simeq 2b_{e\mu}\simeq 8.4\%$. So, we have in this case:
\begin{equation}
\label{Anumb} \left\{ \begin{array}{l} A_{ee}=0.548;\\ A_{\mu\mu}=0.645,\;b_{\mu\mu}=0.019,\;c_{\mu\mu}=-0.044;\\
A_{\tau\tau}=0.627,\;b_{\tau\tau}=0.017,\;c_{\tau\tau}=-0.045;\\ A_{e\mu}=0.283,\;b_{e\mu}=0.041;\\
A_{e\tau}=0.265,\;b_{e\tau}=-0.043;\\ A_{\mu\tau}=0.348,\;b_{\mu\tau}=0.0077,\;c_{\mu\tau}=-0.0409. %
\end{array} %
\right.
\end{equation} %
Since $b_{e\mu}-b_{e\tau}\simeq 8.4\%$, the contribution of terms containing $\cos\delta_l$ to the ratio of
probabilities $\langle P(\nu_e\nu_{\mu})\rangle$ and $\langle P(\nu_e\nu_{\tau})\rangle$ is of order of $8.4\%$.
The results of corresponding measurements seem to be very interesting (see Fig.2).

\begin{figure}
\begin{center}
\noindent\includegraphics[width=120mm,
keepaspectratio=true]{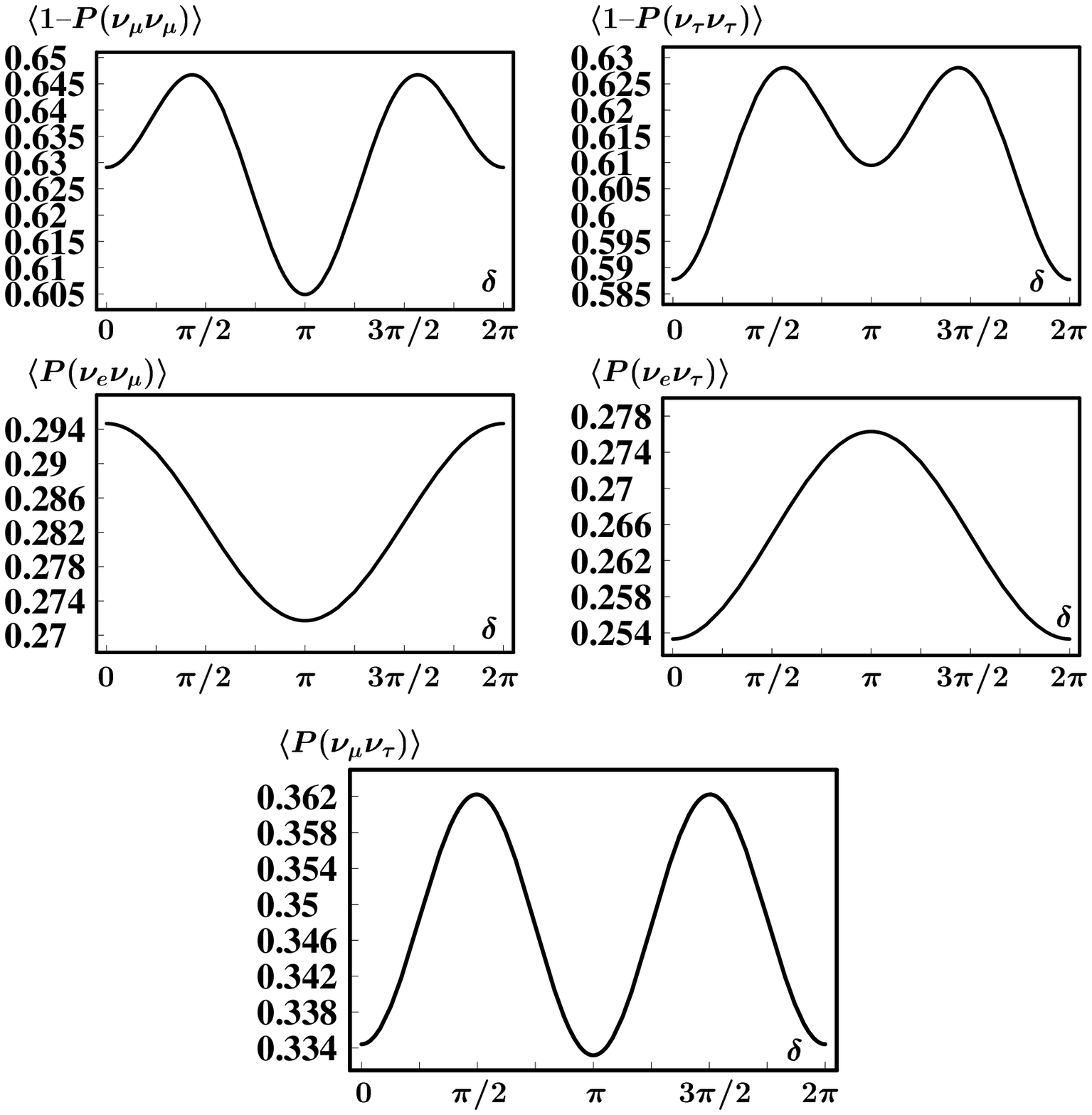}
\end{center}
\caption{%
In this figure the dependences of the probabilities averaging over neutrino oscillations on the leptonic CP-phase are not so
small as in the previous case (Fig.1.) because $\vt_{13}$ is not small
and difference between maximal and minimal values of probabilities
equals approximately $4\%$ in the $\nu_\mu\nu_\mu$ and $\nu_\tau\nu_\tau$ cases, $2.5\%$ in the $\nu_e\nu_\mu$ and
$\nu_e\nu_\tau$ transitions and $2.8\%$ in the $\nu_\mu\nu_\tau$ transitions. Thus the role of $\vt_{13}$ in the
manifestation of leptonic CP-phase in neutrino oscillations is very important.}
\end{figure}

\section{The difference between the $\boldsymbol{\nu_\alpha\to\nu_\beta}$ transitions probability and the
$\boldsymbol{\nu_\beta\to\nu_\alpha}$ transitions probability and leptonic CP-phase}%
Let us consider $\nu_\alpha\to\nu_\beta$ transitions in neutrino oscillations (denote them "forward" transitions) and compare them
with $\nu_\beta\to\nu_\alpha$ transitions (denote them "backward" transitions).
At $\delta_l\ne0$ the probabilities of forward transitions differ from probabilities of
backward transitions. Note right away that the probability of the forward transition
coincides with the probability of the backward transition after averaging over all oscillations, while before
averaging these probabilities are different provided CP-violating, i.e. at $\delta_l\ne 0$. And the difference between
these probabilities is proportional to $\sin\delta_l$. These statements are direct consequences of general
formulas for $P(\nu_e\nu_\mu),P(\nu_e\nu_\tau)$ and $ P(\nu_\mu\nu_\tau)$ which given above. So,
on obtaining probabilities of backward transitions by replacement $\delta_l\to-\delta_l$ we subtract probabilities
of forward transitions from them. As a result we have:
\begin{equation}
\label{P-P} \left\{ \begin{array}{l} P(\nu_{\mu}\nu_e)-P(\nu_e\nu_{\mu})=
a_0(\sin\varphi_{21}+\sin\varphi_{32}-\sin\varphi_{31})\sin\delta_l\\ P(\nu_{\tau}\nu_e)-P(\nu_e\nu_{\tau})=
-a_0(\sin\varphi_{21}+\sin\varphi_{32}-\sin\varphi_{31})\sin\delta_l\\
P(\nu_{\tau}\nu_{\mu})-P(\nu_{\mu}\nu_{\tau})= a_0(\sin\varphi_{21}-2\sin\frac{\varphi_{21}}{2}\cos\frac{\displaystyle
(\varphi_{31}+\varphi_{32})}{\displaystyle 2}) \sin\delta_l \end{array} \right. %
\end{equation} %
where $a_0=\frac{\displaystyle 1}{\displaystyle 2}c_{13}\sin2\vt_{12}\sin 2\vt_{13}\sin 2\vt_{23}$. Here phases
$\varphi_{21},\varphi_{32},\varphi_{31}$ depend on the time of neutrino flight in vacuum $t$ (see~(\ref{vp})) or,
in other words, on distance (base-line) $L$ between points of neutrino birth and neutrino absorption and also on
difference of squared masses $\Delta m_{ij}^2=m_i^2-m_j^2$. %
For the most possible values of neutrino masses and theirs average errors taken from experiment~\cite{ispan,camp}:
\begin{equation}
\label{m}
m_3=(1/17\pm1/50)eV,\; m_2=(1/175\pm 1/300)eV,\; m_1\ll m_2
\end{equation}%
We can see that the distance of the neutrino oscillations, i.e. the base-line for the experimental
devices must not be less $10^3$m. %
The experimental definition of the CP-phase based on the correlation~(\ref{P-P}) would be the most natural, however now it
is practically impossible because beams of different types (for example $\nu_e$ and $\nu_\mu$) of neutrino (that is,
obtained in different reactions) but with the same energy are required for the experiment. This problem
possibly will be solved in the future, but until now all experimental data were obtained only for beams of the
neutrino with the continuous energy spectra. The cause of this problem consists, in particular, in very small
cross-sections of neutrino interactions.

In current experiments we deal only with probabilities of transitions of neutrino with continuous energy spectra
in the initial beam, i.e. with all phases~(\ref{vp}) averaging over oscillations~(\ref{Pav}),(\ref{A}) which also depend
on the CP-violating phase $\delta_l$. The main idea of my talk consists in a suggestion to find the value of the CP-phase
using data of experiments with large base-line and formulas~(\ref{Pav}),(\ref{A}).

Note that coefficient $a_0$ defining the
value of the effect of $t-$symmetry violating~(\ref{P-P}) is not too small in both $a)-$ and $b)-$cases: $$
a)\;a_0=0.07\qquad b)\;a_0=0.23 $$ Moreover in b-case it is large. Therefore measurements of this effect are possible
although they are difficult.


\section{Conclusions.}
So, the main results of our work are the following:\\%
\begin{itemize}
\item[$\bullet$] The expressions for the probabilities of neutrino oscillations were obtained in the explicit form with
regard for the leptonic CP-phase.
\item[$\bullet$] The manifestation of the leptonic CP-phase in neutrino oscillations was investigated
by the example of the probabilities averaging over oscillations. Using modern experimental data the model calculations and
numerical estimates were done.
\item[$\bullet$] The question of the $t-$symmetry violation for neutrino oscillation was analyzed. And there was
established that the difference between the "forward" probabilities and the "backward" probabilities
was proportional to sine of the leptonic CP-phase.
\end{itemize}


%
\newcommand{\lag}{{\cal L}}
\newcommand{\slD}{\slashed{D}}
\newcommand{\slcD}{\slashed{\cal D}}

\title*{Possibility of an Additional Source of Time Reversal %
Violation for Neutrinos}
\author{%
R. Erdem\thanks{E-mail:erdem@likya.iyte.edu.tr}}
\institute{%
Department of Physics
\.Izmir Institute of Technology 
G\"ulbah\c ce K\"oy\"u, Urla 35437 Izmir, Turkey}

\authorrunning{R. Erdem}
\titlerunning{Possibility of an Additional Source of Time Reversal %
Violation for Neutrinos}
\maketitle

\begin{abstract} 
We show that neutral fermions may have an additional source of 
time reversal violation by associating time reversal with gauge 
group representations of fermions through the method of group 
extensions as in the case of parity. This provides a new source 
of time reversal violation for neutral particles. 
\end{abstract}

\section{Introduction} 

Topic of time reversal violation in neutrino processes (especially in
oscillations) has gained much theoretical attention~\cite{ref1} and there are
proposals to study these effects experimentally~\cite{ref2}. The main interest
in these studies is to use T-violation as the measure of CP violation
through fermion mixing matrices. In this study we show that the only
source of time reversal violation in neutral fermion (hence in
neutrino) processes is not the one which is related (through CPT
theorem) to CP violation from complex phases in fermion mass matrices.
There is an additional possible source of time reversal violation for
neutral fermions arising from another kind of fermion mixing other
than the Dirac and Majarona ones as we shall see below.

\section{Four-component fermions and parity} 

Let us consider two 2-component fermions; $\chi_1$, $\chi_2$  
which transform under the same representation of SL(2,C) 
\begin{equation}\label{eq1}
\chi_1 \to e^{\frac{i}{2}\cdot (\vec{\theta}-i\vec{\nu})}\chi_1,
\qquad
\chi_2 \to e^{\frac{i}{2}\cdot (\vec{\theta}-i\vec{\nu})}\chi_2.
\end{equation}
Assume that $\chi_1$, $\chi_2$ belong to different gauge groups or to 
different representations of the same gauge group. The 
simplest Lagrangian which can be constructed out of $\chi_1$, $\chi_2$ 
with related kinetic terms and quadratic interaction terms is 
\begin{equation} \label{eq2}
\lag = i \chi_1^\dagger\sigma_\mu D_\mu^{(1)}\chi_1
     + i \chi_2^\dagger\bar{\sigma}_\mu D_\mu^{(2)}\chi_2
     + m(\chi_1^\dagger i\sigma_2\chi_2^* + \chi_2^\dagger
     i\sigma_2\chi_1^*)
  + h.c.
\end{equation}
where h.c. stands for taking the complex conjugate of the terms
preceeding it and 
$\sigma_\mu D_\mu = D_0 \ \vec{\sigma}\cdot\vec{D}$,
$D_\mu^{(1)}= \partial_\mu + ig_1B_\mu^{(1)}$,
$D_\mu^{(2)}= \partial_\mu + ig_2B_\mu^{(2)}$
with $B_\mu^{(1)}$, $B_\mu^{(2)}$ 
being the gauge fields coupling to $\chi_1$ and $\chi_2$,
respectively. We assume an unbroken electromagnetic gauge 
group $U(1)_{em}$ . So $\chi_1$ , $\chi_2$ have opposite electric
charges. Instead of identifying Eq.(\ref{eq2}) as a Lagrangian of two 
(two-component) fermions interacting through a quadratic interaction term 
one may pass to a $4$-component formulation so that Eq.(\ref{eq2}) becomes 
\begin{equation}\label{eq3}
\lag = i\bar{\psi}\slD^{(1)}P_L\psi + i\bar{\psi}\slD^{(2)}P_R\psi
       + m\bar{\psi}\psi + h.c.
\end{equation}
Here
\begin{equation}\label{eq4}
\psi = \begin{pmatrix} \chi_1 \\ i\sigma_2\chi_2^* \end{pmatrix} 
     = \begin{pmatrix} \chi_L \\ \chi_R \end{pmatrix},
\quad \bar{\psi}=\psi^\dagger\gamma^0, 
\quad P_L = \frac{1}{2} (1+\gamma_5),
\quad P_R = \frac{1}{2} (1-\gamma_5)
\end{equation} 
where in the chiral representation employed 
\begin{equation}\label{eq5} 
\gamma_5=
\begin{pmatrix}
 I & 0 \\
 0 & -I
\end{pmatrix}
\end{equation}
The difference between Eq.(\ref{eq2}) and Eq.(\ref{eq3}) is that, 
in Eq.(\ref{eq3}) there is a single $4$-component massive fermion 
while in Eq.(\ref{eq2}) there are
two interacting $2$-component massless fermions. So the $4$-component
formulation simplifies the field theoretic calculations but the price
to be paid is that the upper and lower two components transform under
different gauge interactions. In fact this is one of the basic
observations behind the parity violation in the standard model of
electroweak interactions. Mathematically speaking the procedure given
above where space reflection is associated with a $Z_2$ group corresponding to
the exchange of the gauge group representations of left-handed and
right-handed fermions is known as the method of group extensions~\cite{ref3}.
In particular here the procedure corresponds to the pullback of two 
$Z_2$ groups, one associated with space reflection and the other one associated
with the exchange of the gauge group representations of the
left-handed and right-handed fermions via the isomorphism relating the
two $Z_2$ groups~\cite{ref4}. In the next section we shall employ a similar
procedure for time reversal.  

A field theoretic formulation of these
observations will make the picture clearer and more precise. First we
take two $2$-component fermion fields, $\chi_1$ and $\chi_2$ (coupling
through a quadratic interaction), with~\cite{ref5}
\begin{eqnarray}
 \chi_1 = \int \frac{d^3p}{(2\pi)^{3/2}} \frac{1}{\sqrt{2E_p}}
             [u_1(\vec{p})b_1(\vec{p}) e^{-ip\cdot x}
             +v_1(\vec{p})d_1^\dagger(\vec{p}) e^{ip\cdot x}]
  \nonumber \\
 \chi_2 = \int \frac{d^3p}{(2\pi)^{3/2}} \frac{1}{\sqrt{2E_p}}
             [u_2(\vec{p})b_2(\vec{p}) e^{-ip\cdot x}
             +v_2(\vec{p})d_2^\dagger(\vec{p}) e^{ip\cdot x}]
 \label{eq6}
\end{eqnarray}
where $u_1$, $u_2$, $v_1$, $v_2$ are $2$-component
spinors with $v_1\sim i\sigma_2u_1^*$, $v_2\sim i\sigma_2u_2^*$,
$u_2\sim i\sigma_2u_1^*$, $v_2\sim i\sigma_2u_2^*$
and $b_1$, $b_2$, $d_1^\dagger$, $d_2^\dagger$ are the
annihilation, creation operators for $\chi_1$ ,$\chi_2$ and 
$p\cdot x = E\cdot t - \vec{p}\cdot\vec{x}$. 
In the $4$-component formulation $\chi_1$ ,$\chi_2$ are replaced
by a single $4$-component fermion with 
\begin{equation}\label{eq7}
\psi = \int \frac{d^3p}{(2\pi)^{3/2}} \frac{m}{\sqrt{E_p}}
             [u(\vec{p},\sigma)b(\vec{p},\sigma) e^{-ip\cdot x}
             + v(\vec{p},\sigma)d^\dagger(\vec{p},\sigma) e^{ip\cdot x}]
\end{equation}
where $\sigma$ stands for either of helicity or spin. In the chiral 
basis~\cite{ref5,ref6,ref7}
\begin{eqnarray}
  u(\vec{p},1/2) = \begin{pmatrix} u_1 \\ 0 \end{pmatrix},
  \quad u(\vec{p},-1/2) = \begin{pmatrix} 0 \\ i\sigma_2v_2^* \end{pmatrix},
\nonumber \\
  v(\vec{p},1/2) = \begin{pmatrix} v_1 \\ 0 \end{pmatrix},
  \quad v(\vec{p},-1/2) = \begin{pmatrix} 0 \\ i\sigma_2u_2^*
  \end{pmatrix}
  \label{eq8} \\
 b(\vec{p},1/2)=b_1(\vec{p}), \quad b(\vec{p},-1/2)=d_2(\vec{p}),
 \nonumber \\
 d(\vec{p},1/2)=d_1(\vec{p}), 
 \quad d(\vec{p},-1/2) =b_2(\vec{p})
 \nonumber
\end{eqnarray}
where $\pm1/2$ stands for different helicity states. In the mass basis 
(which is the solution basis for the Dirac
equation)~\cite{ref8} one may write 
\begin{eqnarray}
[u(\vec{p},\sigma)b(\vec{p},\sigma) e^{-ip\cdot x}
 + v(\vec{p},\sigma)d^\dagger(\vec{p},\sigma) e^{ip\cdot x}]
 \nonumber \\
 =
\begin{pmatrix}
 (u_1+i\sigma_2u_1^*)b_1 + (v_2+i\sigma_2v_2^*)d_2 \\
 (i\sigma_2v_2^*-v_2)d_2 + (i\sigma_2u_1^*-u_1)b_1
\end{pmatrix}
e^{-ip\cdot x} 
\label{eq9} \\
+ \begin{pmatrix}
(-v_1+i\sigma_2v_1)d_1^\dagger + (u_2-i\sigma_2u_2^*)b_2^\dagger \\
(u_2+i\sigma_2u_2^*)b_2^\dagger + (v_1+i\sigma_2v_1^*)d_1^\dagger
\end{pmatrix}
e^{ip\cdot x}
\nonumber
\end{eqnarray}
which is a sum of two independent spin states.  

The lower and upper two components of $\psi$ in the chiral representation,
$\chi_L$ and $\chi_R$ are related by space reflection and interchange 
of $\chi_L$ and $\chi_R$
amounts to a space reflection operation plus interchange of the gauge group
representations of $\chi_L$ and $\chi_R$~\cite{ref9}. This is the case 
because in the $4$-component formulation $\chi_L$ and $\chi_R$ form 
a single system $\psi$ so their
interchanges are mutually related to each other. In the next section
we shall investigate such a possibility for time reversal. We shall
couple two $4$-component fermions (with different gauge group
representations) through a quadratic coupling where each $4$-component
fermion in the coupling can be redefined so that each is the time
reversal of the other. Then the result is a single system i.e. an
$8$-component fermion consisting of the original $4$-component fermions.
In this way time reversal is associated with the interchange of the
gauge group representations of the original two $4$-component fermions.
This, in turn, enables one to break the time reversal invariance in a
way other than the complex phases in the fermion mixing matrices.

\section{$8$-component fermions and time reversal}

In the previous section we have seen that one can interprete a
quadratic interaction term between two $2$-component massless fermions
as a fermion mass term by passing to $4$-component fermion formulation.
In this section we raise the question if one can find a Lorentz
invariant quadratic term, other than the mass term for $4$-component
fermions, between two $4$-component fermions, which can be interpreted
as a mass term for $8$-component fermions. We shall see that this
possible provided the fermion is neutral and massless ( or almost
massless) and the upper and lower two components of these $8$-component
fermions are related by time reversal. This observation opens up the
possibility of an additional source of time reversal violation for
neutral fermions in a way similar to the case of parity for
$4$-component formulation.

We consider two $4$-component fermions given by 
\begin{eqnarray}
\psi_1 &=& \sum_\sigma \int \frac{d^3p}{(2\pi)^{3/2}} \frac{1}{\sqrt{2E_p}}
             [u_1(\vec{p})b_1(\vec{p}) e^{-ip\cdot x}
             +v_1(\vec{p})d_1^\dagger(\vec{p}) e^{ip\cdot x}]
  \nonumber \\
\psi_2 &=& \sum_\sigma \int \frac{d^3p}{(2\pi)^{3/2}} \frac{1}{\sqrt{2E_p}}
             [u_2(\vec{p})b_2(\vec{p}) e^{-ip\cdot x}
             +v_2(\vec{p})d_2^\dagger(\vec{p}) e^{ip\cdot x}]
 \label{eq10}
\end{eqnarray}
$u_1$, $u_2$, $v_1$, $v_2$ are, this time, $4$-component spinors with 
the spinor parts of $\psi_1$, $\psi_2$ being each others time reversals i.e.  
\begin{equation}\label{eq11}
u_2(\vec{p},\sigma)\sim i \gamma^1\gamma^3 u_1^*(\vec{p},\sigma),
\quad
v_2(\vec{p},\sigma)\sim i \gamma^1\gamma^3 v_1^*(\vec{p},\sigma)
\end{equation}
The only possible Lorentz invariant quadratic interaction term
between $\psi_1$ and $\psi_2$ is~\cite{ref10} 
\begin{equation}\label{eq12}
M(\psi_1^\dagger\psi_2+\psi_2^\dagger\psi_1)
\end{equation}
The simplest possible Lagrangian reads 
\begin{equation}\label{eq13}
\lag = i\bar{\psi}_1\slD^{(1)}\psi_1 - i\bar{\psi}_2\gamma^0\psi_2
       + m\bar{\psi}_1\bar{\psi}_1 + m\bar{\psi}_2\bar{\psi}_2
       + M(\psi_1^\dagger\psi_2+\psi_2^\dagger\psi_1) + h.c.
\end{equation}
where $\slD^{(1)} = \gamma^\mu(\partial_\mu-ig_1B_\mu^{(1)})$,
$\slD^{(2)} = \gamma^\mu(\partial_\mu-ig_2B_\mu^{(2)})$.
We take both $\psi_1$ and $\psi_2$ be massless (for example through a chiral
symmetry). Then~(\ref{eq13}) can be put into a simple $8$-component
form 
\begin{equation}\label{eq14}
\lag = i\bar{\Psi}\slcD^{(1)}P_U\Psi + i\bar{\Psi}\slcD^{(2)}P_D\Psi
       + M\bar{\Psi}\Psi + h.c.
\end{equation}
where 
\begin{eqnarray}
{\cal D} = \begin{pmatrix} 0 & -\slD\gamma^0 \\
                           \gamma^0\slD & 0 
           \end{pmatrix},
\quad
P_U = \begin{pmatrix} I_4 & 0 \\  0 & 0 \end{pmatrix},
\quad 
P_D = \begin{pmatrix} 0 & 0 \\ 0 & I_4 \end{pmatrix},
\nonumber \\
\Psi = \begin{pmatrix} \psi_1 \\ \psi_2 \end{pmatrix},
\qquad
\bar{\Psi} = \Psi^\dagger\tilde{\Gamma},
\qquad
\tilde{\Gamma} = \begin{pmatrix} 0 & I_4 \\ i_4 & 0 \end{pmatrix}
\label{eq15}
\end{eqnarray}
So $\Psi$ acts as a single fermion whose upper and lower
$4$-components couple to different gauge groups, and the spinor 
parts of the lower and the upper components are related by time
reversal. Because the spinor parts of  $\psi_1$ , $\psi_2$ are 
related by time reversal $\psi_2$ has the same electric charge 
as $\psi_1$ . However because time reversal is violated (for example 
through weak interactions) there will be an effective charge 
asymmetry induced between $\chi_1$ and $\chi_2$ which forbids  
$\psi_1^\dagger\psi_2$ type of terms. This requires that $\Psi$ 
should be a neutral particle in order to allow $\psi_1^\dagger\psi_2$ 
type of terms (provided the electric charge is a conserved quantity). 
In summary the Lagrangian~(\ref{eq14}) is applicable only for neutral 
particles and in that case there is an additional source of time 
reversal violation other than the one due to the complex phases in 
the fermion mixing matrix. If one assigns one of the  $\psi_1$, 
$\psi_2$ to the singlet and the other to a nontrivial representation 
of the gauge group then the time reversal becomes maximal in the same 
way as in the standard model of electroweak interactions. The
propagator of $\Psi$ corresponding to Eq.~(\ref{eq14}) is 
\begin{equation}\label{eq16}
\frac{i}{\Gamma^\mu p_\mu - M}\qquad\mbox{where}\qquad
\Gamma^\mu = \begin{pmatrix} 0 & -\gamma^\mu\gamma^0 \\
                             \gamma^0\gamma^\mu & 0 
             \end{pmatrix}
\end{equation}
If we take $m_1 = m_2 = m\simeq 0$ then the mass terms of $\psi_1$,
$\psi_2$ behave as quadratic interaction terms which modify the 
propagator self energy contribution which changes $MI_8$ to 
$MI_8 + \gamma^0 m\tilde{\Gamma}$ where $I_8$ stands for 
$8\times 8$ unit matrix. In 
this case $\Psi$ can be still considered as a single fermionic 
system with some perturbative self interaction term. 

The parity and time reversal operators in this case can be infered 
more easily by considering the Dirac and the 
Dirac-like equations corresponding to the Lagrangians (\ref{eq3}) 
and (\ref{eq14})
\begin{eqnarray} 
i\slD^{(1)}P_L\psi + m P_R\psi = 0, \qquad
i\slD^{(2)}P_R\psi + m P_L\psi = 0
\label{eq17} \\
i\slD^{(1)}P_U\psi + M P_D\psi = 0, \qquad
i\slD^{(2)}P_D\psi + M P_U\psi = 0
\label{eq18} 
\end{eqnarray}
One can determine the parity and time reversal operators by letting 
$\vec{x}\to -\vec{x}$ and $t \to -t$, respectively, in the equations 
given above. One notices that there are more than one operator acting 
on the fields corresponding to each of these 
transformations. For example, for parity one may take either of 
\begin{equation}\label{eq19}
\begin{pmatrix}  \gamma^0 & 0 \\ 0 & \gamma^0 \end{pmatrix},
\qquad
\Gamma^0 = \begin{pmatrix} 0 & -I \\ I & 0 \end{pmatrix}
\end{equation}
for time reversal either of 
\begin{equation}\label{eq20}
\begin{pmatrix} i\gamma^1\gamma^3\;K & 0 \\ 
                0 & i\gamma^1\gamma^3\;K
\end{pmatrix},
\qquad
\tilde{\Gamma} =
\begin{pmatrix} 0 & I \\ I & 0 \end{pmatrix}
\end{equation}
where $K$ stands for complex conjugation of c-numbers on its right 
hand side. Here the first operators in (\ref{eq19}) and 
(\ref{eq20}) correspond to the direct extension of the usual parity 
and time reversal operators for the $8$-component case. 
One notices that the second time reversal operator in (\ref{eq20}) 
is unitary. At first sight this seems to contradict with 
the Wigner's requirement that time reversal is a symmetry of the 
physical systems so it should be anti-unitary~\cite{ref11}. 
However Wigner's argument assumes that the time reversal of 
an operator corresponding to a physical observable $Q$ 
at time $t$, $Q(t)=exp(iHt) Q exp(-iHt)$ is given by 
$exp(-iHt) Q exp(iHt)$. In other words he implicitly assumes that 
the Hamiltonian $H$ is invariant under time reversal. Because the 
additional time reversal operator in (\ref{eq20}) only applies 
to neutrinos (whose time reversal properties are not well known) and 
it does not leave the Hamiltonian invariant in 
general (its violation may be even maximal) so the Wigner's argument 
does not apply to the present case. 

While both operators in (\ref{eq19}) or (\ref{eq20}) correspond to the
same space-time reflection the physical operations to perform these
transformations are different. In other words the experiments designed
to study parity or time reversal in a physical process are different
in each case. The parity transformation corresponding to the first
operator in (\ref{eq19}) is accomplished by interchange of the left-handed and
right-handed components of the $4$-component fermions.  Because this
interchange is associated with space reflection and each of left-handed and
right-handed components are associated with opposite helicities one
can check parity invariance under this operation by checking the space
reflection in a given fermionic process. However the second operation 
in (\ref{eq19})
does not correspond to a simple exchange of the left-handed and
right-handed components so it corresponds to preparing a different
experiment where the state vectors for the fermions are transformed
accordingly. In other words it is not possible to test the invariance
of parity due to the second transformation in (\ref{eq19}) by checking space
reflection invariance in the same fermionic process, i.e. in this case one
should first study the original process and then the process
corresponding to the parity transformed one.  A similar argument is
true for the time reversal operators in Eq.(\ref{eq20}). The first (the one
which corresponds to the usual time reversal) corresponds to preparing
a different process which corresponds to the time reversed of the
original process while the test of time reversal for the second
operator can be achieved, in principle, in the same process by
comparing the intensity of the events in the direction forward in time
and the events in the direction backward in time. While this is not
applicable in practice one may test the time reversal in the same
process in this case in indirect ways. The spinor part of $\Psi$, 
$U$ can be written in the following way
\begin{equation}\label{eq21}
U = \begin{pmatrix} U_1 \\ U_2 \end{pmatrix},
\qquad
U_1 = \begin{pmatrix} u_1^{(1)} \\ u_2^{(1)} \end{pmatrix},
\qquad
U_2 = \begin{pmatrix} u_1^{(2)} \\ u_2^{(2)} \end{pmatrix},
\end{equation}
In the chiral basis the components of $U$ may be related 
as~\cite{ref9,ref12} 
\begin{eqnarray}
u_1^{(1)}=\frac{1}{m} (p_0+\vec{\sigma}\cdot\vec{p})u_2^{(1)},
\qquad
u_1^{(2)}=\frac{1}{m} (p_0-\vec{\sigma}\cdot\vec{p})u_2^{(2)}
\nonumber \\
u_1^{(1)}=\frac{1}{m} (-p_0+\vec{\sigma}\cdot\vec{p})u_1^{(2)},
\qquad
u_2^{(1)}=\frac{1}{m} (-p_0-\vec{\sigma}\cdot\vec{p})u_2^{(2)}
\label{eq22}
\end{eqnarray}
We notice that both $\psi_1$ and  $\psi_2$ have exactly the same 
helicities. So if two kinds of neutrinos are produced which have 
the same helicities and which do not interact with one another through 
gauge interactions (due to Lorentz invariance) 
one may suspect if these additional neutrinos are  $\psi_2$'s 
(provided  $\psi_1$'s are identified by the usual neutrinos). If the 
interaction of $\Psi$'s is time reversal invariant under electroweak 
interactions then the presence of additional neutrinos 
(i.e.  $\psi_2$'s) are seen as an extra factor of $2$ in the
production cross sections of neutrinos when compared with their 
interaction with $Z$ bosons. If there is a time reversal violation 
then this may either be due to the usual source of 
time reversal violation arising from complex phase(s) in fermion 
mixing matrices or it may be due to non-invariance 
under the interchanges of $\psi_1$ and $\psi_2$. Seeking for any 
asymmetry between the gauge interactions of $\psi_1$'s and $\psi_2$'s 
in the same process amounts to seeking a time reversal violation due 
to the second operator in Eq(\ref{eq20}). However it may 
not be easy to disentagle these two types of time reversal violation. 

\section{Conclusion} 

We have seen that as in the case of parity one may extend time
reversal operator by a $Z_2$ group associated with the interchange of
the gauge group representations of two coupled fermions whose spinor
parts are time reversals of each other. In this way one gets the
possibility of an additional source of time reversal violation for
neutral fermions (e.g for neutrinos). This time reversal violation is
similiar to the parity violation and it may be even maximal as in the
case of parity. If this additional time reversal has a comparable
magnitude as the usual time reversal violation resulting from complex
phase(s) in fermion mass matrices then the mixing phase can not be
direcly derived from the magnitude of time reversal violation and vica
versa. The presence of the extra time reversal violation can be
infered by comparing the magnitudes of the time reversal and the usual
CP violation in a neutrino process (for example in neutrino
oscillations) in future experiments.

\title*{Quark-Lepton Masses and the Neutrino Puzzle in the AGUT Model}
\author{%
C.D. Froggatt}
\institute{%
Department of Physics and Astronomy, Glasgow University,
Glasgow G12 8QQ, Scotland, UK}

\authorrunning{C.D. Froggatt}
\titlerunning{Quark-Lepton Masses and the Neutrino Puzzle in the AGUT Model}
\maketitle

\section{Introduction}
I reviewed the general problem of the quark-lepton 
mass spectrum at the first Bled workshop on ``What 
comes beyond the Standard Model'' \cite{bled1}. 
So, in this talk, I will mainly concentrate on two 
topics: the Lightest Flavour Mass Generation model 
and the Neutrino Mass and Mixing problem in the 
Anti-Grand Unification Theory (AGUT). 


\section{Lightest Flavour Mass Generation Model}

A commonly accepted framework for discussing the flavour problem is based on
the picture that, in the absence of flavour mixing, only the particles
belonging to the third generation $t$, $b$ and $\tau $ have non-zero masses.
All other masses and the mixing angles then appear as a result of
the tree-level mixings of families, related to some underlying family
symmetry breaking. 
Recently, a new mechanism of flavour mixing, which we call Lightest Family
Mass Generation (LFMG), was proposed \cite{lfm}. 
According to LFMG the whole of flavour
mixing for quarks is basically determined by the mechanism responsible for
generating the physical masses of the {\it up} and {\it down} quarks, 
$m_{u}$ and $m_{d}$ respectively. So, in the chiral symmetry limit, when 
$m_{u}$ and $m_{d}$ vanish, all the quark mixing angles vanish. 
Therefore, the masses (more precisely
any of the diagonal elements of the quark and
charged lepton mass matrices)
of the second and third families are practically the same in
the gauge (unrotated) and physical bases.
The proposed flavour mixing mechanism, driven solely by the generation of
the lightest family mass, could actually be realized in two generic ways.

The first basic alternative (I) is when the lightest family mass ($m_{u}$ or 
$m_{d} $) appears as a result of the complex flavour mixing of all three
families. It ``runs along the main diagonal'' of the corresponding $3\times
3 $ mass matrix $M$, from the basic dominant element $M_{33}$ to the element 
$M_{22}$ (via a rotation in the 2-3 sub-block of $M$) and then to the
primordially texture zero element $M_{11}$ (via a rotation in the 1-2
sub-block). The direct flavour mixing of the first and third quark and 
lepton families is supposed to be absent or negligibly small in $M$.

The second alternative (II), on the contrary, presupposes direct flavour
mixing of just the first and third families. There is no involvement of the
second family in the mixing. In this case, the lightest mass appears in the
primordially texture zero $M_{11}$ element ``walking round the corner'' (via
a rotation in the 1-3 sub-block of the mass matrix $M$). Certainly, this
second version of the LFMG mechanism cannot be used for both the up and the
down quark families simultaneously, since mixing with the second family
members is a basic part of the CKM quark mixing 
phenomenology (Cabibbo mixing, non-zero $V_{cb}$ 
element, CP violation). However, the alternative II could work for
the up quark family provided that the down quarks follow the alternative I.

Here we will just consider the latter scenario.

\subsection{Quark Sector}
We propose that the mass matrix for the down quarks ($D$ = $d$, $s$, $b$) 
is Hermitian with three texture zeros of the following 
alternative I form:

\begin{equation}
M_{D}=\begin{pmatrix}
 0 & a_D & 0 \\ a_D^{\ast} & A_D & b_D \\
 0 & b_D^{\ast} & B_D 
\end{pmatrix}  \label{LFM1}
\end{equation}
It is, of course, necessary to assume some hierarchy between the elements,
which we take to be: $B_{D}\gg A_{D}\sim \left| b_{D}\right| \gg \left|
a_{D}\right| $. The zero in the $\left( M_{D}\right)
_{11}$ element corresponds to the commonly accepted conjecture
that the lightest family masses appear as a direct result of flavour
mixings. The zero in $\left( M_{D}\right) _{13}$ means that only minimal
``nearest neighbour'' interactions occur, giving a tridiagonal matrix
structure.

Now our main hypothesis, that the second and third family diagonal mass
matrix elements are practically the same in the gauge and physical
quark-lepton bases, means that : 
\begin{equation}
B_{D}=m_{b}+\delta_{D} \qquad A_{D}= m_{s} + \delta_{D}^{\prime }  
\label{BA}
\end{equation}
The components $\delta_{D} $ and $\delta_{D}^{\prime }$ are supposed to be
much less than the masses of the particles in the next lightest family,
meaning: 
\begin{equation}
|\delta_D |\ll m_{s} \qquad |\delta _{D}^{\prime }|\ll m_{d}  
\label{deldelp}
\end{equation}
Since the trace and determinant of the Hermitian matrix $M_{D}$ gives the
sum and product of its eigenvalues, it follows that 
\begin{equation}
\delta _{D}\simeq - m_{d}  \label{del}
\end{equation}
while $\delta _{D}^{\prime }$ is vanishingly small and can be neglected
in further considerations.

It may easily be shown that our hypothesis and related equations (\ref{BA} - 
\ref{del}) are entirely equivalent to the condition that the diagonal
elements ($A_{D}$, $B_{D}$) of $M_{D}$ are proportional
to the modulus square of the off-diagonal elements ($a_{D}$, $b_{D}$): 
\begin{equation}
\frac{A_{D}}{B_{D}}=\left| \frac{a_{D}}{b_{D}}\right| ^{2}
\label{ABab}
\end{equation}

Using the conservation of the trace, determinant and sum of principal minors
of the Hermitian matrices $M_{D}$ under unitary transformations, we are led
to a complete determination of the moduli of all their elements, which
can be expressed to high accuracy as follows: 
\begin{equation}
\left| M_{D} \right| = 
\begin{pmatrix} 0 & \sqrt{m_d m_s} & 0 \\ 
\sqrt{m_d m_s} & m_s & \sqrt{m_d m_b} \\
 0 & \sqrt{m_d m_b} & m_b - m_d 
\end{pmatrix}  \label{LFM1A}
\end{equation}

Now the Hermitian mass matrix for the up quarks is taken to be 
of the following alternative II form: 
\begin{equation}
M_{U}=
\begin{pmatrix} 
0 & 0 & c_U \\ 0 & A_U & 0 \\ c_U^{\ast} & 0 & B_ U 
\end{pmatrix}
\label{LFM2}
\end{equation}
The moduli of all the elements of $M_{U}$ can also be readily 
determined in terms of the physical masses as follows: 
\begin{equation}
\left| M_{U} \right| = 
\begin{pmatrix}
 0 & 0 & \sqrt{m_u m_t} \\ 
 0 & m_c & 0 \\ 
 \sqrt{m_u m_t} & 0 & m_t - m_u 
\end{pmatrix}
\label{LFM2A}
\end{equation}

The CKM quark mixing matrix elements can now be readily calculated 
by diagonalising the mass matrices $M_D$ and $M_U$. They are 
given by the following simple and compact formulae in terms of 
quark mass ratios:  
\begin{eqnarray}
\left| V_{us}\right| = \sqrt{\frac{m_{d}}{m_{s}}} = 0.222 \pm 0.004 
\qquad \left| V_{us}\right|_{exp} = 0.221 \pm 0.003 \\
\left|V_{cb}\right| = \sqrt{\frac{m_{d}}{m_{b}}} = 0.038 \pm 0.004 
\qquad \left|V_{cb}\right|_{exp} = 0.039 \pm 0.003 \\ 
\left|V_{ub}\right| = \sqrt{\frac{m_{u}}{m_{t}}} = 0.0036 \pm 0.0006
\qquad \left|V_{ub}\right|_{exp} = 0.0036 \pm 0.0006   \label{angles}
\end{eqnarray}
As can be seen, they are in impressive agreement with the experimental 
values.

\subsection{Lepton Sector}

The MNS lepton mixing matrix is defined analogously to the CKM 
quark mixing matrix: 
\begin{equation}
U=U_{\nu }U{_{E}}^{\dagger }  \label{U}
\end{equation}
Here $U_E$ and $U_{\nu}$ are the unitary matrices which diagonalise 
the charged lepton mass matrix $M_E$ and the effective neutrino mass 
matrix $M_{\nu}$ respectively. Assuming the charged lepton masses 
follow alternative I, like the down quarks, the LFMG model 
predicts the charged lepton mixing angles in the matrix $U_{E}$ 
to be: 
\begin{equation}
\sin \theta _{e\mu }=\sqrt{\frac{m_{e}}{m_{\mu }}}\qquad \sin \theta _{\mu
\tau }=\sqrt{\frac{m_{e}}{m_{\tau }}}\qquad \sin \theta _{e\tau }\simeq 0
\label{charged}
\end{equation}
These small charged lepton mixing angles will not
markedly effect atmospheric neutrino oscillations, which
appear to require maximal mixing $\sin ^{2}2\theta _{atm}\simeq
1 $. Similarly, in the case of the large mixing angle (LMA) MSW solution  
of the solar neutrino problem, they are essentially negligible. 
It follows then that the large neutrino mixings should mainly come 
from the $U_{\nu }$
matrix associated with the neutrino mass matrix. 

According to the ``see-saw''
mechanism, the effective mass-matrix $M_{\nu}$ for physical neutrinos has the
form 
\begin{equation}
M_{\nu }=-M_{N}^{T}M_{NN}^{-1}M_{N}  \label{nuCDF}
\end{equation}
where $M_{N}$ is their Dirac mass matrix, while $M_{NN}$ is the Majorana
mass matrix of their right-handed components. 
Matsuda et al \cite{refs} have extended the alternative I LFMG texture 
to the Dirac $M_N$ and Majorana $M_{NN}$ matrices. 

The eigenvalues of the neutrino Dirac mass matrix $M_{N}$ are
taken to have a hierarchy similar to that for the charged leptons (and down
quarks)

\begin{equation}
M_{N3}:M_{N2}:M_{N1}\simeq 1:y^{2}:y^{4},\quad y\approx 0.1  \label{h1}
\end{equation}
and the eigenvalues of the Majorana mass matrix $M_{NN}$ are taken to have a
stronger hierarchy

\begin{equation}
M_{NN3}:M_{NN2}:M_{NN1}\simeq 1:y^{4}:y^{6}  \label{h2}
\end{equation}
One then readily determines the general LFMG matrices $M_{N}$ and $M_{NN}$
to be of the type

\begin{equation}
M_{N}\simeq M_{N3}\left( 
\begin{array}{lll}
0 & \alpha y^{3} & 0 \\ 
\alpha y^{3} & y^{2} & \alpha y^{2} \\ 
0 & \alpha y^{2} & 1
\end{array}
\right)  \label{m0}
\end{equation}
and

\begin{equation}
M_{NN}\simeq M_{NN3}\left( 
\begin{array}{lll}
0 & \beta y^{5} & 0 \\ 
\beta y^{5} & y^{4} & \beta y^{3} \\ 
0 & \beta y^{3} & 1
\end{array}
\right).  \label{m00}
\end{equation}
We further take an extra condition of
the type

\begin{equation}
\left| \Delta -1\right| \le y^{2}\qquad (\Delta \equiv
\alpha ,\beta )  \label{natur}
\end{equation}
for both the order-one parameters $\alpha $ and $\beta $ contained 
in the matrices $M_{N}$ and $M_{NN}$, 
according to which they are supposed to be equal to unity with a few percent
accuracy. Substitution in the seesaw formula (\ref{nuCDF}) generates an
effective physical neutrino mass matrix $M_{\nu }$ of the form:

\begin{equation}
M_{\nu }\simeq -\frac{M_{N3}^{2}}{M_{NN3}}\left( 
\begin{array}{lll}
0 & y & 0 \\ 
y & 1+(y-y^{2})^{2} & 1-(y-y^{2}) \\ 
0 & 1-(y-y^{2}) & 1
\end{array}
\right)  \label{matr1}
\end{equation}
The physical neutrino masses are then given by:
\begin{eqnarray}
m_{\nu 1} &\simeq &(\frac{1}{2}-\frac{\sqrt{3}}{2})\frac{M_{N3}^{2}}{M_{NN3}}%
\cdot y, \nonumber \\
m_{\nu 2} &\simeq &(\frac{1}{2}+\frac{\sqrt{3}}{2})\frac{M_{N3}^{2}}{M_{NN3}}%
\cdot y,\quad m_{\nu 3}\simeq (2-y)\frac{M_{N3}^{2}}{M_{NN3}}  
\end{eqnarray}
The predicted values of the neutrino oscillation parameters are: 
\begin{equation}
\sin ^{2}2\theta _{atm}\simeq 1,\quad \sin ^{2}2\theta _{sun}\simeq \frac{2}{%
3},\quad U_{e3}\simeq \frac{1}{2\sqrt{2}}y,\quad \frac{\Delta m_{sun}^{2}}{%
\Delta m_{atm}^{2}}\simeq \frac{\sqrt{3}}{4}y^{2}  \label{pred1}
\end{equation}
in agreement with atmospheric and LMA-MSW solar neutrino oscillation 
data.

The proportionality condition (\ref{ABab}), which leads to the LFMG 
texture, is not so easy to generate 
from an underlying symmetry beyond the Standard Model. 
However Jon Chkareuli, Holger Nielsen and myself have recently shown 
\cite{SU3} that it is possible to give a natural realisation of the 
LFMG texture in a local chiral $SU(3)$ family symmetry model. 

\section{Fermion Masses in the AGUT Model}

The AGUT model is based on a non-simple extension of the Standard 
Model (SM) with three copies of the SM gauge group---one for each 
family---and, in the absence of right-handed neutrinos, 
one extra abelian factor: $G = SMG^3 \times U(1)_f$, where 
$SMG \equiv SU(3) \times SU(2) \times U(1)$. This AGUT gauge 
group is broken down by four Higgs fields $S$, $W$, $T$ and 
$\xi$ to the usual SM gauge group, identified as the 
diagonal subgroup of $SMG^3$. The Higgs field $S$ has a 
vacuum expectation value (VEV) taken to be unity in fundamental 
(Planck) mass units, while $W$, $T$ and $\xi$ have VEVs an 
order of magnitude smaller. So the pure SM is essentially 
valid, without supersymmetry, up to energies close to the 
Planck scale. The AGUT gauge group $SMG^3 \times U(1)_f$  
only becomes effective near the Planck scale, where the 
$i$'th proto-family couples to just the $i$'th $SMG$ factor 
and $U(1)_f$. The $U(1)_f$ charges assigned to the quarks 
and leptons are determined, by anomaly cancellation constraints, 
to be zero for the first family and all left-handed fermions, 
and for the remaining right-handed states to be as follows:
\begin{equation}
Q_f(\tau_r) = Q_f(b_R) = Q_f(c_R) =1 \qquad 
Q_f(\mu_r) = Q_f(s_R) = Q_f(t_R) = -1
\end{equation}   
I refer to the review of the AGUT model by Holger and myself 
at the first Bled workshop \cite{bled2} for more details.

The quarks and leptons are mass protected by the approximately 
conserved AGUT chiral gauge charges \cite{fn}. The quantum numbers 
of the Weinberg-Salam Higgs field $\phi_{WS}$ are chosen so 
that the $t$ quark mass is not suppressed, whereas the $b$ 
quark and $\tau$ lepton are suppressed. This is done by 
taking the four abelian charges, expressed as a charge 
vector $\vec{Q} = (y_1/2, y_2/2, y_3/2, Q_f)$, for $\phi_{WS}$ 
to be given by:
\begin{equation}
\vec{Q}_{\phi_{WS}} = \vec{Q}_{c_R} - \vec{Q}_{t_L} 
=(0,2/3,0,1) - (0,0,1/6,0) = (0,2/3,-1/6,1) 
\end{equation}
We assume that, like the quark and lepton fields, the Higgs 
fields belong to singlet or fundamental representations of 
all the non-abelian groups. Then, by imposing the usual SM 
charge quantisation rule for each of the $SMG$ factors, the 
non-abelian representations are determined from the weak 
hypercharge quantum numbers $y_i$. The abelian quantum 
numbers of the other Higgs fields are chosen as follows:
\begin{eqnarray}
\vec{Q}_W = (0,-1/2,1/2,-4/3) \qquad \vec{Q}_T = 
(0,-1/6,1/6,-2/3) \\
\vec{Q}_{\xi} = (1/6,-1/6,0,0) \qquad \vec{Q}_S = 
(1/6,-1/6,0,-1)
\end{eqnarray}   
Since we have $<S>=1$ in Planck units, the Higgs field $S$ 
does not suppress the fermion masses and the quantum numbers 
of the other Higgs fields $W$, $T$, $\xi$ and $\phi_{WS}$ 
given above are only determined modulo those of $S$.

The effective SM Yukawa coupling matrices in this AGUT model can 
now be calculated in terms of the VEVs of the fields $W$, $T$ 
and $\xi$ in Planck units---up to ``random`` complex order unity 
factors multiplying all the matrix elements---for the quarks:
\begin{equation}
Y_U \sim \left ( \begin{array}{ccc}
        W T^2 \xi^2 & W T^2 \xi & W^2 T \xi \\
        W T^2 \xi^3 & W T^2 & W^2 T\\
        \xi^3  & 1 & W T
        \end{array} \right ) \label{H_U} 
\end{equation}
\begin{equation}
Y_D  \sim  \left ( \begin{array}{ccc}
        W (T^2 \xi^2 & W T^2 \xi & T^3 \xi \\
        W T^2 \xi & W T^2 & T^3 \\
        W^2 T^4 \xi & W^2 T^4 & W T 
        \end{array} \right ) \label{H_D} 
\end{equation}
and the charged leptons:
\begin{equation}
Y_E  \sim  \left ( \begin{array}{ccc}
        W T^2  \xi^2 & W T^2 \xi^3 & W T^4 \xi \\
        W T^2 \xi^5 & W T^2 & W T^4 \xi^2 \\
        W T^5 \xi^3 & W^2 T^4 & W T
        \end{array} \right ) \label{H_E}
\end{equation}
A good order of magnitude fit is then obtained \cite{bled2} 
to the charged fermion masses with the following values for 
the Higgs field VEVs in Planck units:
\begin{equation}
W = 0.179, \qquad T = 0.071 \qquad \xi = 0.099.
\end{equation}

We now consider the neutrino mass matrix in the AGUT model.

\section{Neutrino Mass and Mixing Problem} 

Without introducing new physics below the AGUT scale, the 
effective light neutrino mass matrix $M_{\nu}$ is generated 
by tree level diagrams involving the exchange of two 
Weinberg-Salam Higgs tadpoles and the appropriate 
combination of $W$, $T$, $\xi$ and $S$ Higgs tadpoles. 
In this way we obtain:
\begin{equation}
\label{eq:Mnuminagut}
M_{\nu} \simeq \frac{\langle {\phi}_{WS} \rangle^2}{M_{Pl}} 
\left ( \begin{array}{ccc}
        W^2 {\xi}^4 T^4  & W^2 {\xi} T^4  & W^2 {\xi}^3 T \\
        W^2 \xi T^4  & W T^5 & W^2 T \\
        W^2 {\xi}^3 T  & W^2 T & W^2 T^2 {\xi}^2 
\end{array} \right ),
\end{equation}
The off-diagonal element $(M_{\nu})_{23} = (M_{\nu})_{32}$ dominates 
the matrix, giving large $\nu_{\tau}-\nu_{\mu}$ mixing with the 
following two neutrino masses and mixing angle:
\begin{equation}
m_2 \sim m_3 \sim \frac{\langle {\phi}_{WS} \rangle^2}{M_{Planck}}W^2T 
\qquad \sin^2 2\theta_{\mu \tau} \simeq 1
\end{equation} 
Although the large mixing angle $\sin^2 2\theta_{\mu \tau}$ is 
suitable for atmospheric neutrino oscillations, there are two 
problems associated with the neutrino masses. Firstly the ratio of 
neutrino mass squared differences $\Delta m_{23}^2/\Delta m_{12}^2 
\sim 2 T \xi^2 \sim 1.4 \times 10^{-3}$, whereas the small mixing 
angle (SMA) MSW solution to the solar neutrino problem requires 
$\Delta m_{23}^2/\Delta m_{12}^2 \sim 10^{-2}$. Secondly the 
predicted overall absolute mass scale for the neutrinos 
$\langle {\phi}_{WS} \rangle^2/M_{Planck} \sim 3 \times 10^{-6}$ eV 
is far too small. 

We conclude it is necessary to introduce a new mass scale into the 
AGUT model. Two ways have been suggested of obtaining realistic 
neutrino masses and mixings in the AGUT model:
\begin{enumerate}
\item By extending the AGUT Higgs spectrum to include a weak 
isotriplet Higgs field $\Delta$ with SM weak hypercharge 
$y/2 =-1$ and a VEV $\langle \Delta^0 \rangle \sim 1$ eV; 
also a new Higgs field $\psi$ giving large $\mu - \tau$ mixing 
in the charged lepton Yukawa coupling matrix $Y_E$ is required.
\item By including right-handed neutrinos and extending the AGUT 
gauge group to $G_{extended} = (SMG \times U(1)_{B-L})^3$; also 
two new Higgs fields $\phi_{B-L}$ and $\chi$ are introduced to 
provide a see-saw mass scale and structure to the Majorana 
right-handed neutrino mass matrix. 
\end{enumerate}

Yasutaka Takanishi reported on the second approach 
\cite{yasutaka} at the workshop; so I will report on the 
first approach \cite{mark} here. 
We must therefore consider the introduction of a new Higgs field 
$\psi$, which can yield large mixing from the charged lepton mass 
matrix without adversely affecting the quark mass matrices.
With the following choice of charges for the $\psi$ field
\begin{equation}
\vec{Q}_{\psi}  =  3\vec{Q}_{\xi} + \vec{Q}_W + 4\vec{Q}_T 
 =  \left ( \frac{1}{2}, -\frac{5}{3}, \frac{7}{6}, -4
\right),
\end{equation}
we obtain new expresssions for the quark Yukawa matrices:
\begin{equation}
\label{eq:psimat}
Y_U  =  \left(\begin{array}{ccc} WT^2\xi^2 & WT^2\xi & W^2T\xi \\
                                   WT^2\xi^3 & WT^2    & W^2T    \\
                                   \xi^3     & 1       & WT \end{array}
          \right) 
\end{equation}
\begin{equation}
Y_D  =  \left(\begin{array}{ccc} WT^2\xi^2 & WT^2\xi & T^3\xi  \\
                                   WT^2\xi   & WT^2    & T^3     \\
                                   W^2T\psi & W^2T\xi\psi  & WT \end{array}
          \right) \label{eq:H_DLSND}
\end{equation}
and the charged lepton Yukawa matrix:
\begin{equation}
Y_E  =  \left(\begin{array}{ccc} WT^2\xi^2 & W^2T^2\psi & \xi^4\psi \\
                                   W^4T\xi\psi^2 & WT^2    & \xi\psi \\
                                   W^3\xi^2\psi & W^2T\xi\psi  & WT 
\end{array} \right).
\end{equation}
As we can see from the charged lepton matrix we will indeed have
large mixing if $\left\langle \psi \right\rangle = O(0.1)$, so that
$(Y_E)_{23} \sim (Y_E)_{33}$. In the following
discussion we shall take $\psi$ to have a vacuum expectation
value of $\left\langle \psi \right\rangle = 0.1$ for definiteness. 
The effect of the field $\psi$
on the charged fermion masses is then small, since the elements involving 
$\psi$
do not make any significant contribution to the determinant, or the sum
of the minors, of the mass matrices, or the trace of the squares
$Y Y^{\dagger}$
of the Yukawa matrices. The mixings of the quarks is essentially unaffected
by the terms involving $\psi$, and the only 
significant effect is on the mixing matrix $U_E$, which is now given by:
\begin{equation}
\label{eq:UEX}
U_E  \sim  \left( \begin{array}{ccc} 
1 & \frac{\xi \psi^2X}{T} &  \frac{\xi^3}{X} \\
-W\psi & \frac{WT}{\xi\psi X} & \frac{1}{X} \\
\frac{\xi \psi^2}{T} & -\frac{1}{X} & \frac{WT}{\xi\psi X}
\end{array}\right)
 \sim  \left( \begin{array}{ccc}
1 & 0.021 & 6.4 \times 10^{-4} \\
- 0.016 & 0.75 & 0.66 \\
0.014 & - 0.66 & 0.75 \end{array}\right)
\end{equation}
where
$X = \sqrt{1 + W^2T^2/\xi^2 \psi^2} \sim 1.51$
This gives the large mixing required, 
\begin{equation}
\sin^2 2\theta_{atm} \sim 1,
\end{equation}
for the atmospheric neutrino oscillations.

We can further obtain a solution with vacuum oscillations
for the solar neutrinos by choosing appropriate charges for 
the isotriplet Higgs field $\Delta$.
We require a large off-diagonal $(1, 2)$ element for the neutrino mass
matrix and hence we choose the charges on $\Delta$ to be
\begin{equation}
 \vec{Q}_{\Delta} = ( -\frac{1}{2}, -1,\frac{1}{2}, \frac{5}{3})
\end{equation}
We then obtain the neutrino mass matrix,
\begin{equation}
M_{\nu} \sim \langle\Delta^0\rangle
\left ( \begin{array}{ccc} W \xi^6 &  W  \xi^3 & T \xi^2 \psi\\
W \xi^3 & W & T \xi \psi\\
 T \xi^2 \psi & T \xi \psi &  T^2 \xi \psi \end{array} \right).
\end{equation}
This has the eigenvalues,
\begin{equation}
m_1  \sim   \langle \Delta^0 \rangle \left(-T \xi^2 \psi + 
\frac{T^2 \xi \psi}{2} \right),\quad
m_2  \sim  \langle \Delta^0 \rangle W, \quad
m_3  \sim  \langle \Delta^0 \rangle \left( T \xi^2 \psi + 
\frac{T^2 \xi \psi}{2}
\right)
\end{equation}
where the splitting between $m_1$ and $m_2$ comes from the
mass matrix element $(M_{\nu})_{33}$.  
The neutrino mixing matrix is then given by,
\begin{equation}
U_{\nu} \sim \left ( \begin{array}{ccc}
\frac{1}{\sqrt{2}}(1 + \frac{T}{4\xi}) & \xi^3 &
\frac{1}{\sqrt{2}}(1 - \frac{T}{4\xi}) \\
\frac{T\xi\psi}{\sqrt{2}W}(1 - \frac{T}{4\xi}) & 1 &
-\frac{T\xi\psi}{\sqrt{2}W}(1 + \frac{T}{4\xi}) \\
-\frac{1}{\sqrt{2}}(1 - \frac{T}{4\xi}) & \frac{T\xi\psi}{W} &
\frac{1}{\sqrt{2}}(1 + \frac{T}{4\xi})
\end{array} \right).
\end{equation}
Hence, using  $U_E$ from eqn.~\ref{eq:UEX}, we have the 
lepton mixing matrix $U = U_E^{\dagger} U_{\nu}$:  
\begin{equation}
U \sim \left ( \begin{array}{ccc}
\frac{1}{\sqrt{2}}(1 + \frac{T}{4\xi}) & -W\xi &
\frac{1}{\sqrt{2}}(1 - \frac{T}{4\xi})\\
\frac{1}{\sqrt{2}X}(1 - \frac{T}{4\xi}) & \frac{WT}{\xi \psi X} &
-\frac{1}{\sqrt{2}}(1 + \frac{T}{4\xi})\\
-\frac{WT(1 - \frac{T}{4\xi})}{\sqrt{2}\xi\psi X} & \frac{1}{X} &
\frac{WT(1 + \frac{T}{4\xi})}{\sqrt{2}\xi\psi X}
\end{array} \right) 
\sim  \left ( \begin{array}{ccc}
0.83 & -0.016 & 0.58\\
0.38 & 0.75 & -0.55\\
-0.43 & 0.66 & 0.62
\end{array} \right)
\end{equation}
which, as we can see, has large electron neutrino mixing, as we require
for a vacuum oscillation solution to the solar neutrino problem.  
 
We also have the mass hierarchy,
\begin{equation}
\frac{\Delta m^2_{13}}{\Delta m^2_{23}} \sim 2 \frac{T^3 \xi^3 \psi^2}{W^2}
\sim 3 \times 10^{-7}.
\end{equation}
Hence, if we then take $\langle \Delta^0 \rangle \sim 0.2$ eV, 
so that we have an
overall mass scale suitable for the atmospheric neutrino problem, then we
will also have,
\begin{equation}
\Delta m^2_{13} \sim 2 \langle\Delta^0\rangle^2 T^3 \xi^3 \psi^2 \sim
3 \times 10^{-10} \mbox{eV}^2.
\end{equation}
With such a hierarchy of $\Delta m^2$s we effectively
have two-neutrino oscillations for the solar neutrinos, with
the mixing angle given by,
\begin{equation}
\sin^2 2\theta_{sun} = 4 U^2_{e1} U^2_{e3} \sim 0.9.
\end{equation}
So, we have the `just-so' vacuum oscillation solution to the
solar neutrino problem with large electron neutrino mixing. 
We remark that $U_{e2} = -0.016$ 
satisfies the CHOOZ electron neutrino survival probability 
bound ($U_{e2}$ is the relevant mixing matrix element, since 
$\Delta m_{12}^2 \sim \Delta m_{23}^2 \gg \Delta m_{13}^2$).

It is also possible to obtain a small mixing angle SMA-MSW 
solution to the solar neutrino problem, with a different 
choice of charges for $\Delta$:
\begin{equation}
\vec{Q}_{\Delta} = (-\frac{1}{2}, -\frac{2}{3}, -\frac{1}{6},0)
\end{equation} 
which gives the quasi-diagonal neutrino mass matrix
\begin{equation}
M_{\nu} \sim \langle\Delta^0\rangle\left ( \begin{array}{ccc}
W^4 T \xi^2 \psi^2 & \: W T^2 \xi^3  &  \: T^3 \xi^2 \psi \\
W T^2 \xi^3  & W T^2  & W T \xi^2 \\
T^3 \xi^2 \psi & W T \xi^2  &  \xi \psi \\
\end{array}
\right).
\end{equation}
The mixing matrix $U_{\nu}$ for this mass matrix is given by,
\begin{equation}
U_{\nu} \sim \left ( \begin{array}{ccc}
1 & \xi^3 & - T^3 \xi \\
-\xi^3 & 1 & \frac{WT\xi}{\psi}\\
-T^3 \xi & - \frac{WT\xi}{\psi} & 1 \\
\end{array} \right).
\end{equation}
Thus  we obtain the lepton mixing matrix:
\begin{equation}
U = U_E^{\dagger} U_{\nu}  \sim \left( \begin{array}{ccc}
1 & -W\psi & \frac{\xi \psi^2}{T} \\
 \frac{\xi \psi^2 X}{T} & \frac{WT}{\xi\psi X} & -\frac{1}{X} \\
-\frac{\xi^3}{X}  & \frac{1}{X}  & \frac{WT}{\xi\psi X} \end{array}\right)
\sim \left( \begin{array}{ccc}
1 & 0.016 & 0.014\\
0.021 & 0.75 & -0.66\\
6 \times 10^{-4} & 0.66 & 0.75 \end{array}\right).
\end{equation}
Taking $\langle\Delta^0\rangle \sim 3$ eV, we then obtain 
suitable masses and mixings for the solution
of both the solar and atmospheric neutrino problems:
\begin{equation}
\sin^2 2\theta_{atm} \sim 1  \ \Delta m_{23}^2 \sim 
1 \times 10^{-3} \mbox{eV}^2 \quad
\sin^2 2\theta_{sun} \sim 1  \ \Delta m_{12}^2 \sim 
6 \times 10^{-6} \mbox{eV}^2
\end{equation}

We did not manage to find an LMA-MSW solution, which is 
favoured by the latest solar neutrino data from Sudbury 
and SuperKamiokande, using this approach. However, during this 
workshop, Holger, Yasutaka and I constructed a promising   
LMA-MSW solution \cite{fnt} using the extended version of the AGUT 
model with right-handed neutrinos and the usual see-saw 
mechanism. 

\section*{Acknowledgements}

I should like to thank PPARC for a travel grant and my collaborators, 
Jon Chkareuli, Holger Bech Nielsen and Yasutaka Takanishi, for many 
discussions.

\def\cdfsleq{\raisebox{-.6ex}{${\textstyle\stackrel{<}{\sim}}$}}
\def\cdfsgeq{\raisebox{-.6ex}{${\textstyle\stackrel{>}{\sim}}$}}
\def\cdfsss{\scriptscriptstyle}
\def\cdfnn{\hspace{2mm}}
\def\cdfsVEV#1{\left\langle #1\right\rangle}
\newcommand{\cdfMeV}{\mbox{\rm MeV}}
\newcommand{\cdfGeV}{\mbox{\rm GeV}}
\newcommand{\cdfeV}{\mbox{\rm eV}}

\title*{Neutrinos in the Family Replicated Gauge Group Model}
\author{C.D. Froggatt}
\institute{%
Department of Physics and Astronomy, Glasgow University,
Glasgow G12 8QQ, Scotland, UK}

\titlerunning{Neutrinos in the Family Replicated Gauge Group Model}
\authorrunning{C.D. Froggatt}
\maketitle

\section{Introduction}
In this paper, I will update my report \cite{bled2001}
on fermion masses in the Anti-Grand Unification Theory
(AGUT) at the Bled 2001 workshop. There are two versions
of the AGUT model based on three family replicated
copies of the Standard Model (SM) gauge group,
selected according to whether or not each family is
supplemented by a right-handed neutrino. We note that
neither model introduces supersymmetry.
In the absence of right-handed neutrinos, the AGUT gauge
group is $G_1 = SMG^3\times U(1)_f$, where
$SMG \equiv SU(3)\times SU(2) \times U(1)$. With the
inclusion of three right-handed neutrinos, the AGUT
gauge group is extended to
$G_2 = (SMG \times U(1)_{B-L})^3$, where the three
copies of the SM gauge group are supplemented by an
abelian $(B-L)$ (= baryon number minus lepton
number) gauge group for each family.
In each case, the AGUT gauge group $G_1$ ($G_2$)
is the largest anomaly free group \cite{trento}
transforming the
known 45 Weyl fermions (and the additional three
right-handed neutrinos for $G_2$) into each other
unitarily, which does NOT unify the irreducible
representations under the SM gauge group.

Here we present good order of magnitude fits, with four and five
adjustable parameters respectively, to the quark and lepton masses
and mixing angles for the two versions of the AGUT model. In each
case the fit to the charged fermion masses and quark mixings is
arranged to essentially reproduce the original three parameter
AGUT fit \cite{bled2001}. It is necessary to introduce a new mass
scale into the theory, in order to obtain realistic neutrino
masses. For the $G_1 = SMG^3\times U(1)_f$ model we introduce a
weak isotriplet Higgs field  and obtain a vacuum oscillation
solution to the solar neutrino problem, whereas for the $G_2 =
(SMG \times U(1)_{B-L})^3$ model we introduce the usual see-saw
mass scale for the right-handed neutrinos and obtain a large
mixing angle (LMA) MSW solution. During the last year, further
data \cite{Sudbury} from the Sudbury Neutrino Observatory have
confirmed that the LMA-MSW solution is strongly favoured, with the
Vacuum Oscillation and LOW solutions now allowed at the $3 \sigma$
level while the SMA-MSW solution seems to be completely ruled out.
We refer to \cite{Nir} for a recent review of the phenomenology of
neutrino physics.

\section{The $SMG^3\times U(1)_f$ Model}

The usual SM gauge group is identified as the diagonal subgroup of
$SMG^3$ and the AGUT gauge group $SMG^3\times U(1)_f$ is broken
down to this subgroup by four Higgs fields $S$, $W$, $T$ and
$\xi$. Thus, for example, the SM weak hypercharge $y/2$ is given
by the sum of the weak hypercharge quantum numbers $y_i/2$ for the
three proto-families:
\begin{equation}
\frac{y}{2} = \frac{y_1}{2} + \frac{y_2}{2} + \frac{y_3}{2}
\end{equation}
The spontaneously broken chiral AGUT
gauge quantum numbers of the quarks and leptons protect
the charged fermion masses and generate a mass hierarchy
for them \cite{CDFfn} in terms of the Higgs field vacuum
expectation values (VEVs). However the VEV of the
Higgs field $S$ is taken to be unity in
fundamental (Planck) mass units. Thus only the VEVs of
the other three Higgs fields are used as free parameters,
in the order of magnitude fit to the effective SM
Yukawa coupling matrices $Y_U$, $Y_D$ and $Y_E$ for
the quarks and charged leptons \cite{bled2001}.

In this model, the large $\nu_{\mu}-\nu_{\tau}$
mixing required for atmospheric neutrino oscillations
is generated by introducing a large off-diagonal
element, $(Y_E)_{23} \sim (Y_E)_{33}$, in the charged
lepton Yukawa coupling matrix. This is achieved by
introducing a new Higgs field $\psi$ with a VEV equal
to unity in Planck units and the following set of
$U(1)$ gauge charges:
\begin{equation}
\vec{Q}_{\psi}  =  3\vec{Q}_{\xi} + 2\vec{Q}_W + 4\vec{Q}_T
 =  \left ( \frac{1}{2}, -\frac{13}{6}, \frac{5}{3},
-\frac{16}{3} \right),
\end{equation}
Here we express the abelian gauge charges in the model as a
charge vector $\vec{Q} = (y_1/2, y_2/2, y_3/2, Q_f)$.

We then have the Yukawa matrices for the quarks:
\begin{equation}
Y_U  =  \left(\begin{array}{ccc} WT^2\xi^2 & WT^2\xi & W^2T\xi \\
                                   WT^2\xi^3 & WT^2    & W^2T    \\
                                   \xi^3     & 1       & WT \end{array}
          \right), \quad
Y_D  =  \left(\begin{array}{ccc} WT^2\xi^2 & WT^2\xi & T^3\xi  \\
                                   WT^2\xi   & WT^2    & T^3     \\
                                   W^3T & W^3T\xi  & WT \end{array}
          \right)
\end{equation}
and the charged lepton Yukawa matrix:
\begin{equation}
Y_E  =  \left(\begin{array}{ccc} WT^2\xi^2 & W^3T^2 & W^2T\xi^3 \\
                                   WT^2\xi^5 & WT^2    & W\xi \\
                                   W^4\xi^2 & W^3T\xi  & WT \end{array}
          \right).
\label{eq:yukabel2}
\end{equation}
We still obtain a good order of magnitude phenomenology for the charged
fermion masses and quark mixing angles, similar to the original
AGUT fit \cite{bled2001}, with the following VEVs in Planck units:
\begin{equation}
<W> = 0.179, \qquad <T> = 0.071 \qquad <\xi> = 0.099.
\end{equation}

The unitary matrix $U_E$ which diagonalises $Y_EY_E^{\dagger}$
is then given by
\begin{equation}
\left( \begin{array}{ccc}
1 & \frac{W\xi^4}{T^3} & W \xi^3  \\
-\frac{W\xi^4}{T^3\sqrt{1+\frac{\xi^2}{T^2}}} &
     \frac{1}{\sqrt{1+\frac{\xi^2}{T^2}}}
     & \frac{\xi}{T\sqrt{1+\frac{\xi^2}{T^2}}} \\
\frac{W\xi^5}{T^4\sqrt{1+\frac{\xi^2}{T^2}}}
     & -\frac{\xi}{T\sqrt{1+\frac{\xi^2}{T^2}}} &
    \frac{1}{\sqrt{1+\frac{\xi^2}{T^2}}}
    \end{array}\right)\\
 \sim  \left(\hspace{-2pt} \begin{array}{ccc}
1 & 0.05& 1.7\times 10^{-4} \\
- 0.03& 0.58 & 0.81 \\
0.04 & - 0.81 & 0.58 \end{array} \hspace{-2pt}\right)
\end{equation}
As we can see from the structure of $U_E$, we naturally
obtain the large $\mu-\tau$ mixing required for the atmospheric
neutrinos. We can now obtain suitable mixing and a
suitable hierarchy of neutrino masses for a vacuum oscillation
solution to the solar neutrino problem, by making the following
choice of charges for a weak iso-triplet
Higgs field, $\Delta$.
\begin{equation}
\vec{Q}_{\Delta} = (\frac{1}{2}, \frac{1}{3}, -\frac{5}{6}, 0).
\end{equation}
We then have the neutrino mass matrix,
\begin{equation}
M_{\nu} \sim <\Delta^0> \left (
\begin{array}{ccc}
W\xi^6 & W\xi^3 & WT\xi^2\\
W \xi^3 & W & WT\xi\\
WT\xi^2 & WT\xi & WT^2\xi \end{array} \right)
\end{equation}
This has the hierarchy,
\begin{eqnarray}
\Delta m^2_{12} & \sim & \Delta m^2_{23},\\
\frac{\Delta m^2_{13}}{\Delta m^2_{12}} & \sim & 2 T^3 \xi^3 \sim
7 \times 10^{-7},
\end{eqnarray}
which is just suitable for the atmospheric neutrinos and the
vacuum oscillation solution to the solar neutrino problem. The
electron neutrino mixing is also large enough for the vacuum
oscillation solution to the solar neutrino problem, as we
can see from the matrix $U_{\nu}$ which diagonalises $M_{\nu}$:
\begin{equation}
U_{\nu} \sim \left (
\begin{array}{ccc} \frac{1}{\sqrt{2}}(1 + \frac{T}{4\xi}) & \xi^3
&
\frac{1}{\sqrt{2}}(1 - \frac{T}{4\xi}) \\
\frac{T\xi}{\sqrt{2}}(1 - \frac{T}{4\xi}) & 1 &
-\frac{T\xi}{\sqrt{2}}(1 + \frac{T}{4\xi}) \\
-\frac{1}{\sqrt{2}}(1 - \frac{T}{4\xi}) & T\xi &
\frac{1}{\sqrt{2}}(1 + \frac{T}{4\xi})
\end{array} \right).
\end{equation}
Hence we have the lepton mixing matrix,
\begin{eqnarray}
U = U_E^{\dagger} U_{\nu} & \sim & \left( \begin{array}{ccc}
\frac{1}{\sqrt{2}}(1 + \frac{T}{4\xi}) &
\frac{W \xi^4}{T^3 \sqrt{1+\frac{\xi^2}{T^2}}} &
\frac{1}{\sqrt{2}(1 - \frac{T}{4\xi})}\\
\frac{\xi(1-\frac{T}{4\xi})}{\sqrt{2}T\sqrt{1+\frac{\xi^2}{T^2}}}
& \frac{1}{\sqrt{1+\frac{\xi^2}{T^2}}}
    & -\frac{\xi(1+\frac{T}{4\xi})}{T\sqrt{2(1+\frac{\xi^2}{T^2})}}\\
-\frac{1-\frac{T}{4\xi}}{\sqrt{2(1+\frac{\xi^2}{T^2})}} &
\frac{\xi}{T \sqrt{1+\frac{\xi^2}{T^2}}} &
\frac{1+\frac{T}{4\xi}}{\sqrt{2(1+\frac{\xi^2}{T^2})}}
\end{array} \right)\\
& \sim & \left( \begin{array}{ccc}
0.83 & 2.8 \times 10^{-2} & 0.58\\
0.47 & 0.58 & -0.68\\
-0.34 & 0.813 & 0.49
\end{array} \right).
\end{eqnarray}
If we take $<\Delta^0> \sim 0.18 \ \mbox{eV}$, then we have
\begin{eqnarray}
m_1 & \sim & <\Delta^0> \left(-W T \xi^2 + \frac{T^2 \xi W}{2}
\right)
\sim - 1.4 \times 10^{-5} \ \mbox{eV} \nonumber \\
m_2 & \sim & <\Delta^0> W \sim 3.2 \times 10^{-2}
\ \mbox{eV} \nonumber \\
m_3 & \sim & <\Delta^0> \left(W T \xi^2 + \frac{T^2 \xi W}{2}
\right) \sim  3 \times 10^{-5} \ \mbox{eV}.
\end{eqnarray}
The $\Delta m^2$ and mixing angle for the solar neutrinos
are given by
\begin{equation}
\Delta m^2_{13} \sim 7 \times 10^{-10} \ \mbox{eV}^2,
\qquad \sin^2 2\theta_{\odot} \sim 4 U_{e1}^2 U_{e3}^2 \sim 0.93
\end{equation}
which are compatible with vacuum oscillations for the solar
neutrinos. Similarly the $\Delta m^2$ and mixing angle for
atmospheric neutrinos are given by,
\begin{equation}
\Delta m^2_{23} \sim 1 \times 10^{-3} \ \mbox{eV}^2,
\qquad \sin^2 2\theta_{atm} \sim 0.93.
\end{equation}
Hence we can see that we have large (but not maximal) mixing for
both the solar and atmospheric neutrinos. We also note that the 
CHOOZ electron survival probability bound is readily satisfied by
$U_{e2} \sim 0.028 < 0.16$, which is the relevant mixing matrix element
since $\Delta m_{12}^2 \sim \Delta m_{23}^2 \gg \Delta m_{13}^2$.
Thus we obtain a good order of magnitude fit (agreeing with the 
data to within a factor of 2) to the 17 measured fermion mass and 
mixing angle variables with just 4 free parameters ($W$, $T$, 
$\xi$ and $\Delta^0$), but assuming a vacuum oscillation solution to 
the solar neutrino problem.

We note that we have really only used the abelian gauge quantum
numbers to generate a realistic spectrum of fermion masses; the
non-abelian representations are determined by imposing the usual
SM charge quantisation rule for each of the SMG factors in the
gauge group. Furthermore two of the $U(1)$s (more precisely
two linear combinations of the $U(1)$s) in the gauge group are
spontaneously broken by the Higgs fields $S$ and $\psi$, which
have VEVs $<S> = <\psi> = 1$. Hence these $U(1)$s play
essentially no part in obtaining the spectrum of fermion
masses and mixings. This means that we can construct a model
based on the gauge group $SMG \times U(1)^{\prime}$ with the same
fermion spectrum as above. However it turns out \cite{mg}
that some of the quarks and leptons must have extremely
large (integer) $U(1)^{\prime}$ charges, making
this reduced model rather unattractive. Also three Higgs fields
$W$, $T$ and $\xi$ are responsible for the spontaneous breakdown
to the SM gauge group, $SMG \times U(1)^{\prime} \rightarrow
SMG$, and they have large relatively prime $U(1)^{\prime}$
charges. So we prefer the better motivated $SMG^3\times U(1)_f$
AGUT model.

\section{The $(SMG \times U(1)_{B-L})^3$ Model}

In this extended AGUT model we introduce a right-handed neutrino
and a gauged $B-L$ charge for each family with the associated
abelian gauge groups $U(1)_{\cdfsss B-L,i}$ ($i=1,2,3$). The $U(1)_f$
abelian factor of the $SMG \times U(1)_f$ model in the previous
section gets absorbed as a linear combination of the $B-L$ charge
and the weak hypercharge abelian gauge groups for the different
families (or generations). It is these 6 abelian gauge charges
which are responsible for generating the fermion mass hierarchy
and we list their values in Table \ref{CDFTable1} for the 48 Weyl
proto-fermions in the model.
\begin{table}[!ht]
\caption{All $U(1)$ quantum charges for the proto-fermions in the
$(SMG \times U(1)_{B-L})^3$ model.} \vspace{3mm} \label{CDFTable1}
\begin{center}
\begin{tabular}{|c||c|c|c|c|c|c|} \hline
& $SMG_1$& $SMG_2$ & $SMG_3$ & $U_{\cdfsss B-L,1}$ & $U_{\cdfsss B-L,2}$
& $U_{\cdfsss B-L,3}$ \\ \hline\hline
$u_L,d_L$ &  $\frac{1}{6}$ & $0$ & $0$ & $\frac{1}{3}$ & $0$ & $0$ \\
$u_R$ &  $\frac{2}{3}$ & $0$ & $0$ & $\frac{1}{3}$ & $0$ & $0$ \\
$d_R$ & $-\frac{1}{3}$ & $0$ & $0$ & $\frac{1}{3}$ & $0$ & $0$ \\
$e_L, \nu_{e_{\cdfsss L}}$ & $-\frac{1}{2}$ & $0$ & $0$ & $-1$ & $0$ & $0$ \\
$e_R$ & $-1$ & $0$ & $0$ & $-1$ & $0$ & $0$ \\
$\nu_{e_{\cdfsss R}}$ &  $0$ & $0$ & $0$ & $-1$ & $0$ & $0$ \\ \hline
$c_L,s_L$ & $0$ & $\frac{1}{6}$ & $0$ & $0$ & $\frac{1}{3}$ & $0$ \\
$c_R$ &  $0$ & $\frac{2}{3}$ & $0$ & $0$ & $\frac{1}{3}$ & $0$ \\
$s_R$ & $0$ & $-\frac{1}{3}$ & $0$ & $0$ & $\frac{1}{3}$ & $0$\\
$\mu_L, \nu_{\mu_{\cdfsss L}}$ & $0$ & $-\frac{1}{2}$ & $0$ & $0$ & $-1$ &
$0$\\ $\mu_R$ & $0$ & $-1$ & $0$ & $0$  & $-1$ & $0$ \\
$\nu_{\mu_{\cdfsss R}}$ &  $0$ & $0$ & $0$ & $0$ & $-1$ & $0$ \\ \hline
$t_L,b_L$ & $0$ & $0$ & $\frac{1}{6}$ & $0$ & $0$ & $\frac{1}{3}$ \\
$t_R$ &  $0$ & $0$ & $\frac{2}{3}$ & $0$ & $0$ & $\frac{1}{3}$ \\
$b_R$ & $0$ & $0$ & $-\frac{1}{3}$ & $0$ & $0$ & $\frac{1}{3}$\\
$\tau_L, \nu_{\tau_{\cdfsss L}}$ & $0$ & $0$ & $-\frac{1}{2}$ & $0$ & $0$ &
$-1$\\ $\tau_R$ & $0$ & $0$ & $-1$ & $0$ & $0$ & $-1$\\
$\nu_{\tau_{\cdfsss R}}$ &  $0$ & $0$ & $0$ & $0$ & $0$ & $-1$ \\
\hline \hline
\end{tabular}
\end{center}
\end{table}
The see-saw scale for the right-handed neutrinos is introduced via
the VEV of a new Higgs field $\phi_{\cdfsss SS}$. However, in order
to get an LMA-MSW solution to the solar neutrino problem, we have
to replace \cite{CDFfnt} the AGUT Higgs fields $S$ and $\xi$ by two
new Higgs fields $\rho$ and $\omega$. The abelian gauge quantum
numbers of the new system of Higgs fields for the $(SMG \times
U(1)_{B-L})^3$ model are given in Table \ref{qc}.
\begin{table}[!th]
\caption{All $U(1)$ quantum charges of the Higgs fields in the
$(SMG \times U(1)_{B-L})^3$ model.} \vspace{3mm} \label{qc}
\begin{center}
\begin{tabular}{|c||c|c|c|c|c|c|} \hline
& $SMG_1$& $SMG_2$ & $SMG_3$ & $U_{\cdfsss B-L,1}$ & $U_{\cdfsss B-L,2}$
& $U_{\cdfsss B-L,3}$ \\ \hline\hline
$\omega$ & $\frac{1}{6}$ & $-\frac{1}{6}$ & $0$ & $0$ & $0$ & $0$\\
$\rho$ & $0$ & $0$ & $0$ & $-\frac{1}{3}$ & $\frac{1}{3}$ & $0$\\
$W$ & $0$ & $-\frac{1}{2}$ & $\frac{1}{2}$ & $0$ & $-\frac{1}{3}$
& $\frac{1}{3}$ \\
$T$ & $0$ & $-\frac{1}{6}$ & $\frac{1}{6}$ & $0$ & $0$ & $0$\\
$\phi_{\cdfsss WS}$ & $0$ & $\frac{2}{3}$ & $-\frac{1}{6}$ & $0$
& $\frac{1}{3}$ & $-\frac{1}{3}$ \\
$\phi_{\cdfsss SS}$ & $0$ & $1$ & $-1$ & $0$ & $2$ & $0$ \\
\hline
\end{tabular}
\end{center}
\end{table}

As can be seen from Table~\ref{qc}, the fields $\omega$ and $\rho$
have only non-trivial quantum numbers with respect to the first
and second families. This choice of quantum numbers makes it
possible to express a fermion mass matrix element involving the
first family in terms of the corresponding element involving the
second family, by the inclusion of an appropriate product of
powers of $\rho$ and $\omega$. With the system of quantum numbers
in Table~\ref{qc} one can easily evaluate, for a given mass matrix
element, the numbers of Higgs field VEVs of the different types
needed to perform the transition between the corresponding left-
and right-handed Weyl fields. The results of calculating the
products of Higgs fields needed, and thereby the order of
magnitudes of the mass matrix elements in our model, are presented
in the following mass matrices (where, for clarity, we distinguish
between Higgs fields and their hermitian conjugates):

\noindent the up-type quarks:
\begin{eqnarray}
M_{\cdfsss U} \simeq \frac{\cdfsVEV{(\phi_{\cdfsss\rm WS})^\dagger}}
{\sqrt{2}}\hspace{-0.1cm}
\left(\!\begin{array}{ccc}
        (\omega^\dagger)^3 W^\dagger T^2
        & \omega \rho^\dagger W^\dagger T^2
        & \omega \rho^\dagger (W^\dagger)^2 T\\
        (\omega^\dagger)^4 \rho W^\dagger T^2
        &  W^\dagger T^2
        & (W^\dagger)^2 T\\
        (\omega^\dagger)^4 \rho
        & 1
        & W^\dagger T^\dagger
\end{array} \!\right)\label{CDFM_U}
\end{eqnarray}
\noindent
the down-type quarks:
\begin{eqnarray}
M_{\cdfsss D} \simeq \frac{\cdfsVEV{\phi_{\cdfsss\rm WS}}}{\sqrt{2}}
\hspace{-0.1cm}
\left (\!\begin{array}{ccc}
        \omega^3 W (T^\dagger)^2
      & \omega \rho^\dagger W (T^\dagger)^2
      & \omega \rho^\dagger T^3 \\
        \omega^2 \rho W (T^\dagger)^2
      & W (T^\dagger)^2
      & T^3 \\
        \omega^2 \rho W^2 (T^\dagger)^4
      & W^2 (T^\dagger)^4
      & W T
                        \end{array} \!\right) \label{CDFM_D}
\end{eqnarray}
\noindent %
the charged leptons:
\begin{eqnarray}
M_{\cdfsss E} \simeq \frac{\cdfsVEV{\phi_{\cdfsss\rm WS}}}{\sqrt{2}}
\hspace{-0.1cm}
\left(\hspace{-0.1 cm}\begin{array}{ccc}
    \omega^3 W (T^\dagger)^2
  & (\omega^\dagger)^3 \rho^3 W (T^\dagger)^2
  & (\omega^\dagger)^3 \rho^3 W^4 (T^\dagger)^5  \\
    \omega^6 (\rho^\dagger)^3  W (T^\dagger)^2
  &   W (T^\dagger)^2
  &   W^4 (T^\dagger)^5 \\
    \omega^6 (\rho^\dagger)^3  (W^\dagger)^2 T^4
  & (W^\dagger)^2 T^4
  & WT
\end{array} \hspace{-0.1cm}\right) \label{CDFM_E}
\end{eqnarray}
\noindent
the Dirac neutrinos:
\begin{eqnarray}
M^D_\nu \simeq \frac{\cdfsVEV{(\phi_{\cdfsss\rm WS})^\dagger}}
{\sqrt{2}}\hspace{-0.1cm}
\left(\hspace{-0.1cm}\begin{array}{ccc}
        (\omega^\dagger)^3 W^\dagger T^2
        & (\omega^\dagger)^3 \rho^3 W^\dagger T^2
        & (\omega^\dagger)^3 \rho^3 W^2 (T^\dagger)^7 \\
        (\rho^\dagger)^3 W^\dagger T^2
        &  W^\dagger T^2
        & W^2 (T^\dagger)^7 \\
        (\rho^\dagger)^3 (W^\dagger)^4 T^8
        &   (W^\dagger)^4 T^8
        & W^\dagger T^\dagger
\end{array} \hspace{-0.1 cm}\right)\label{CDFMdirac}
\end{eqnarray}
\noindent %
and the Majorana (right-handed) neutrinos:
\begin{eqnarray}
M_R \simeq \cdfsVEV{\phi_{\cdfsss\rm SS}}\hspace{-0.1cm}
\left (\hspace{-0.1 cm}\begin{array}{ccc}
(\rho^\dagger)^6 T^6
& (\rho^\dagger)^3 T^6
& (\rho^\dagger)^3 W^3 (T^\dagger)^3 \\
(\rho^\dagger)^3 T^6
& T^6 & W^3 (T^\dagger)^3 \\
(\rho^\dagger)^3 W^3 (T^\dagger)^3 & W^3 (T^\dagger)^3
& W^6 (T^\dagger)^{12}
\end{array} \hspace{-0.1 cm}\right ) \label{CDFMmajo}
\end{eqnarray}

Then the light neutrino mass matrix -- effective left-left
transition Majorana mass matrix -- can be obtained via the see-saw
mechanism~\cite{CDFseesaw}:
\begin{equation}
  \label{CDFeq:meff}
  M_{\rm eff} \! \approx \! M^D_\nu\,M_R^{-1}\,(M^D_\nu)^T\cdfnn.
\end{equation}
with an appropriate renormalisation group running from the Planck 
scale to the see-saw scale and then to the electroweak scale. The
experimental quark and lepton masses and mixing angles in Table
\ref{CDFconvbestfit} can now be fitted, by varying just 5 Higgs field
VEVs and averaging over a set of complex order unity random 
numbers, which multiply all the independent mass matrix
elements. The best fit is obtained with the following values for
the VEVs:
\begin{eqnarray}
\label{CDFeq:VEVS} &&\cdfsVEV{\phi_{\cdfsss SS}}=5.25\times10^{15}~\cdfGeV\cdfnn,
\cdfnn\cdfsVEV{\omega}=0.244\cdfnn, \cdfnn\cdfsVEV{\rho}=0.265\cdfnn,\nonumber\\
&&\cdfnn\cdfsVEV{W}=0.157\cdfnn, \cdfnn\cdfsVEV{T}=0.0766\cdfnn,
\end{eqnarray}
where, except for the Higgs field $\cdfsVEV{\phi_{\cdfsss SS}}$, the
VEVs are expressed in Planck units. The resulting 5 parameter 
order of magnitude fit, with an LMA-MSW solution to the 
solar neutrino problem, is presented
in Table \ref{CDFconvbestfit}.

\begin{table}[!t]
\caption{Best fit to conventional experimental data. All masses
are running masses at $1~\cdfGeV$ except the top quark mass which is
the pole mass.}
\begin{displaymath}
\begin{array}{|c|c|c|}
\hline\hline
 & {\rm Fitted} & {\rm Experimental} \\ \hline
m_u & 4.4~\cdfMeV & 4~\cdfMeV \\
m_d & 4.3~\cdfMeV & 9~\cdfMeV \\
m_e & 1.6~\cdfMeV & 0.5~\cdfMeV \\
m_c & 0.64~\cdfGeV & 1.4~\cdfGeV \\
m_s & 295~\cdfMeV & 200~\cdfMeV \\
m_{\mu} & 111~\cdfMeV & 105~\cdfMeV  \\
M_t & 202~\cdfGeV & 180~\cdfGeV \\
m_b & 5.7~\cdfGeV & 6.3~\cdfGeV \\
m_{\tau} & 1.46~\cdfGeV & 1.78~\cdfGeV \\
V_{us} & 0.11 & 0.22 \\
V_{cb} & 0.026 & 0.041 \\
V_{ub} & 0.0027 & 0.0035 \\ \hline
\Delta m^2_{\odot} & 9.0 \times 10^{-5}~\cdfeV^2 &  5.0 \times 10^{-5}~\cdfeV^2 \\
\Delta m^2_{\rm atm} & 1.7 \times 10^{-3}~\cdfeV^2 &  2.5 \times 10^{-3}~\cdfeV^2\\
\tan^2\theta_{\odot} &0.26 & 0.34\\
\tan^2\theta_{\rm atm}& 0.65 & 1.0\\
\tan^2\theta_{\rm chooz}  & 2.9 \times 10^{-2} & \cdfsleq~2.6 \times 10^{-2}\\
\hline\hline
\end{array}
\end{displaymath}
\label{CDFconvbestfit}
\end{table}

Transforming from $\tan^2\theta$ variables to $\sin^22\theta$
variables, our predictions for the neutrino mixing angles become:
\begin{equation}
  \label{CDFeq:sintan}
 \sin^22\theta_{\odot} = 0.66\cdfnn, \quad
 \sin^22\theta_{\rm atm} = 0.96\cdfnn, \quad
 \sin^22\theta_{\rm chooz} = 0.11\cdfnn.
\end{equation}
Note that our fit to the CHOOZ mixing angle lies close to the 
$2\sigma$ Confidence Level
experimental bound. We also give here our predicted hierarchical
left-handed neutrino masses ($m_i$) and the right-handed neutrino
masses ($M_i$) with mass eigenstate indices ($i=1,2,3$):
\begin{equation}
m_1 =  1.4\times10^{-3}~~\cdfeV\cdfnn, \quad
M_1 =  1.0\times10^{6}~~\cdfGeV\cdfnn,
\label{CDFeq:neutrinomass1}
\end{equation}
\begin{equation}
m_2 =  9.6\times10^{-3}~~\cdfeV\cdfnn, \quad
M_2 =  6.1\times10^{9}~~\cdfGeV\cdfnn,
\label{CDFeq:neutrinomass2}
\end{equation}
\begin{equation}
m_3 =  4.2\times10^{-2}~~\cdfeV\cdfnn, \quad
M_3 =  7.8\times10^{9}~~\cdfGeV\cdfnn.
\label{CDFeq:neutrinomass3}
\end{equation}

\section*{Acknowledgements}

I should again like to thank my collaborators, Mark Gibson,
Holger Bech Nielsen and Yasutaka Takanishi, for many
discussions.

\backmatter
\thispagestyle{empty}
\parindent=0pt
\begin{flushleft}
\mbox{}
\vfill
\vrule height 1pt width \textwidth depth 0pt
{\parskip 6pt

{\sc Blejske Delavnice Iz Fizike, \ \ Letnik~3, \v{s}t. 4,} 
\ \ \ \ ISSN 1580--4992

{\sc Bled Workshops in Physics, \ \  Vol.~3, No.~4}

\bigskip

Zbornik delavnic `What comes beyond the Standard model', 2000, 2001 in 2002

Zvezek 2: Zbornik 5.delavnice `What comes beyond the Standard model', 
Bled, 13.~-- 24.~julij 2002

Proceedings to the workshops 'What comes beyond the Standard model', 
2000, 2001, 2002

Volume 2: Proceedings to the 5th Workshop 
`What comes beyond the Standard model', 
Bled, July 13--24,  2002

\bigskip

Uredili Norma Manko\v c Bor\v stnik, Holger Bech Nielsen, 
Colin D. Froggatt in Dragan Lukman 

Publikacijo sofinancira Ministrstvo za \v solstvo, znanost in \v sport 

Tehni\v{c}ni urednik Vladimir Bensa

\bigskip

Zalo\v{z}ilo: DMFA -- zalo\v{z}ni\v{s}tvo, Jadranska 19,
1000 Ljubljana, Slovenija

Natisnila Tiskarna MIGRAF v nakladi 100 izvodov

\bigskip

Publikacija DMFA \v{s}tevilka 1517

\vrule height 1pt width \textwidth depth 0pt}
\end{flushleft}


\end{document}